\documentclass{aa}
\usepackage{txfonts}
\usepackage{natbib}
\bibpunct{(}{)}{;}{a}{}{,}
\usepackage{graphicx}
\usepackage{placeins}
\usepackage[]{rotating}
\usepackage{cases}

\newcommand{\teff}{$T_{\rm{eff}}$}

\newcommand{\lL}{\ifmmode \log \frac{L}{L_{\sun}} \else $\log \frac{L}{L_{\sun}}$\fi}

\newcommand{\myr}{M$_{\sun}$ yr$^{-1}$}

\newcommand{\vinf}{$\varv_{\infty}$}

\newcommand{\kms}{km~s$^{-1}$}
\newcommand{\msun}{M$_{\sun}$}
\newcommand{\zsun}{Z$_{\sun}$}
\newcommand{\lya}{Ly$\alpha$}
\newcommand{\heiiuv}{$\ion{He}{ii}$~1640}
\newcommand{\heiiopt}{$\ion{He}{ii}$~4686}
\newcommand{\civopt}{\ion{C}{iv}~5802-12}
\newcommand{\civuv}{$\ion{C}{iV}$~1550}
\newcommand{\nivuv}{$\ion{N}{iv}$~1720}
\newcommand{\nivuva}{$\ion{N}{iv}$~1486}
\newcommand{\nvuv}{$\ion{N}{v}$~1240}
\newcommand{\ovuv}{$\ion{O}{v}$~1371}

\begin{document}

\title{Very massive stars at low metallicity: \\
evolution, synthetic spectroscopy, and impact on the integrated light of starbursts}

\author{F. Martins\inst{1}
\and A. Palacios\inst{1}
\and D. Schaerer\inst{2}
\and R. Marques-Chaves\inst{2}
}

\institute{
  LUPM, Univ. Montpellier, CNRS, Montpellier, France 
 \and
 Observatoire de Gen\`{e}ve, Universit\'e de Gen\`{e}ve, Chemin Pegasi 51, 1290 Versoix, Switzerland
}

\offprints{Fabrice Martins\\ \email{fabrice.martins@umontpellier.fr}}

\date{Received / Accepted }

\abstract
{Very massive stars (VMS), with a mass in excess of 100~\msun, are known in the Galaxy and the Large Magellanic Cloud (LMC). They are mostly characterised by their strong stellar winds compared to normal massive stars. Their mass loss rates have been calibrated at the metallicity of the LMC. No constraint exist at other metallicities. }
{We aim to study the spectroscopic appearance of VMS and their effect on the integrated light of starbursts at low metallicity. }
{In absence of empirical constraint, we adopt two frameworks for the mass loss rates of VMS: in one case we assume no metallicity dependence, in the other case we assume a linear scaling with metallicity. Under these assumptions we compute evolutionary models for masses 150, 200, 250 and 300~\msun\ at Z=0.2, 0.1 and 0.01~\zsun. We compute the associated synthetic spectra at selected points along the evolutionary tracks. Finally we build population synthesis models including VMS based on our new VMS models.}
{We find that the evolution of VMS depends critically on the assumptions regarding mass loss rates. In case of no metallicity dependence VMS remain hot for all their lifetime. Conversely when mass loss rates are reduced because of lower metallicity VMS follow a classical evolution towards the red part of the Hertzsprung-Russell diagram. VMS display \heiiuv\ emission in most phases of their evolution, except when they become too cool. This line is present in the integrated light of population synthesis models down to 0.1~\zsun\ whatever the star formation history, and is also sometimes seen at Z=0.01~\zsun. \heiiuv\ is weaker in models that include a metallicity scaling of the mass loss rates. The optical spectra of starbursts, especially the Wolf-Rayet bumps, sometimes display VMS signatures when these stars are present. At low metallicity, adding VMS to population synthesis models produces more ionising photons down to $\sim$45~eV. At higher energy the ionising flux depends on age, metallicity, assumption regarding VMS mass loss rates, and on the very short phases at the end of VMS evolution. \ion{He}{ii} ionising fluxes large enough to produce some amount of nebular \ion{He}{ii}~4686 emission can be produced under specific circumstances. Our models are able to reproduce qualitatively and sometimes also quantitatively the UV spectra of star-forming regions. However we are not able to clearly identify which mass loss framework is favoured. }
{VMS can be identified down to 0.1~\zsun, and potentially to 0.01~\zsun\ depending on the mass loss rates metallicity scaling, through their \heiiuv\ emission. Their detailed evolution at these low metallicities, especially their mass loss rates, can be constrained when more UV spectra of star-forming regions at low metallicity are available.  }


\authorrunning{Martins et al.}
\titlerunning{Very Massive Stars at low metallicity}

\maketitle

\section{Introduction}
\label{s_intro}

The mass of the most massive stars is unset. Based on observations of \citet{feitzinger80}, \citet{cassi81} determined a mass of 2500~\msun\ for the R136 object at the core of giant \ion{H}{ii}  region 30~Doradus (30~Dor), in the Large Magellanic Cloud (LMC). High spatial observations subsequently revealed that R136 is a cluster of stars rather than a unique object. A "canonical" upper mass limit of 150~\msun\ was for some time promoted by \citet{figer05}, before new analysis of the R136 objects established that stars with masses up to 200-300~\msun\ exist in that cluster  \citep{crowther10,besten20a,brands22}. So far these mass estimates appear robust enough against further decomposition of the R136 objects into more pieces \citep{kalari22,sabhahit25}. In particular, binarity is excluded by the most recent Hubble Space Telescope observations \citep{shenar23}. The existence of stars with masses in excess of 100~\msun\ is thus robustly established. These stars are referred to as Very Massive Stars (VMS, \citealt{vink15}). 

Because of the shape of the initial mass function VMS are rare. Outside of R136, individual VMS have been observed in the Galactic clusters NGC~3603 \citep{schnurr08,crowther10} and the Arches \citep{arches}. All clusters have masses in excess of about $10^4$~\msun\ \citep[e.g.][]{stolte06}. A few other relatively isolated objects reach the mass limit to be classified as VMS \citep{hamann06,barniske08,besten11}. In spite of being rare objects, \citet{crowther16} showed that VMS can dominate the integrated ultraviolet (UV) light of clusters. In particular the \heiiuv\ emission of R136 is only produced by the few VMS it hosts. This property can be used to infer the presence of VMS in star-forming regions even if they are not resolved into their individual components \citep{martins23}. The presence of VMS is thus considered likely in the clusters NGC3125-A \citep{wofford14,wofford23}, II~Zw-40-A \citep{leitherer18}, NGC5352-5 \citep{smith16}, and MrK71-A \citep{smith23}. Beyond the Local Universe \citet{upad} explain the morphology of the rest-frame UV spectra of intensely star-forming galaxies by the presence of VMS. \citet{senchyna20} argue that accounting for all stellar emission and nebular lines of starburst galaxies in the Local Universe is only feasible with population synthesis models that include VMS. The most powerful starburst galaxies detected at redshift 2-4 also show spectroscopic features that can be attributed to VMS \citep{marques20,marques21,marques22}. 

VMS efficiently process chemical elements in their interior and release new material in their close environment \citep[e.g.][]{higgins23}. This has raised their interest to explain the properties of Nitrogen emitters which are star-forming regions or galaxies showing a ratio of nitrogen to oxygen abundance higher than commonly seen at a given metallicity. These objects have been known since a couple of decades \citep{villar04,james09,patricio16}, but their existence in the early Universe, as discovered by JWST observations \citep{bunker23,isobe23,ji24,castellano24,pascale23,schaerer24,senchyna24}, has placed them in the front line. The surprising nitrogen content has been tentatively explained by normal Wolf-Rayet (WR) stars, supermassive stars, but also by VMS \citep{cameron23,charbonnel23,higgins23,vink23,marques24}.

Besides their mass, VMS are mostly characterised by their high luminosity and large mass loss rates. The resulting rich spectrum of emission lines makes VMS appear as hydrogen-rich WN stars, although they are in fact main sequence stars for most of their lifetime. In hot massive stars winds are radiatively-driven \citep{cak,kud,puls00} and the associated mass loss rates mostly scale with luminosity. For VMS the mass loss rates determined from spectroscopic analysis do not follow the relation established for lower mass O-type stars: they are stronger, sometimes by nearly a factor of 10 \citep{arches,besten14}. This is explained by the proximity to the Eddington limit \citep{gh08,vink11}. Recent studies have established a relation between the mass loss rates of VMS and the Eddington factor \citep{besten20b,graef21}. These relations are calibrated for the metallicity of the LMC since they rely on observational data in 30~Dor. In a previous study we adopted one such calibration to study the spectroscopic evolution of VMS in the LMC \citep{mp22}. We showed that we could reproduce the observed features and that the inclusion of VMS spectra in population synthesis models produced for the first time the observed \heiiuv\ emission of young star-forming regions. \citet{schaerer25} used these models to quantify the impact of VMS on the ionising properties of star-forming regions.

The behaviour of VMS mass loss rates at different metallicities is unknown. The number of VMS observed in the Galaxy is small, and they are presently not seen at sub-LMC metallicity because of observational limitations (distance, spatial resolution of current observing facilities). Consequently no scaling relation for their mass loss rates exists outside of the LMC. Theoretical models for the winds of VMS do not exist at present. Predictions for hydrogen-poor and less massive Wolf-Rayet stars indicate a complex relation between mass loss, metallicity and Eddington parameter \citep{sander20b}. These models are only partly relevant for VMS which are usually hydrogen-rich. Large hydrogen fractions are known to affect line-driving \citep{gh08}. \citet{sabhahit23} developed a framework to predict VMS mass loss rates at different metallicities, but empirical confirmation of the predicted trends is pending. In the present study we explore different assumptions regarding VMS mass loss rates to investigate their spectroscopic appearance at low metallicity. 

The remaining of the paper is organised as follows. Sect.~\ref{s_method} describes our method and assumptions. In Sect.~\ref{s_res} we describe our results in terms of evolutionary models and synthetic spectra. The latter are used in Sect.~\ref{s_popsyn} to build population synthesis models that are tested against observations of low metallicity star-forming regions. We discuss our findings in Sect.~\ref{s_disc} and conclude in Sect.~\ref{s_conc}.

\section{Method}
\label{s_method}

The method we use is the same as that presented in \citet{mp22}. We first compute evolutionary models with mass loss rate recipes adapted for VMS. We then compute atmospheric models and synthetic spectra along the evolutionary tracks. Both evolutionary and atmospheric models are consistent in the sense that the output of evolutionary models are used as inputs for the computation of atmospheric models. The resulting synthetic spectra are subsequently used to build population synthesis models that include VMS. 

We perform our computations for three different metallicities: 0.2, 0.1 and 0.01~\zsun. This choice is mostly driven to cover the range of metallicity of local and high redshift star-forming galaxies. Fig.~6 of \citet{vanzella23} illustrates that Z varies between $\sim$0.5 and 0.02~\zsun\ for a compilation of these objects, with a clear floor at 0.01~\zsun\footnote{The measurements reported in \citet{vanzella23} are obtained by the electron temperature-direct method and the uncertainties on the derived 12+log(O/H) are of the order 0.1-0.2~dex.}. Recent JWST observations \citep[e.g.][]{nakajima23,chemerynska24} indicate that metallicities below 1\% of the solar metallicity are so far not observed, except in very rare candidates \citep{vanzella23,fujimoto25}. We provided models at Z=0.4~\zsun\ in \citet{mp22}. In the present study we aim at sampling the metallicity range described above, with 0.2~\zsun\ corresponding roughly to the Small Magellanic Cloud value and to the bulk of the samples shown in \citet{vanzella23}, 0.1~\zsun\ being a representative value of low metallicity star-forming galaxies and 0.01~\zsun\ setting the lower limit of the metallicity range.

\subsection{Mass loss recipe at low Z}
\label{s_mdotZ}

The first step is to select the mass loss recipe relevant for VMS mass loss rates at low metallicity. As described in Sect.~\ref{s_intro} no empirical recipe exists, and theoretical predictions are either partial or uncalibrated. We thus adopt a pragmatic approach.

 Empirical studies have shown that the winds of normal WR stars vary with Z. \citet{crowther02} report a reduction of mass loss rates of WC stars in the LMC compared to Galactic counterparts. The analysis of Galactic, LMC and SMC WR stars by \citet{sander12,sander14,hainich14,hainich15,hamann19} reveal a global metallicity dependence of the form $\dot{M} \propto Z^{1.2\pm0.1}$. But other parameters controlling the wind strength, such as the Eddington factor and/or the mass to light ratio, are hidden in this global scaling. For the specific case of VMS, \citet{smith23} present UV spectroscopy of the massive cluster A in the galaxy MrK~71. Its metallicity is estimated to be 0.16~\zsun\ from nebular lines \citep{chen23}. The weakness of $\ion{C}{iv}$~1550 confirms a relatively low metal content. However \heiiuv\ appears to be strong with a shape similar to that of VMS in R136. From this observation \cite{smith23} conclude that the winds of VMS do not depend on Z.
\citet{sabhahit23} present a theoretical framework for mass loss of VMS at low metallicity. They describe how the transition between normal and boosted mass loss rate should change with Z. No empirical confirmation of this scheme has been performed so far. \citet{sander20b} performed hydrodynamical simulations of stellar winds of WR stars with no hydrogen and studied the dependence of the mass loss with the Eddington factor ($\Gamma_e$) and the metal content. They find that for a given $\Gamma_e$ the metallicity dependence is weak, but that the transition $\Gamma_e$ from thin to thick winds varies strongly with Z, implying that low Z WR stars with relatively small $\Gamma_e$ have reduced mass loss rates. However, as demonstrated early on by \citet{gh08} mass loss rates of VMS depend sensitively on the hydrogen mass fraction.

In this context we decided to adopt a pragmatic and empirical approach for the present study. Since there is no clear mass loss recipe for VMS at metallicities lower than that of the LMC, we simply assumed two representative cases. In the first one, we follow \citet{smith23} and assume no Z-dependence. We thus adopt the recipe of \citet{graef21} as in our previous study at Z=1/2.5~\zsun\ \citep{mp22}. In the second case, we assume a Z scaling close to that pointed out by \citet{hainich15} for classical WR stars, i.e. we scale the mass loss rates of \citet{graef21} according to $\dot{M} \propto (\frac{Z}{Z_{LMC}})^{1.0}$, since the recipe of \citet{graef21} is calibrated for metallicity of the LMC. Regarding the transition between optically thin and thick winds, we stick to the recipe of \citet{graef21} and assume no variation with metallicity. Our models should thus produce two relatively different sets of spectroscopic sequences for VMS at low Z that can be compared to observed spectra of star-forming regions hosting VMS.

\subsection{Stellar evolution models}
\label{s_evmod}
 As in our previous work \citep{mp22}, we computed stellar evolution models with the STAREVOL code adopting the following physical ingredients: a grey atmosphere as outer boundary to the stellar structure equations, mixing length theory to treat the energy transport in convective regions with $\alpha_{MLT} = 1.6304$, step core overshooting with $\alpha_{ov} = 0.1 H_p$, and $H_p$ the pressure scale height, and solar scaled chemical compositions for Z = 0.002216 (0.2 Z$_\odot$), Z = 0.0013446 (0.1 Z$_\odot$) and Z = 0.00013446 (0.01 Z$_\odot$) with respect to the solar reference chemical composition by \citet{asplund09}. No $\alpha$-elements enhancement was considered\footnote{At the metallicities considered in the present work, [$\alpha$/Fe] is expected to be enhanced and $\approx 0.3$ dex according to the recent studies in the Milky Way. We thus performed a test and checked that increasing the $\alpha$-elements abundance to [$\alpha$/Fe] = +0.3 dex has a very modest impact on the evolutionary tracks hence the spectral appearance of the stars (\teff\ varies at most by 1000~K on the main sequence and L is barely affected).}. We used the OPAL opacity tables for the corresponding set of solar-scaled abundances, and nuclear rates extracted from the NetGen server and including NACRE II rates for the light elements up to oxygen \citep{xu13}.
 
 The mass loss recipe was from \citet{graef21} as in \citet{mp22} with a different treatment of the optically thin and thick winds and a metallicity-dependence of the optically thick winds as follows:
 
\begin{eqnarray}
 \log(\dot{M}_{\rm thin}) = & -& 6.697 ~(\pm 0.061)\\ \nonumber
 &+& 2.194 ~(\pm 0.021) \times \log\left(\frac{L_*}{10^5}\right)\\\nonumber
&-&1.313 ~(\pm 0.046) \times  \log\left(\frac{M_*}{30}\right)\\\nonumber
&-&1.226 ~(\pm 0.037) \times \log\left(\frac{v_\infty/v_{esc}}{2}\right)\\\nonumber
&+&0 .933 ~(\pm 0.064) \times \log\left(\frac{T_{\rm eff}}{40 000}\right)\\\nonumber
&-&10.92 ~(\pm 0.90) \times {\log\left(\frac{T_{\rm eff}}{40 000}\right)}^2\\\nonumber
&+&0.85 ~(\pm 0.10) \times \log\left(\frac{Z}{Z_\odot}\right)\\
\log(\dot{M}_{\rm thick}) &=& \Biggl(5.22 \log(\Gamma_{\rm e}) - 0.5\log(D) -2.6\Biggr) \left(\frac{Z}{Z_{\rm LMC}}\right)^{x}
 \end{eqnarray}
 
where $\dot{M}$ is in $M_\odot$/yr, $L_*$ and $M_*$ are the luminosity and mass of the star in solar units, $T_{\rm eff}$ is the effective temperature in Kelvin, the ratio between terminal and escape velocities $v_\infty /v_{\rm esc}$ is taken equal to 2.6 as in Eq. (24) of \cite{vink01}, Z$_\odot = 0.019$, $D = 10 $ the wind clumping factor, $x$ is taken equal to 0 (mass loss rate independent of metallicity) or to 1 (Z-scale for the mass loss rates of VMS), and $\Gamma_{\rm e}$ is the Eddington factor:
\[\Gamma_e = 10^{-4.813}  (1+X) \frac{L}{L_\odot}\frac{M_\odot}{M}\]
The switch between the two mass loss regimes is assumed to occur under the same circumstances as those prescribed by \citet{graef21} for the LMC, that is to say :
\begin{eqnarray*}
~~~\text{if} ~\tau_s < \frac{2}{3} &\text{then} &\text{optically thin regime} \rightarrow  \dot{M} \equiv \dot{M}_{\rm thin}\\
~~~\text{if}~ \tau_s > 1 &\text{then} &\text{optically thick regime} \rightarrow \dot{M} \equiv \dot{M}_{\rm thick}
\end{eqnarray*}
with $\tau_s$ the sonic-point optical depth $\tau_s$ in the wind which, for a stellar model with surface mass $M$, luminosity $L$, radius $R$, and hydrogen mass fraction $X$, writes :
 \[
 \tau_s \approx \frac{\dot{M} v_\infty}{L/c} \left(1 + \frac{v^2_{\rm esc}}{v^2_\infty}\right).
 \]
  In the intermediate regime ($2/3 \leq \tau_s \leq 1$), the mass loss is obtained by a linear interpolation between $\dot{M}_{\rm thin}$ and $\dot{M}_{\rm thick}$ to ensure a smooth transition.
 
We focused on models with initial masses 150, 200, 250 and 300 M$_\odot$ at these metallicities and for each mass, we computed the evolution from the zero-age main sequence (ZAMS) to the end of core He burning for most models\footnote{Models at Z = 0.2 Z$_\odot$ are only evolved up to the beginning of the core He fusion phase due to numerical difficulties.
However this does not affect the results of the present work since these models have already reached 2.5 Myr, corresponding to more than 90\% of their total evolutionary lifetime. We do not consider more advanced phases in the present work (see also Sect.~\ref{s_postMS}).} including mass loss without Z-scaling in the optically thick wind regime. This represents more than 99\% of the total lifetime of the modelled stars.

\subsection{Atmosphere models}
\label{s_atmod}

We computed atmosphere models at selected points along evolutionary tracks corresponding to ages of 0, 0.5, 1, 1.5 and 2~Myr. An additional point was added in most tracks, corresponding either to 2.5 Myr or to a representative age between 2.0 and 2.5 Myr for the most massive models not reaching 2.5~Myr. The position of the selected points on the stellar tracks is shown by the symbols in Fig.~\ref{hrd}, and the corresponding physical parameters are listed in Tables~\ref{tab_gr_zsmc} to \ref{tab_grz_z0p1}.

We used the code CMFGEN \citep{hm98} to compute the atmosphere models. CMFGEN solves the radiative transfer equation under non-Local Thermodynamical Equilibrium conditions. It uses a spherical geometry to include expanding stellar winds. A quasi-hydrostatic solution of the momentum equation is connected to a velocity law of the form $v = v_{\infty} \times (1-\frac{R}{r})^{\beta}$ with \vinf\ the maximum velocity at the top of the atmosphere, $r$ the radial coordinate and $\beta$ a parameter fixed to 1.0 in our calculations. The density structure follows from the velocity structure and the equation of mass conservation. Thousands of lines from various species and their ions are included for a realistic atmospheric structure and emergent spectrum. The final spectrum results from a formal solution of the radiative transfer with the atmospheric structure fixed. Detailed line profiles are used and a micro-turbulent velocity varying from 10 to 0.1$\times$\vinf\ \kms\ is included to account for the observed extra-broadening of lines. 

The surface properties of the evolutionary models at the selected points are used as input to the atmospheric computations, ensuring full consistency. In particular the surface abundances predicted by the evolutionary models are used to predict the corresponding spectral appearance. As already demonstrated in our previous study \citep{mp22} this is important to correctly predict the strength of key VMS lines, especially \heiiuv.

\section{Results}
\label{s_res}

We present our results first focusing on the evolutionary paths in the Hertzsprung-Russell (HR) diagram. We then describe the spectroscopic appearance of our VMS models in the UV and optical range, as well as the shape of their ionising flux.

\subsection{Stellar evolution and VMS mass loss rates}
\label{s_evres}

\begin{figure*}[h]
\centering
\includegraphics[width=0.33\textwidth]{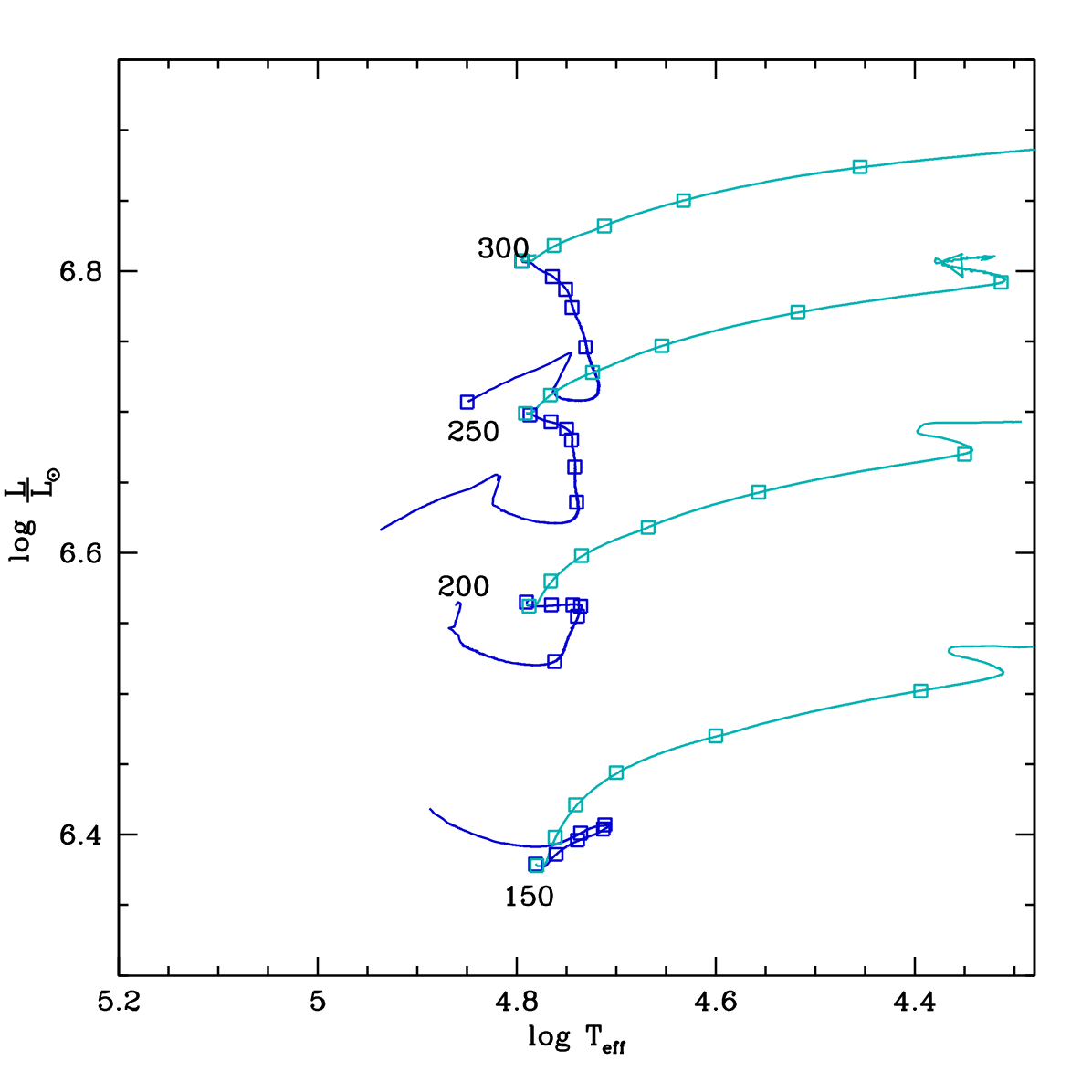}
\includegraphics[width=0.33\textwidth]{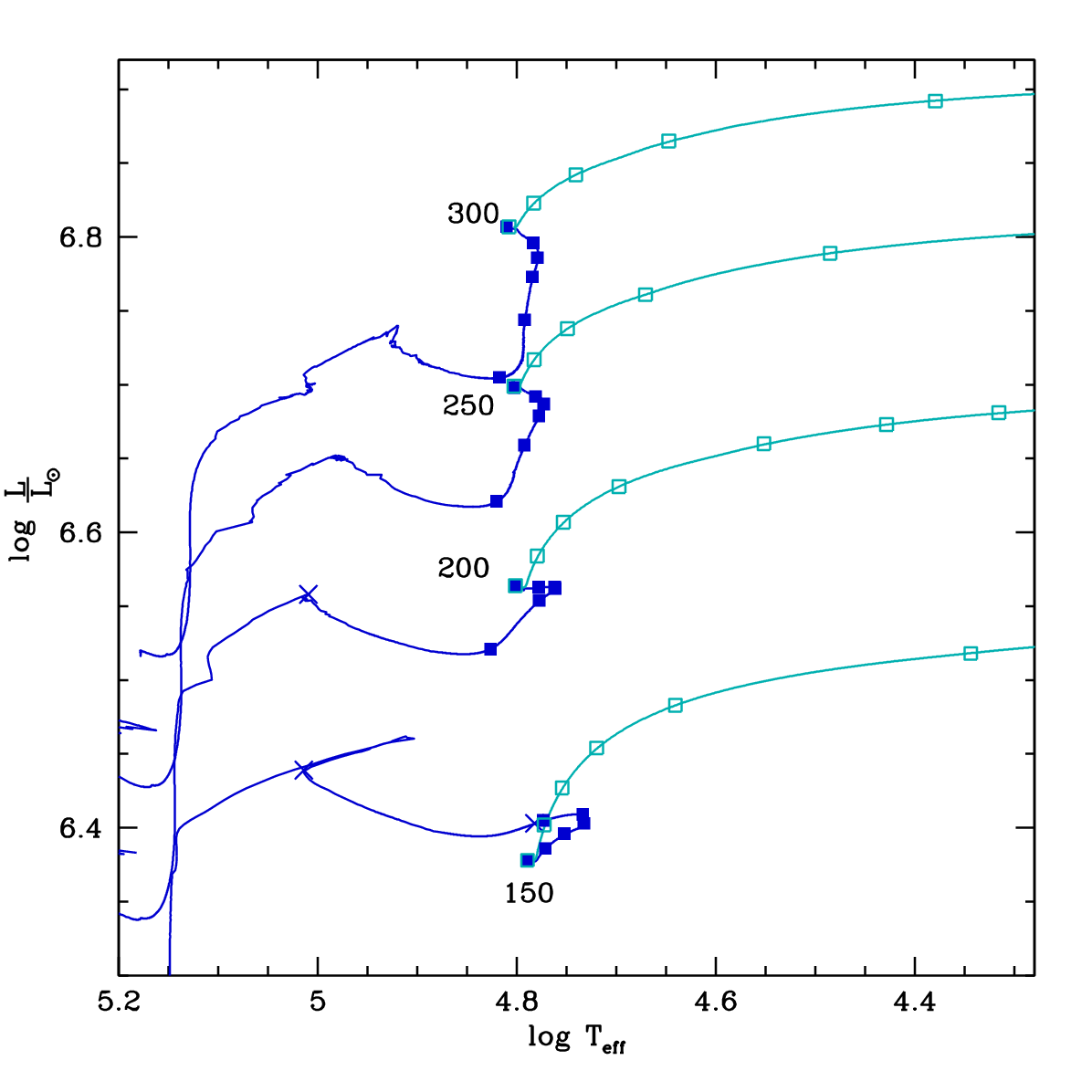}
\includegraphics[width=0.33\textwidth]{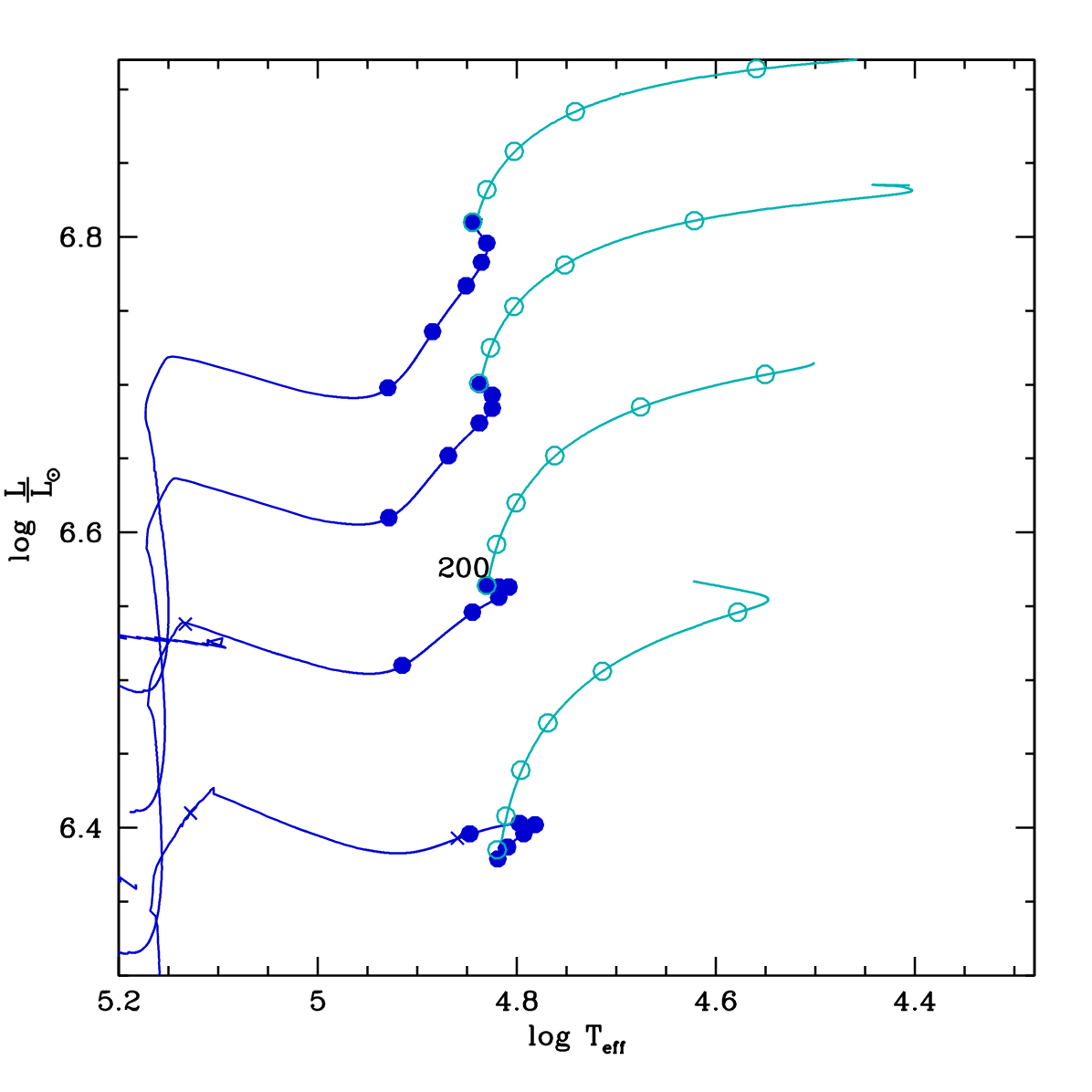}
\caption{HR diagram for models at  Z~=~0.2(0.1, 0.01)~\zsun\ in the left (middle, right) panel. In each panel the blue (cyan) lines are models without (with) metallicity scaling of the VMS mass loss rates. Squares and circles correspond to the points where atmosphere models and synthetic spectra are calculated. Crosses correspond to the models discussed in Sect.~\ref{s_postMS}.}
\label{hrd}
\end{figure*}

We show in Fig.~\ref{hrd} the evolutionary tracks of the various
models.  When decreasing the metallicity, the ZAMS point barely moves, the effective temperature at the beginning of core H
burning differs by at most 5\% in the mass range considered, meaning
that luminosity almost directly scales with mass independently of the
temperature, which is expected for such stars \citep{yusof13}. The
shape of the evolutionary tracks for a given mass loss recipe is also
independent of the initial mass, the tracks simply being shifted to
higher luminosities for larger initial masses. 

\begin{figure*}[h]
\centering
\includegraphics[width=0.47\textwidth]{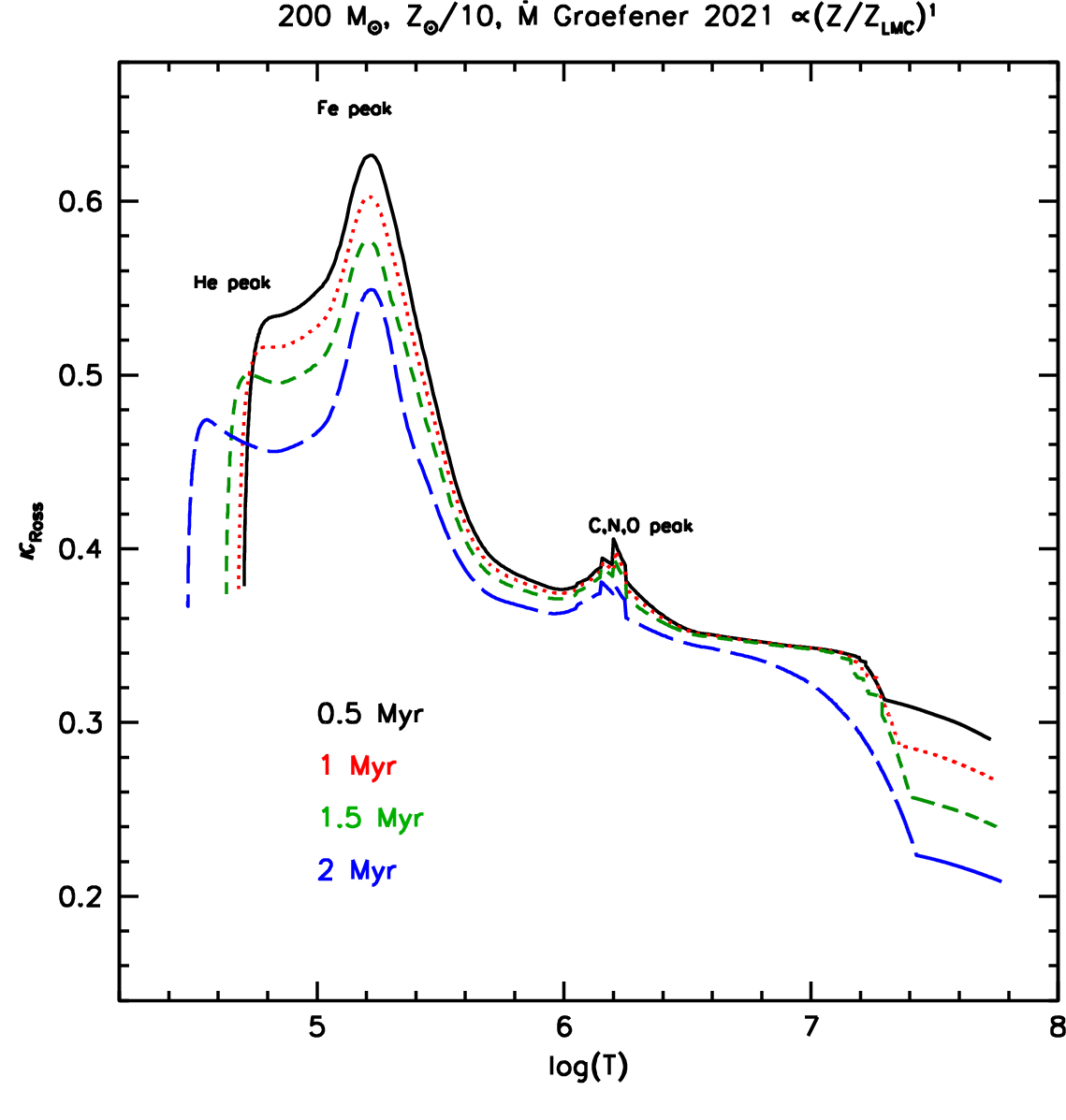}
\includegraphics[width=0.47\textwidth]{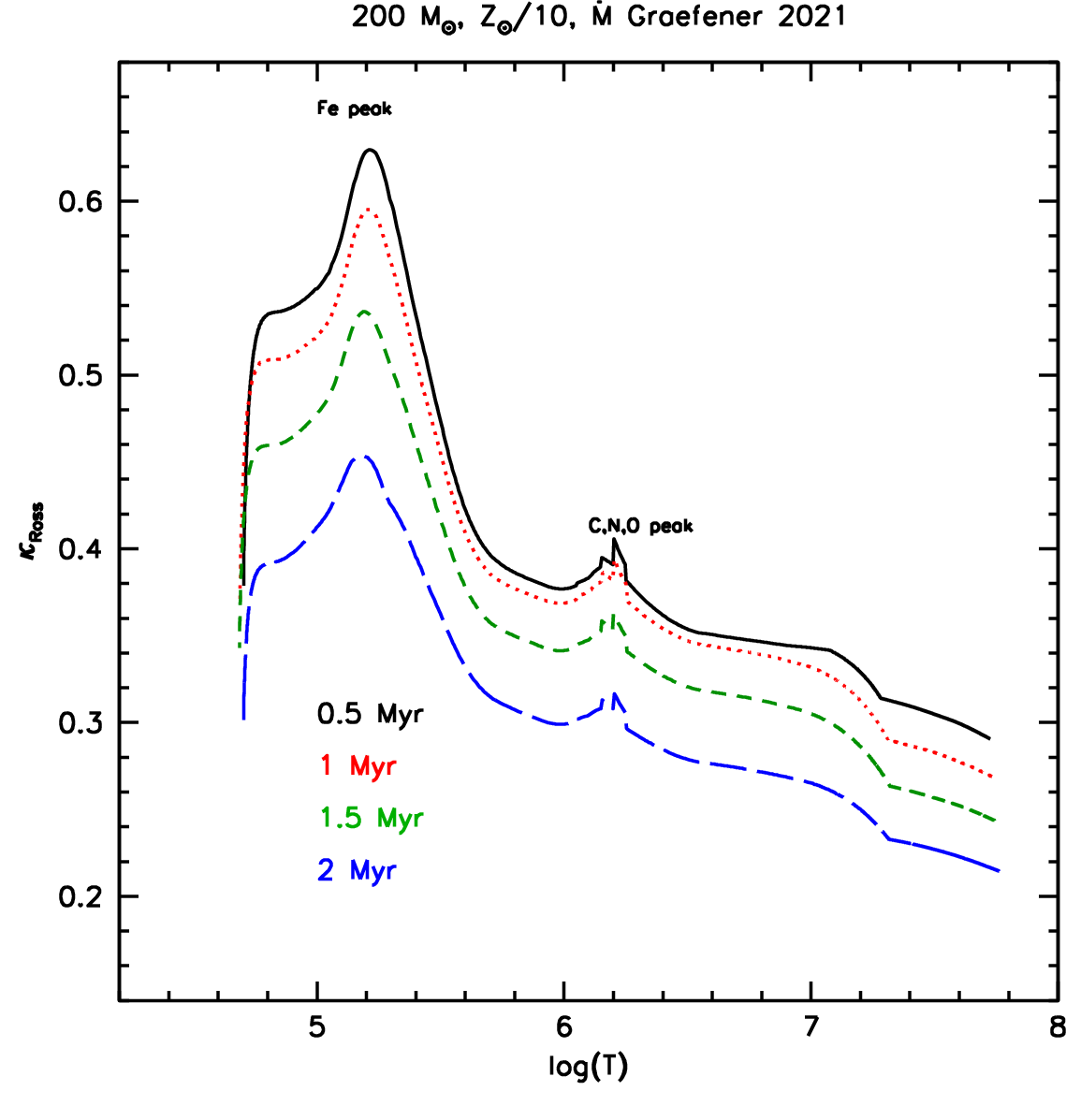}
\caption{Mean Rosseland opacity profiles as a function of temperature at 4 different ages on the main sequence for the 200\msun model at 0.1 Z$_\odot$ with (left) and without (right) Z-scaling of the VMS mass loss rates. The opacity bumps of CNO nuclei, Fe and \ion{He}{ii} are indicated.}
\label{fig:opacity}
\end{figure*}

The major difference encountered between all tracks is then due to the metallicity scaling applied to the mass loss recipe for the optically thick winds regime. 
The models that have reduced mass loss rates because of the metallicity scaling evolve classically towards the red part of the HR diagram. To better understand this evolution, and the difference with the other mass loss scenario we explore, we show in Fig.~\ref{fig:opacity} the opacity profile of our models at different ages. During evolution the total mass is reduced at a relatively slow pace during the main sequence evolution so that the luminosity increase is important. As the surface temperature slowly decreases during the early main sequence evolution, it reaches values lower than that associated to HeII ionisation ($\log\left(\rm{T}_{\rm HeII~bump}\right) \approx 4.6$) -- see left panel , Fig.~\ref{fig:opacity}. A HeII opacity bump is built below the surface that triggers convection and causes a further radius inflation \citep{cant09,grass21} that drives the evolution to the red in the HR diagram for these models. 

Such a bump never appears in the models maintaining a higher mass loss (see Fig.~\ref{fig:opacity}, right panel). More specifically, these models remain close to the zero-age main sequence (ZAMS) for most of the main sequence evolution, which lasts approximately 2.5 Myr. They subsequently evolve to the blue in the core He fusion phase (which lasts for $\approx 10$\% of the main sequence lifetime) where they remain for the very last phases (that represent less than 0.5\% of the main sequence lifetime). The main sequence duration is longer in these models despite their convective core being less massive. This is essentially due to the lack of a heavy envelope on top of that core, which reaches lower temperatures than in models of the other family discussed above, thus causing a slower evolution on the main sequence. Indeed for these models, the star is stripped by mass loss with M$_\star$ decreasing by $\approx$ 55\% to 62\% (for the 150 \msun\ and 300 \msun\ at the three metallicities considered) of its initial value over the duration of the main sequence (see Tables~\ref{tab_gr_zsmc} to \ref{tab_grz_z0p1}). As the envelope mass decreases, the portion of the total mass occupied by the H core is very large (80\% to 90\%) and does not evolve much during the main sequence. Consequently these models have a quasi-homogeneous evolution at this phase, and are thus expected to evolve to the blue \citep{maeder87,langer92,farrell20}. As H is processed into He in the core, the overall mean molecular weight of the stellar plasma $\bar{\mu}$ increases during the main sequence (from 0.65 at the ZAMS to 1.33 at 2 Myr for the 200~\msun~ model at 0.1 Z$_\odot$). A rough estimate gives $L_\star \propto M_\star^3 \bar{\mu}_\star^4$: the increase of $\bar{\mu}$ balances the strong decrease of M$_\star$, which leads to a very modest luminosity variation during the main sequence.

\subsection{Spectroscopic evolution}
\label{s_specres}

\begin{figure*}[h]
\centering
\includegraphics[width=0.47\textwidth]{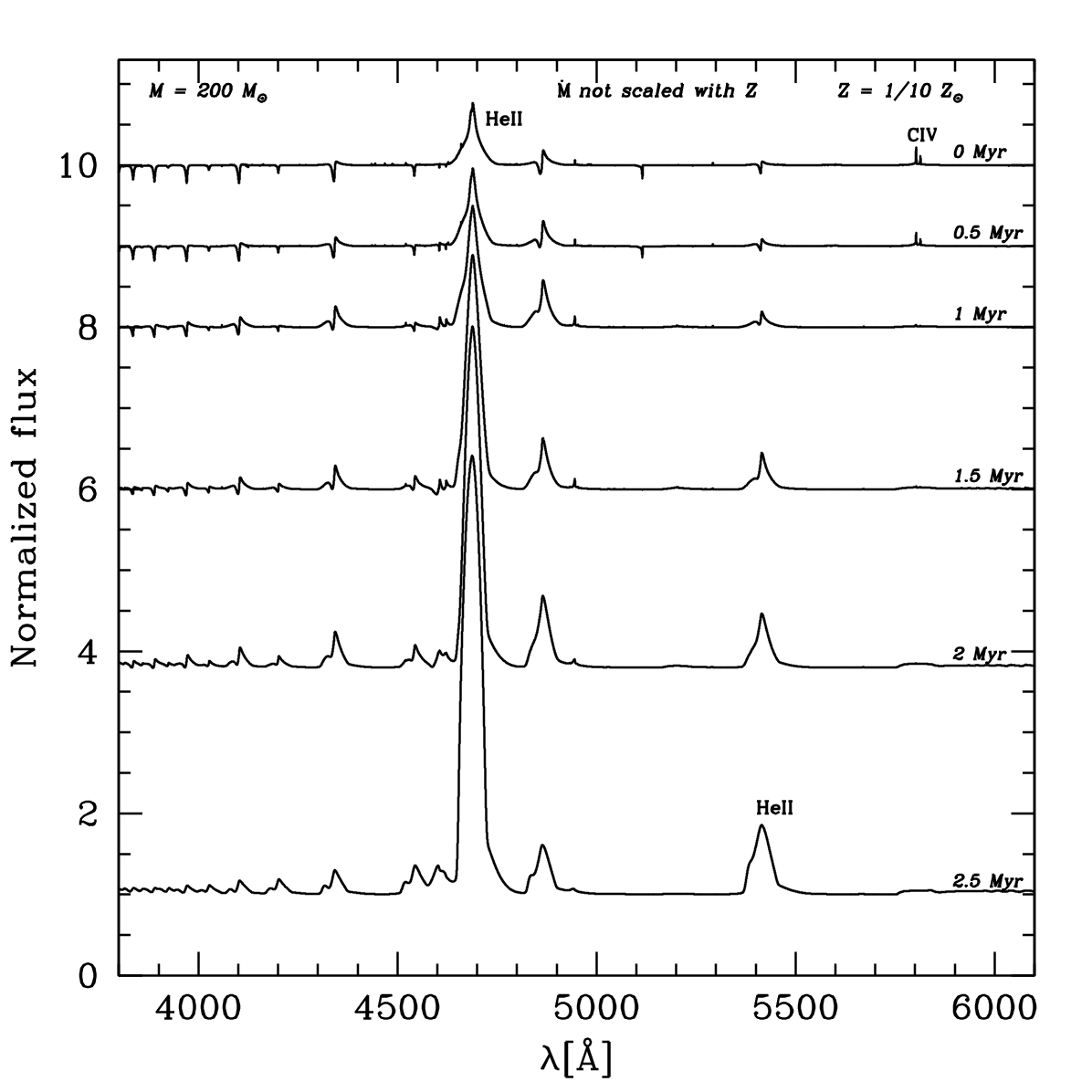}
\includegraphics[width=0.47\textwidth]{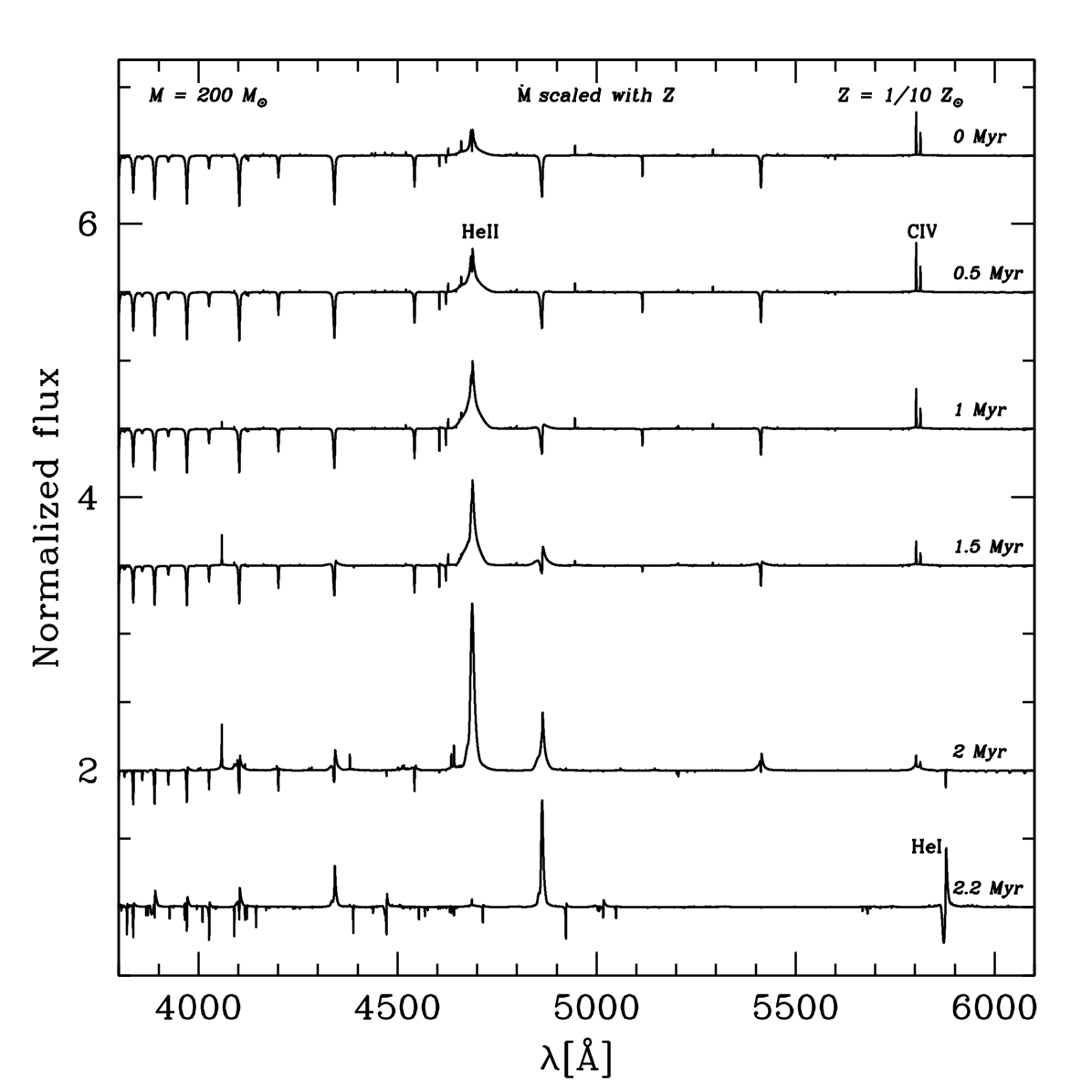}\\
\includegraphics[width=0.47\textwidth]{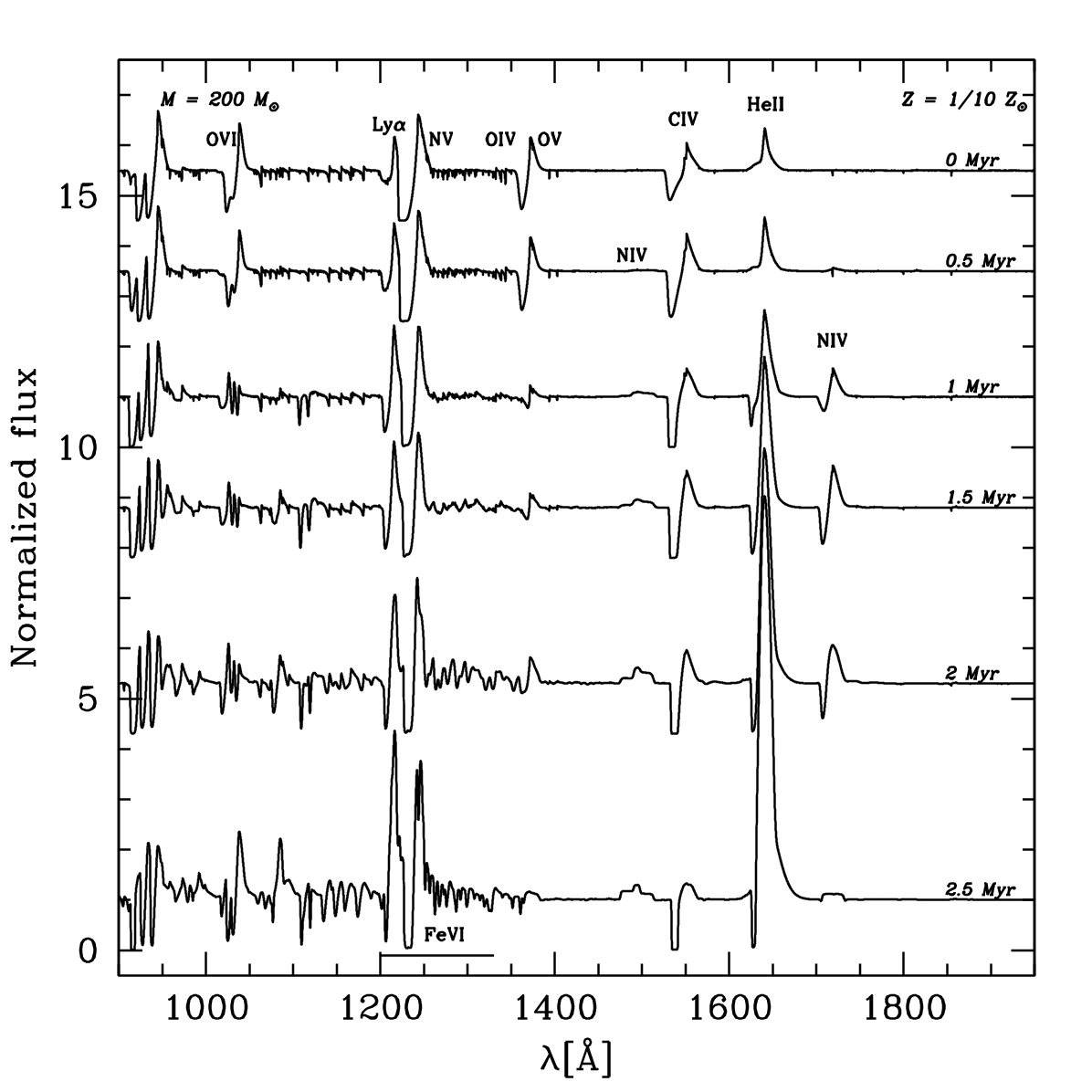}
\includegraphics[width=0.47\textwidth]{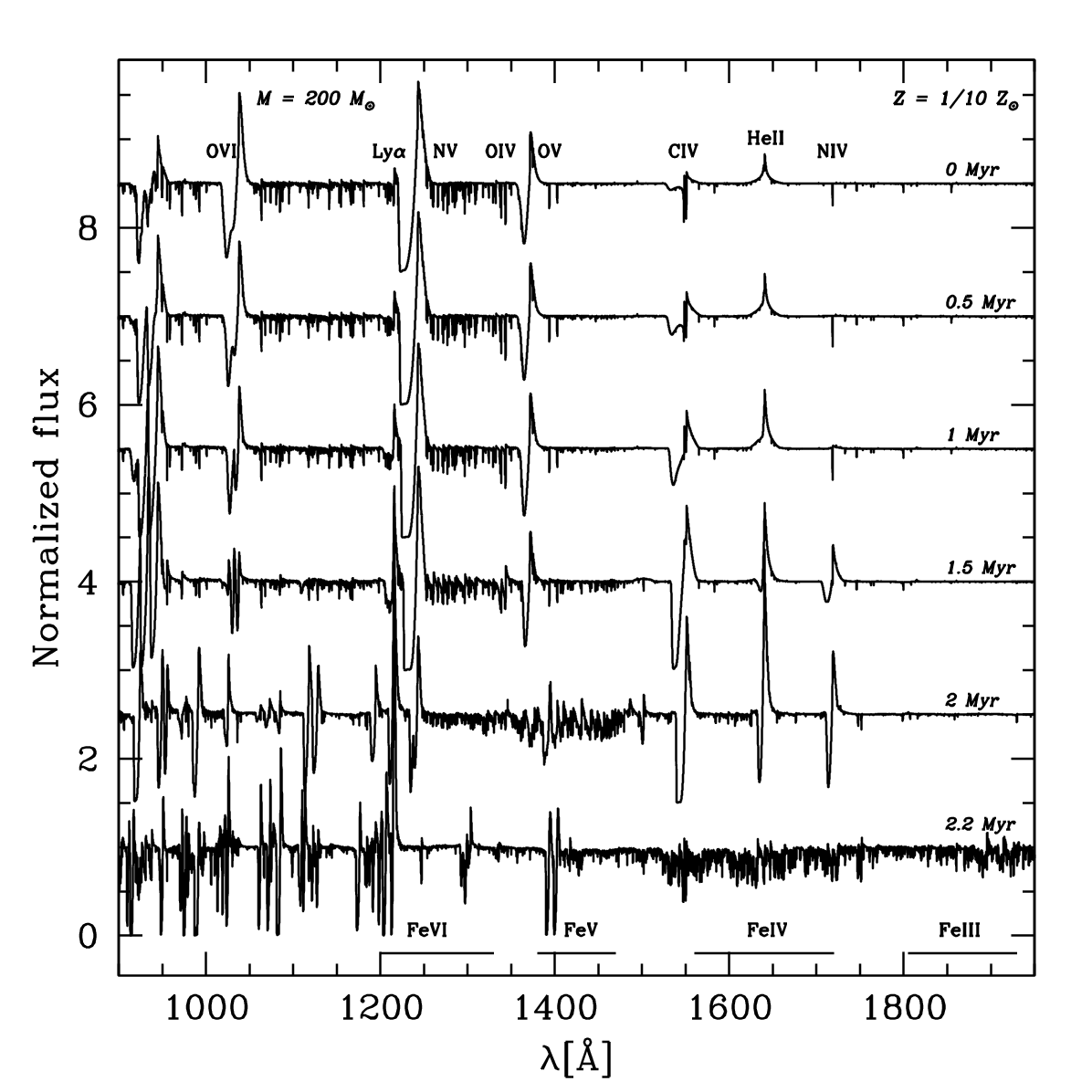}
\caption{Synthetic optical (top) and UV (bottom) spectra of the 200~\msun\ models at Z~=~0.1~\zsun. The left (right) panels correspond to models without (with) metallicity scaling of the mass loss rates. In all panels, the spectra from top to bottom correspond to models with 0.01, 0.5, 1, 1.5, 2 and 2.5 Myr as marked by squared on the evolutionary track in Fig.~\ref{hrd} and with parameters given in Tables~\ref{tab_gr_z0p1} and ~\ref{tab_grz_z0p1}. Spectra are shifted vertically for clarity.}
\label{sv200_z0p1}
\end{figure*}

We now proceed with the description of the spectroscopic evolution of low metallicity VMS. We illustrate the general behaviour of our models with the 200~\msun\ model at Z=0.1~\zsun\footnote{Spectral sequences at other metallicities are gathered in Appendix~\ref{s_ap}}. The optical and UV spectra at different evolutionary points along the tracks plotted in Fig.~\ref{hrd} are shown in Fig.~\ref{sv200_z0p1}. The left panels illustrate the evolution when no metallicity scaling of the VMS mass loss rates is applied. In this case the star remains hot at all times, and barely evolves from its ZAMS position as discussed in the previous section. Consequently the spectral morphology is little affected by changes in ionisation and the spectra exhibit almost the same lines at different evolutionary phases. However variations in surface abundances and mass loss rate affect the strength of these lines. In the optical, this is seen in \heiiopt\ that is the main feature. \heiiuv\ follows the same trend in the UV: it is already present on the ZAMS and becomes stronger when the star evolves, moving from a P-Cygni profile to a pure emission profile in the last model. We also note the strengthening of \ion{N}{iv}~1486 and \ion{N}{iv}~1720 as evolution proceeds. The physical reasons for this temporal evolution have been described in \citet{mp22}. As evolution proceeds, VMS are more and more chemically processed at their surface, showing the products of CNO burning, in particular nitrogen and helium enrichment (see Tables \ref{tab_gr_z0p1}). At the same time mass loss rates are increased because of the larger Eddington factor, driven by the larger luminosity-to-mass ratio. These two effects produce the strong helium and nitrogen lines observed in the spectra of VMS. 

The evolutionary sequence is quite different when VMS mass loss rates are scaled with metallicity. The spectra are thus affected (right panels of Fig.~\ref{sv200_z0p1}). Because of the reduced wind density many lines are now in absorption. The Balmer lines turn into emission at the end of the evolutionary sequence because of the reduction in temperature and increase of wind density. \heiiopt\ is always in emission but is weaker than in the case where VMS mass loss rates are not scaled with Z. We note the presence of a weak \ion{C}{iv}~5802-12 doublet in emission for most models. In the UV the decrease in temperature along the evolution is clearly seen in the shift of the dominating iron ion, from \ion{Fe}{vi} on the ZAMS (upper spectrum) to \ion{Fe}{iv} in the most evolved model (lower spectrum). \ion{O}{v}~1371 is present on the ZAMS and disappears, whereas \ion{Si}{iv}~1393-1403 develops into a strong double P-Cygni profile in the most evolved phases.

The behaviour of spectra at Z = 0.2~\zsun\ is qualitatively similar to that of Z = 0.1~\zsun\ models (see Appendix \ref{s_ap}). There are some quantitative differences though. Obviously with a larger metal content the iron line forests in the UV are stronger at Z = 0.2~\zsun\ (see Figs~\ref{sv200_z0p1} and \ref{sv200_zsmc}). In the optical we have seen that \civopt\ shows a double-peaked profile in Z = 0.1~\zsun\ models. This was also seen at higher metallicity \citep{mp22}. This is also true of Z = 0.2~\zsun\ models close to the ZAMS, but more evolved models with no Z scaling of mass loss rates show a broad profile. This is attributed to a still relatively large carbon content and at the same time a hot temperature and a high mass loss rate. These conditions are gathered only in these models. At higher Z evolved models are cooler, and at lower Z the carbon content is reduced. The morphology of \civopt\ has been used as a criterion to identify VMS in star-forming galaxies \citep{martins23}. We will get back to this in Sect.~\ref{s_bumps}.

At Z = 0.01~\zsun\ the changes are more pronounced. The lower metal content translates into an almost complete disappearance of the iron forests in the UV. In the high mass loss rate scenario \heiiuv\ remains clearly present and dominates this wavelength range for most of the evolution. Longward of 1200~\AA\ only \lya\ and \nvuv\ remain visible past the first 0.5~Myr of evolution. This is due to a combination of hot temperature and chemical processing at the surface that for instance reduce the strength of \civuv. For mass loss rates scaled with metallicity the UV spectrum resembles, for the first time in our modelling of VMS, that of O-type stars. \heiiuv\ is still present but is weak and does not dominate the UV spectrum. Similarly \heiiopt\ is mostly in absorption and does not stand out as a peculiar line. If VMS mass loss rates scale linearly with metallicity VMS cannot be distinguished from normal main sequence massive stars at very low metallicity.

\begin{figure}[h]
\centering
\includegraphics[width=0.47\textwidth]{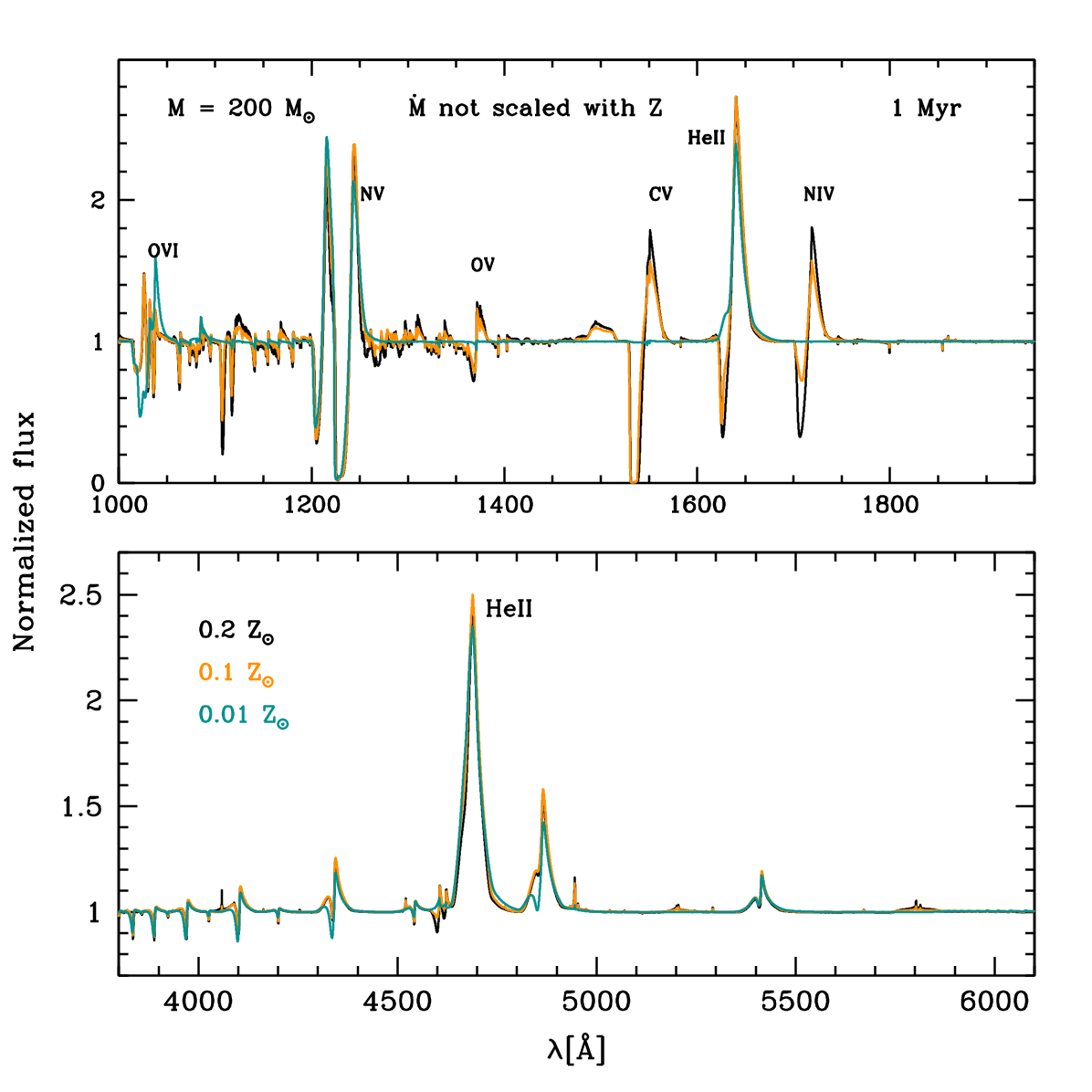}
\caption{Spectra of the 200~\msun\ models at 1~Myr for Z=0.2~\zsun\ (black), 0.1~\zsun\ (orange) and 0.01~\zsun\ (cyan). Models have no metallicity scaling of the mass loss rates. The top (bottom) panel shows the UV (optical) range.}
\label{sv200_zall}
\end{figure}

Fig.~\ref{sv200_zall} illustrates the spectral differences in the 200~\msun\ models after 1~Myr of evolution, at the three metallicities considered in this work. The models correspond to the case of no Z scaling of mass loss rates. The optical range is dominated by \ion{He}{ii} lines and their morphologies are rather similar regardless of the metallicity. In the UV lines are mostly similar between Z=0.2 and 0.1~\zsun, while most metallic lines disappear at Z=0.01~\zsun. From Table ~\ref{tab_gr_zsmc}, \ref{tab_gr_z0p1} and \ref{tab_gr_z0p01} we see that the helium content is the same in all three models (Y$\sim$0.28) but the nitrogen and carbon content decreases by more than a factor of 10 as Z decreases, leading to the disappearance of \civuv\ and \nivuv\ for instance. But metallicity also has an indirect effect: \teff\ is higher at lower Z, thus affecting the ionisation balance. This translates into a weakening of \ion{N}{iv} lines, while \nvuv\ remains strong in spite of the reduced nitrogen content.

Metallicity has an impact of the spectroscopy of VMS. The largest effects are those related to the scaling or not of VMS mass loss rates with Z, since this affects both stellar evolution and spectroscopic appearance. 
However the main conclusion is that down to Z = 0.1~\zsun\ whatever the exact Z scaling, \heiiuv\ is always seen in emission and dominates the UV spectrum. This feature is the most emblematic of VMS. It is still the strongest emission at Z = 0.01~\zsun\ if VMS mass loss rates do not scale with Z. Otherwise, \heiiuv\ emission vanishes. We thus expect very young stellar population hosting VMS to show \heiiuv\ in emission over a wide range of metallicity. We further investigate this in Sect.~\ref{s_popsyn}.

\begin{figure}[h]
\centering
\includegraphics[width=0.47\textwidth]{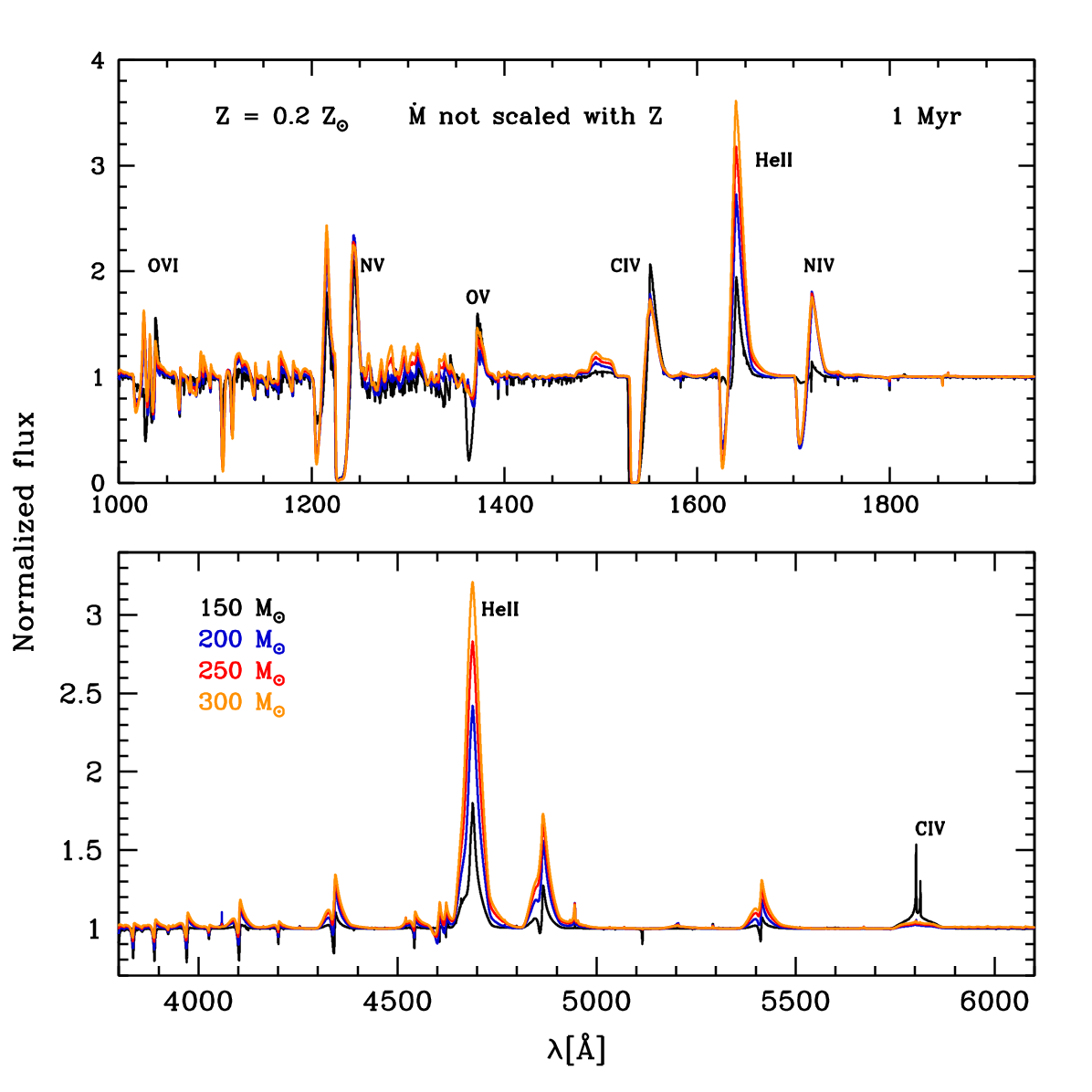}
\caption{Spectra of the Z=0.2~\zsun\ models after 1~Myr, for the four initial masses considered in this work. Models have no metallicity scaling of the mass loss rates. The top (bottom) panel shows the UV (optical) range.}
\label{svall_zsmc}
\end{figure}

In Sect.~\ref{s_evres} we saw that the shape of evolutionary tracks is, to first order, qualitatively the same for different initial masses, at a given metallicity. One may thus wonder whether one can use the synthetic spectra computed along one track and simply scale them according to luminosity to obtain those of another mass. Fig.~\ref{svall_zsmc} shows that this is not possible. The four synthetic spectra correspond to four different initial masses, but the same age (1~Myr). They have similar \teff\ ($\sim$55 to 56~kK) so that the ionisation is rather comparable in the four models. In spite of this lines are not the same in all four spectra. In particular \heiiuv\ and \heiiopt\ show clear sequences of increased strength with mass. This is explained by the higher helium mass fraction and larger mass loss rates when mass increases. Inspection of Table~\ref{tab_gr_zsmc} indicates that Y ranges from 0.251 to 0.369 for initial masses between 150 and 300~\msun, at 1~Myr. At the same time the mass loss rates increases from $10^{-4.73}$ to $10^{-4.14}$ \myr, because the Eddington factor increases. In Fig.~\ref{svall_zsmc} we also notice that \nivuva\ is stronger in higher mass models, and \civopt\ is weaker. This is also a consequence of chemical mixing that exposes the products of CNO burning at the surface, i.e. nitrogen enrichment and carbon depletion (see Table.~\ref{tab_gr_zsmc}).
The conclusion from this comparison is that it is not possible to scale synthetic spectra according to their luminosity to obtain the spectra of more luminous stars, since not only mass loss rates but also surface composition changes. Proper modelling of spectra along all tracks with different initial masses is necessary for quantitative predictions of the spectral appearance of VMS.

\section{Population synthesis}
\label{s_popsyn}

We now proceed to the inclusion of VMS models into population synthesis models. We describe the method and then discuss the impact of VMS on the integrated light of starbursts. We highlight the effects of metallicity. We confront our models to observations of three clusters suspected to host VMS.

\subsection{Model set up}
\label{s_popsynsetup}

We proceeded as in \citet{mp22} to build population synthesis models including VMS. For normal stars with masses below 100~\msun\ we retrieved the BPASS models of \citet{bpass}. We selected models with an upper IMF slope of -2.35 and no binaries. We chose the BPASS models with metallicities Z=0.002, 0.001 and $10^{-4}$ since they are the closest ones for our Z=0.2, 0.1 and 0.01~\zsun\ VMS models respectively. We then extended the mass function up to 225~\msun\ using our 150 and 200~\msun\ models as representative of stars in the mass bins 100-175 and 175-225~\msun. For test models we extended the upper mass cut-off up to 300~\msun, with the VMS 250 and 300~\msun\ models representing stars in the mass range 225-275 and 275-300~\msun\ respectively. We proceeded as in \citet{upad} to correct for the bug in the formal description of the mass function presented by \citet{bpass}. For all ages the contribution of VMS was added to the BPASS models, and we re-normalised the final fluxes so they correspond to a total mass of $10^6$ \msun. BPASS does not provide burst models for ages 0 and 0.5 Myr, so for these ages we added the 1~Myr BPASS models to our 0 and 0.5 Myr populations of VMS. Most of our models reach 2.5~Myr, but some stop just short of it (e.g. the 200~\msun\ model at Z=0.2~\zsun\ and scaled mass loss rates, for which the age is 2.4~Myr - see Table~\ref{tab_grz_zsmc}). In that case we make the assumption that these models with ages $\lesssim$2.5~Myr are still representative at 2.5~Myr and we use them to produce the population spectrum at that age. 
For constant star formation (CSF) models we simply added the contributions of bursts weighted by the duration of the age range they represent. In practice we divided time into the following ranges (expressed in Myr): 0-0.25, 0.25-0.75, 0.75-1.25, 1.25-1.75, 1.75-2.25, 2.25-2.75, 2.75-3.0, 3.0-4.0, 4.0-5.0, 5.0-6.0, 6.0-8.0, 8.0-10.0, 10.0-12.5, 12.5-16.0, 16.0-20.0, 20.0-25.0, 25.0-32.0, 32.0-40.0 and 40.0-50.0. For each of these ranges we used respectively the bursts with age: 0, 0.5, 1.0, 1.5, 2.0, 2.5, 3.0, 4.0, 5.0, 6.0, 8.0, 10.0, 12.5, 16.0, 20.0, 25.0, 32.0, 40.0 and 50.0~Myr.

\subsection{Spectral morphology}
\label{s_popsynmorpho}

\begin{figure*}[h]
\centering
\includegraphics[width=0.33\textwidth]{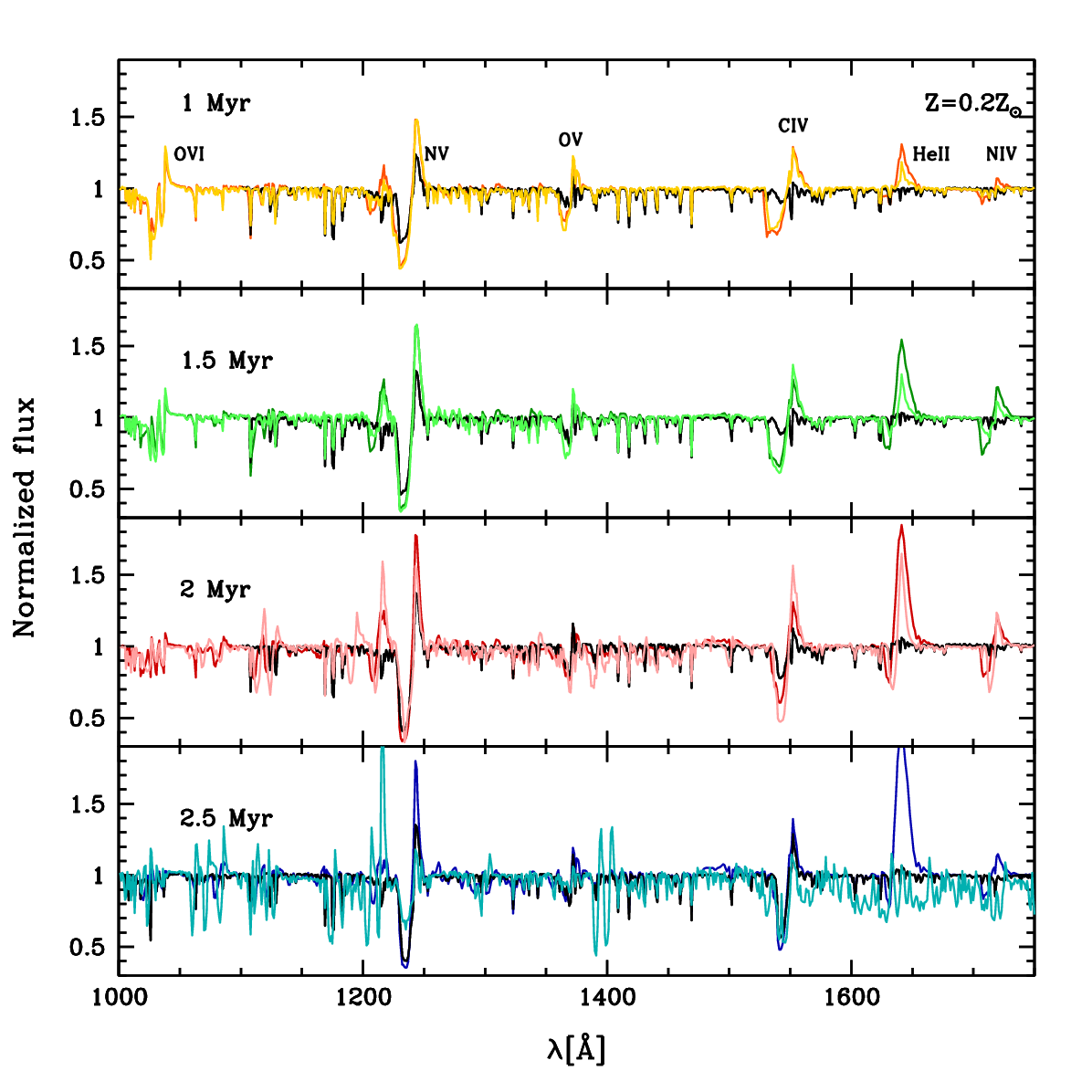}
\includegraphics[width=0.33\textwidth]{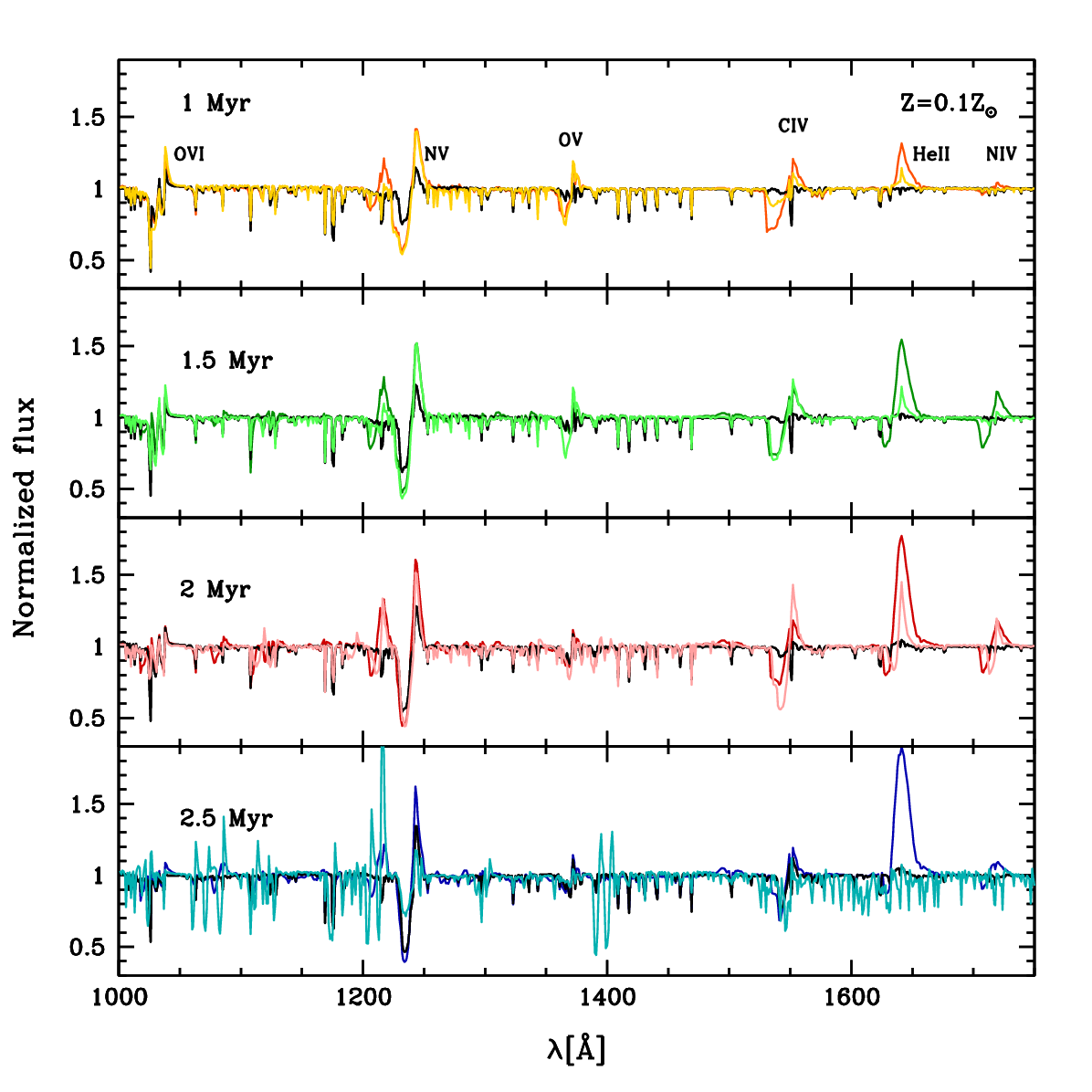}
\includegraphics[width=0.33\textwidth]{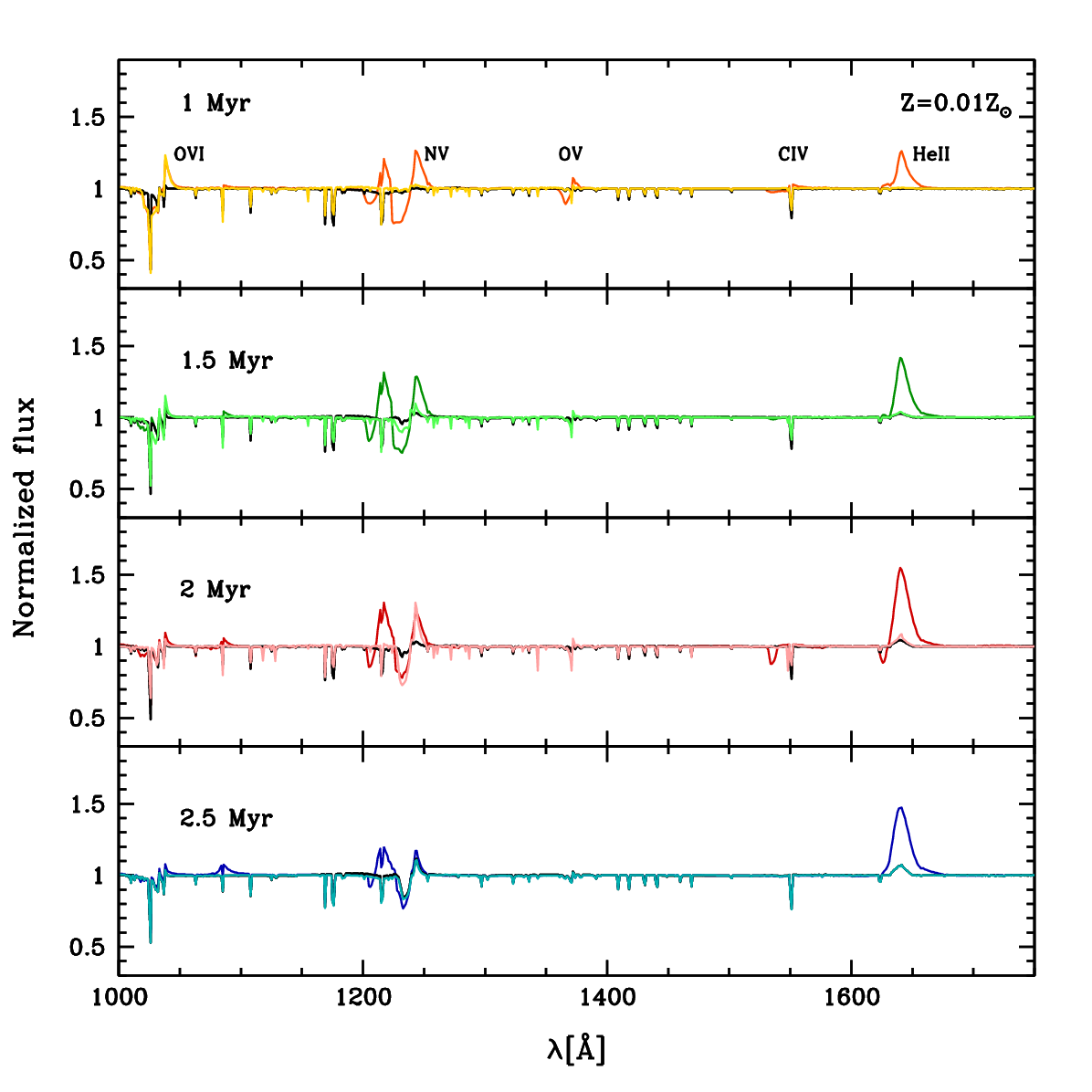}
\caption{UV spectra of burst models for ages between 1 and 2.5 Myr. Coloured lines are models including VMS, black lines models without VMS (i.e. BPASS models). The dark (light) colours correspond to mass loss rates not scaled (scaled) with metallicity. The left (middle, right) panel is for Z=0.2(0.1, 0.01)~\zsun.}
\label{comp_bursts}
\end{figure*}

We show how our models compare to BPASS models without VMS in Fig.~\ref{comp_bursts}. As already found for Z=1/2.5~\zsun\ \citep{mp22} \heiiuv\ emission is produced only by models that include VMS, making this feature a unique tracer of their presence in young stellar populations. \heiiuv\ is stronger in models where the VMS mass loss rates are not scaled with metallicity, which is expected. In that case it is present at all times. If a metallicity scaling is applied, \heiiuv\ disappears at 2.5~Myr because of the different evolution: VMS become cool so that \ion{He}{ii} recombines into \ion{He}{i} - see bottom panels of Fig.~\ref{comp_bursts}, cyan lines. 
At low metallicity \nivuv\ is strongly affected by the presence of VMS, as was the case at Z=1/2.5~\zsun. The presence of a broad emission around \ion{N}{iv}~1486 is also caused by VMS. We thus confirm previous findings that VMS uniquely affect the UV spectra of young starbursts. We also note that other features are strengthened by the presence of VMS: \nvuv, \ovuv, and \civuv. In models with metallicity scaling of mass loss rates \ion{Si}{iv}~1393-1403 appears as a strong doublet at 2.5~Myr, and is almost entirely due to VMS. As anticipated from the discussion of spectra of individual stars, these conclusions vanish in the case where Z = 0.01~\zsun\ and VMS mass loss rates scale with Z (left panel of Fig.~\ref{comp_bursts}, light colours). Here, VMS have little effect on the integrated spectrum. Except for this extreme case, VMS thus manifest themselves in the integrated spectrum of young starbursts over a wide range of metallicities, regardless of their exact mass loss rates. This indicates that UV spectroscopy should be able to probe the presence of VMS over a wide range of metallicity, and in most star-forming regions known to date (see discussion in Sect.~\ref{s_method}).

This conclusion holds as long as only VMS are present or dominate completely the light emitted by massive stars. If in addition to them classical WR stars co-exist in sufficient number, things may get more complicated. Although the current generation of population synthesis models that include WR stars is not able to produce any \heiiuv\ in emission as observed, empirical results on 30~Dor show that normal WR stars may still contribute some flux \citep{crowther24}. This calls for a revision of the inclusion of WR stars in population synthesis models. 
In any case, the relative role of VMS and normal WR stars can be distinguished based on the morphology of the optical features we describe in Sect.~\ref{s_bumps} (see also \citealt{martins23}).

Altogether these results indicate that the characteristic features of VMS at Z=1/2.5~\zsun\ remain  
 tracers of their presence in starbursts at lower metallicity. We now examine in more details the effects of metallicity on these and other spectroscopic quantities.

\subsection{\heiiuv}
\label{Z_1640}

\begin{figure}[h]
\centering
\includegraphics[width=0.47\textwidth]{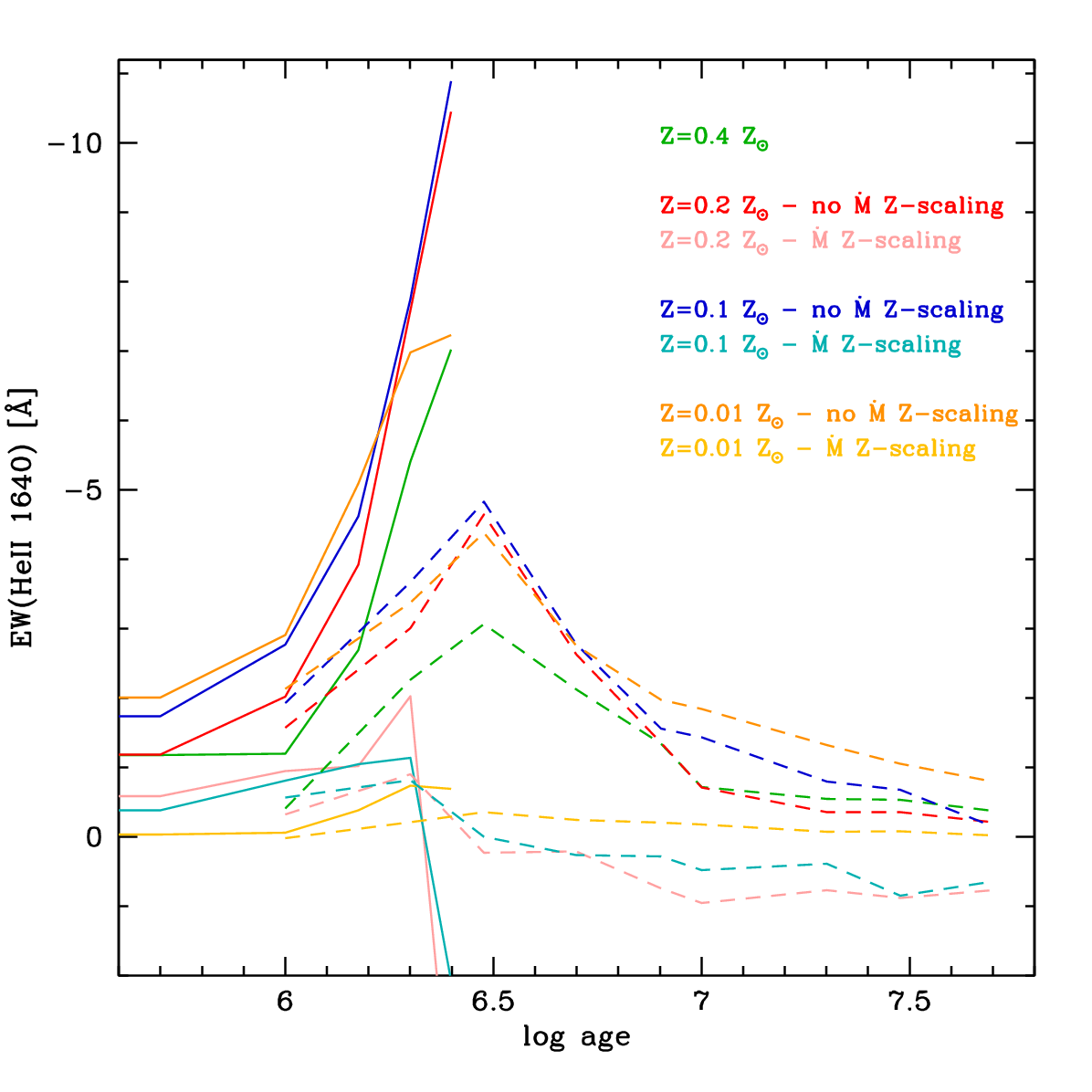}
\caption{EW(\heiiuv) as a function of age for burst models (solid lines) and CSF models (dashed lines). Models with Z=0.2(0.1, 0.01)~\zsun\ are shown in red and pink (blue and cyan, orange and yellow). Light colors (pink, cyan, and yellow) are models with a metallicity scaling of mass loss rate, while blue, red, and orange are models with no Z scaling. Green lines are models from \citet{mp22} at Z=0.4~\zsun.}
\label{ew1640_age}
\end{figure}

\begin{figure}[h]
\centering
\includegraphics[width=0.47\textwidth]{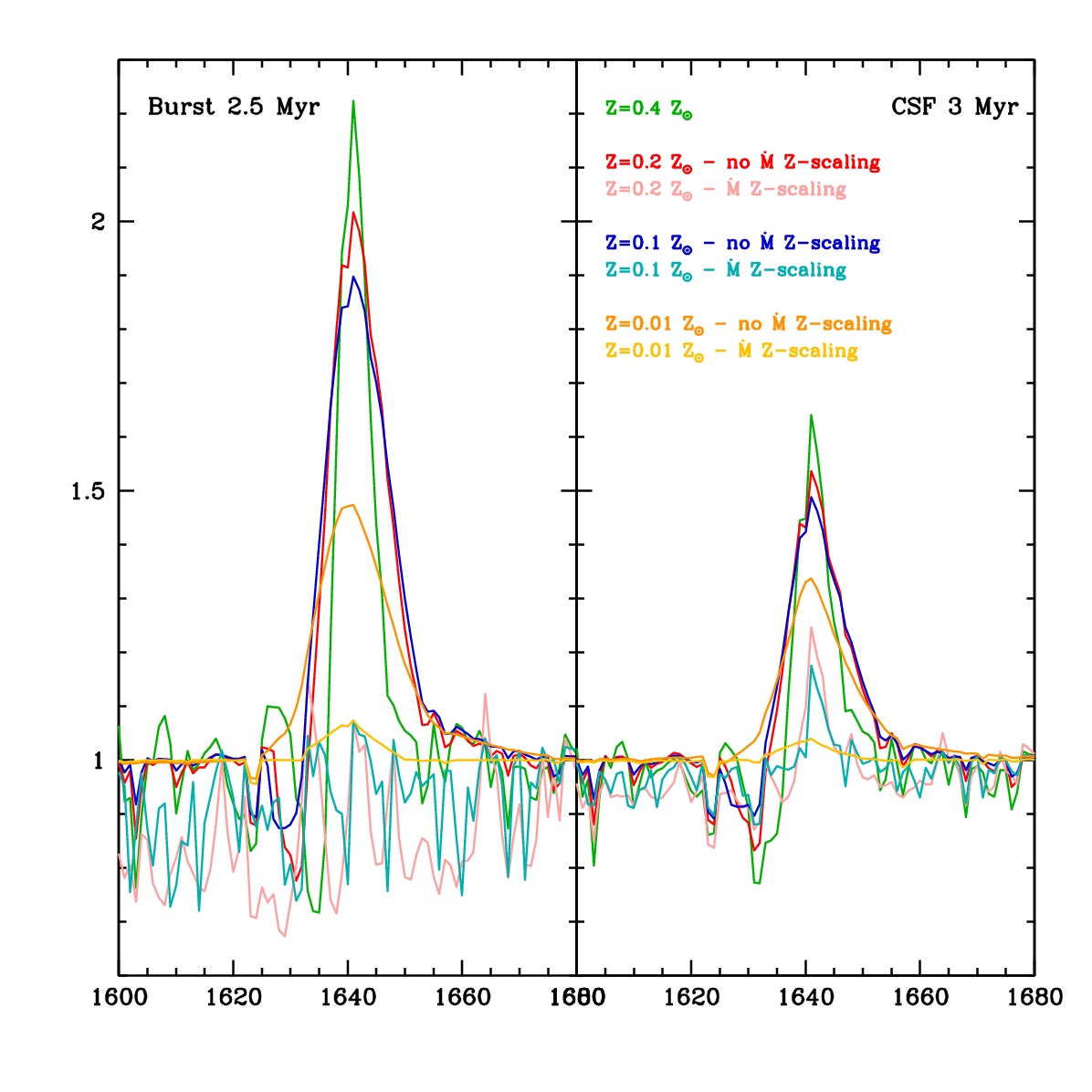}
\caption{Profile of \heiiuv\ for burst models at 2.5 Myr (left panel) and 3~Myr CSF models (right panel). Different colours correspond to different metallicities and assumptions regarding VMS mass loss scaling with Z, and are explained in the right panel. }
\label{prof_heii1640}
\end{figure}

In Sect.~\ref{s_specres} we saw that \heiiuv\ was produced in emission in most of our low metallicity models of individual VMS. We now investigate how the strength of this line varies with Z in young stellar populations. For that we calculate its equivalent width (EW) over the wavelength range 1625-1655~\AA. This range encompasses the widest extension encountered in the \heiiuv\ profile of all our models. We plot EW(\heiiuv) versus age in Fig.~\ref{ew1640_age}. In addition to the models calculated for the present paper we also add those of \citet{mp22} at Z=0.4~\zsun. For these latter models there is only one recipe for VMS mass loss rate, that of \citet{graef21}.

Fig.~\ref{ew1640_age} shows how EW(\heiiuv) varies with age for burst and CSF models. Since \heiiuv\ strengthens as VMS evolve off the ZAMS, we see an increase in its EW with age in all models up to 2.5~Myr. After that VMS and their \heiiuv\ emission disappear. Consequently EW(\heiiuv) decreases in CSF  models after 2.5~Myr. The models with VMS mass loss rates that are not scaled with metallicity show the highest EWs, up to 10~\AA\ in burst models and 4~\AA\ in CSF models.
In bursts models with the strongest mass loss rates EW(\heiiuv) is on average larger at lower metallicity, except at 2.5 Myr. The reasons for this behaviour are rooted in the shape of the line profiles that are illustrated in Fig.~\ref{prof_heii1640}. When metallicity decreases the absorption lines that are superimposed on \heiiuv\ are weaker so the emission near 1640~\AA\ is higher. In addition at lower metallicity \heiiuv\ shifts from a P-Cygni profile to an emission profile. Consequently the absorption part disappears thus pushing the EW to higher values. The reason for the change of profile of \heiiuv\ is that stars are less compact at higher metallicity, leading to a smaller escape velocity at their surface. Since the maximum velocity of the wind is directly proportional to the escape velocity, it is reduced at higher metallicity. This affects the wind density and thus the shape of the stellar lines formed in the wind. This is clearly seen in Fig.~\ref{prof_heii1640} where the dark red profile is more centrally peaked, with a stronger absorption dip in the P-Cygni profile. On the contrary the orange profile (Z=0.01~\zsun) is purely in emission.

In burst models with VMS mass loss rates scaling with metallicity \heiiuv\ emission is reduced. EW(\heiiuv) still reaches a maximum of 1.5~\AA\ at Z=0.2~\zsun. Below that metallicity there is almost no emission. In CSF models the same trends are observed qualitatively. Low metallicity models produce slightly higher EW than the highest Z models, and models with reduced mass loss rates have EW(\heiiuv) close to zero after $\sim$3~Myr. This does not mean that there is no \heiiuv\ emission, as illustrated by the pink and cyan lines in the right panel of Fig.~\ref{prof_heii1640}: a weak emission is detected. But over the wavelength range considered for the EW measurement, absorption lines compensate for this emission.

\begin{figure}[h]
\centering
\includegraphics[width=0.47\textwidth]{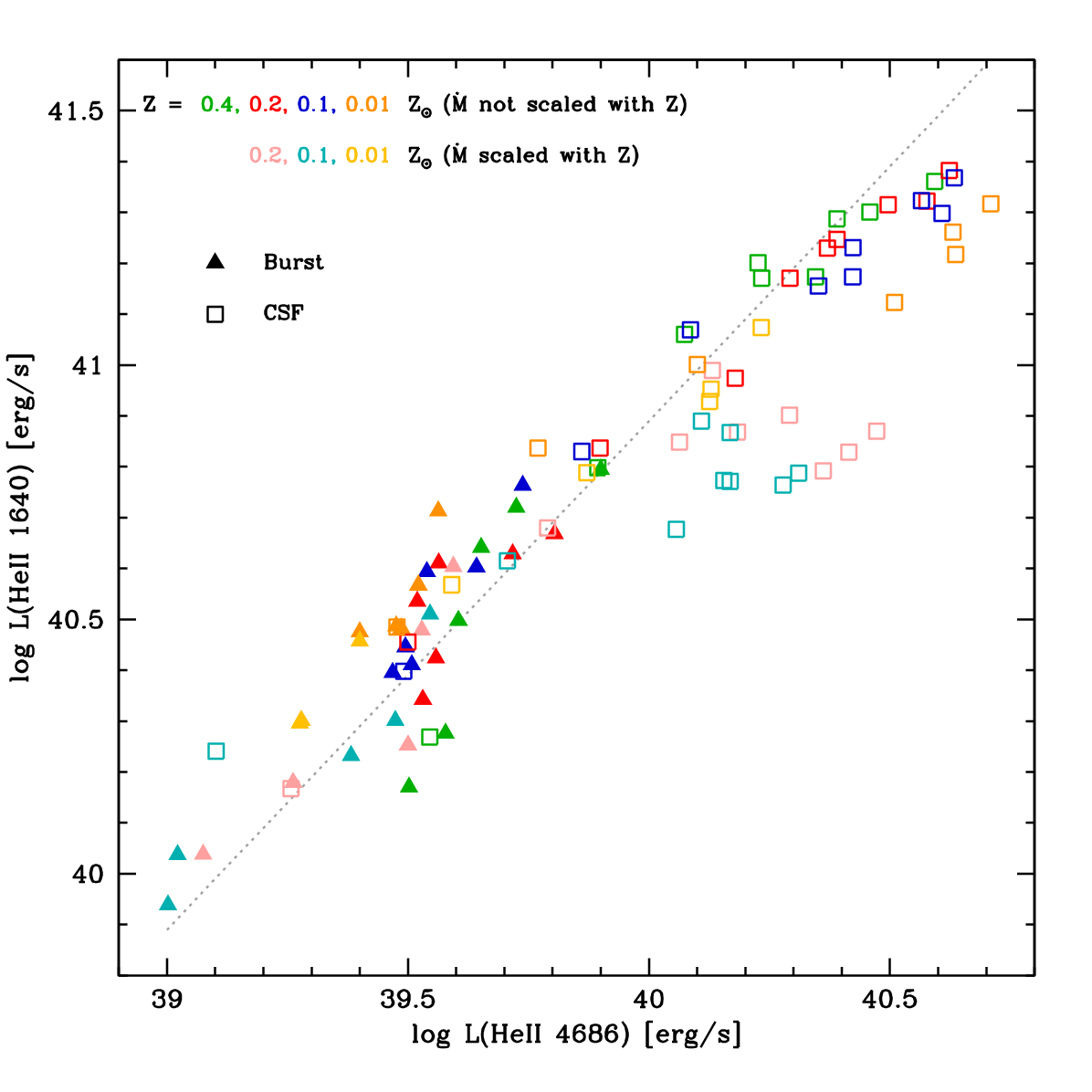}
\caption{Luminosity of \heiiuv\ as a function of luminosity of \heiiopt\ for burst (triangles) and CSF (open squares) models at various metallicities. Metallicities and assumptions regarding mass loss rates are indicated in the upper left part of the figure. The dotted line shows the relation reported by \citet{leitherer19}.}
\label{L1640_4686}
\end{figure}

\citet{cm90} proposed a method to determine dust extinction based on the relative strength of \heiiuv\ and \heiiopt\ in WR stars. Their study is based on the assumption that both lines arise from recombination processes in optically thin nebulae so that their strength mostly relies on atomic physics. Thus in standard physical conditions this ratio is equal to 7-8 \citep{seaton78,humstor87}. \citet{leitherer19} measured line flux ratios in Galactic and LMC WR stars and show that a correlation exists with a line ratio of 7.76. Thus comparing observed ratios to dust-free ratios allows a determination of dust attenuation, as done in \citet{masch24}. In Fig.~\ref{L1640_4686} we show the luminosity of \heiiuv\ compared to \heiiopt. Only models that clearly show both lines in emission are considered for this figure. The wavelength region over which the luminosity is calculated is tailored for each model, in order to take into account the full line width. A correlation between L(\heiiuv) and L(\heiiopt) is seen. It follows relatively well the relation empirically determined by \citet{leitherer19}. The points that deviate the most correspond to CSF models with mass loss rates that scale with metallicity, and at low metallicity. In those cases lines are weak and measurements are more uncertain, especially for \heiiopt.

\subsection{Optical bumps}
\label{s_bumps}

\citet{martins23} show that distinguishing young star forming regions dominated by VMS from those where classical Wolf-Rayet stars are present can be done using the morphology of the emission features near 4630-4690~\AA\ and 5800-5820~\AA, the so-called WR bumps. When only VMS are present the blue feature shows a \heiiopt\ emission with no or little \ion{N}{iii}~4634-42 emission. The red bump is also different depending on the population: VMS produce a double narrow emission sometimes on top of a broader but weak emission, while WR stars form a broad and strong emission feature without narrow doublet.

\begin{figure*}[h]
\centering
\includegraphics[width=0.45\textwidth]{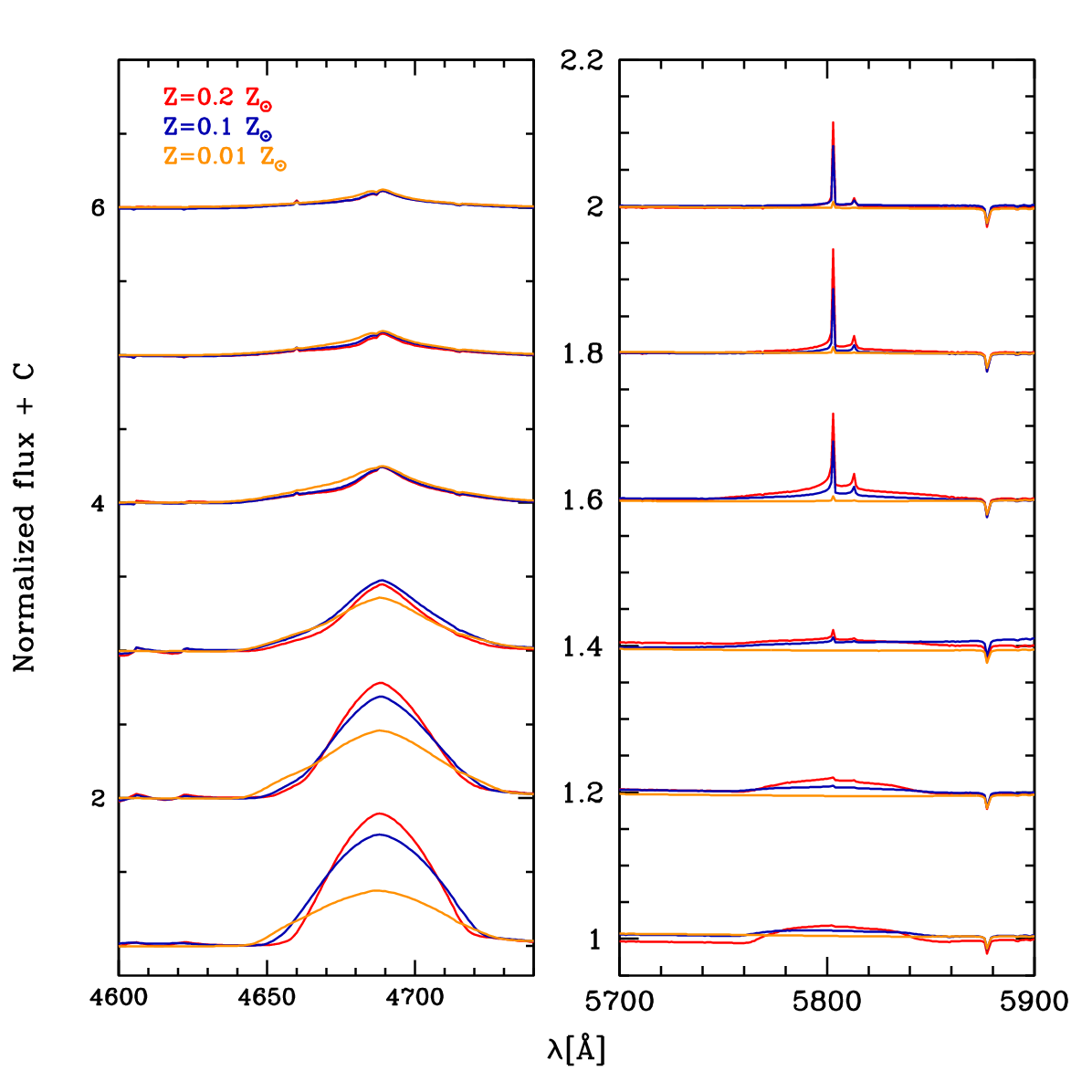} 
\includegraphics[width=0.45\textwidth]{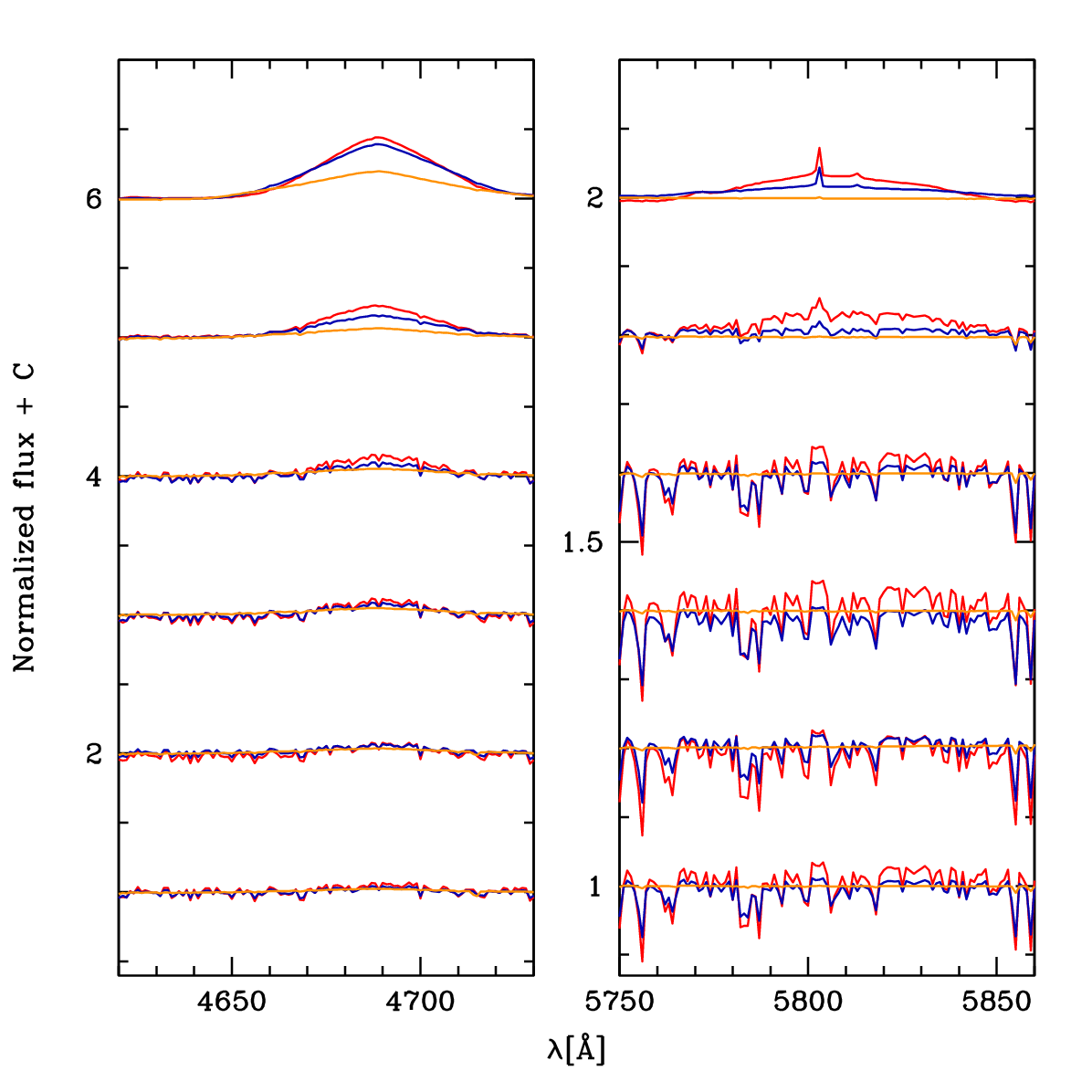} \\
\includegraphics[width=0.45\textwidth]{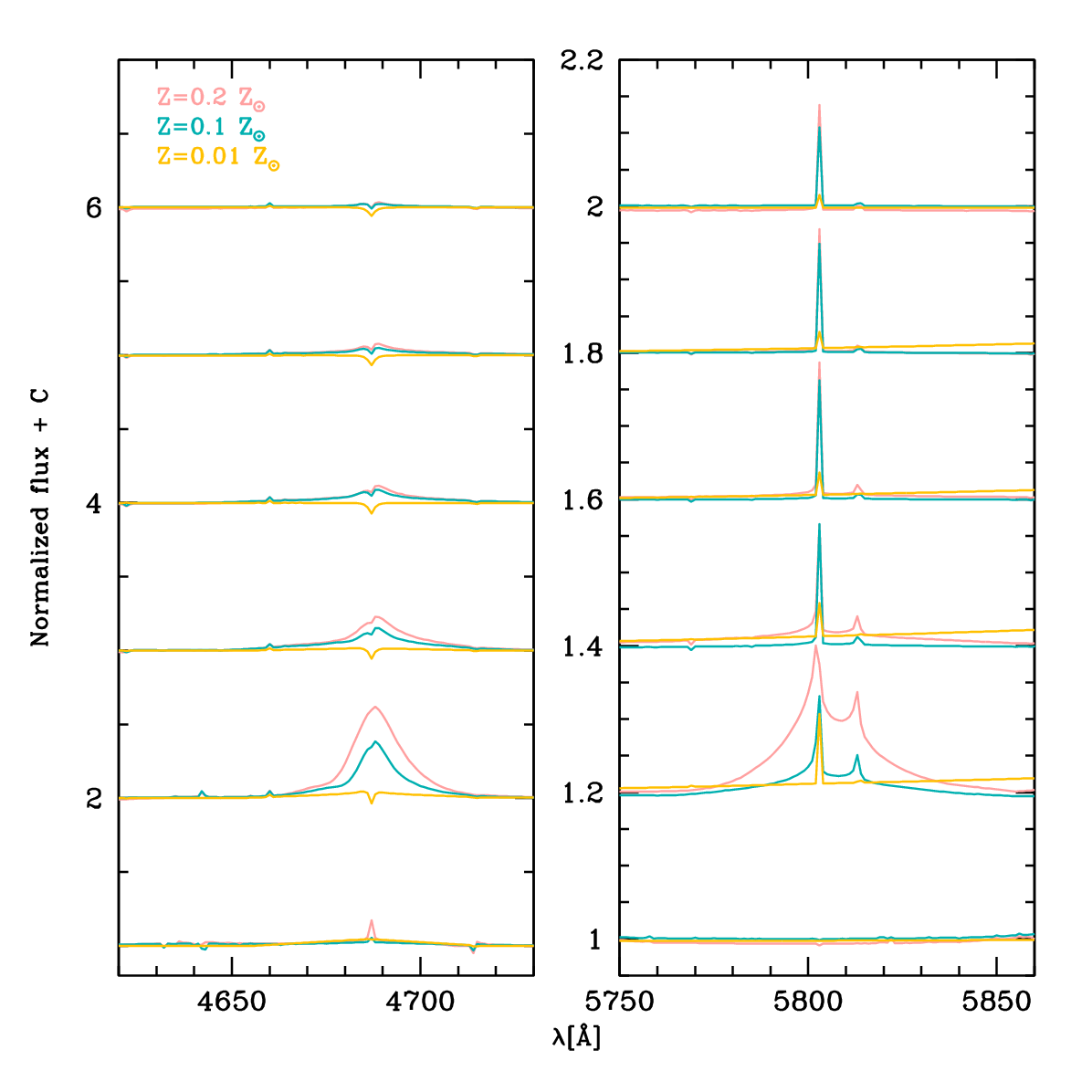}
\includegraphics[width=0.45\textwidth]{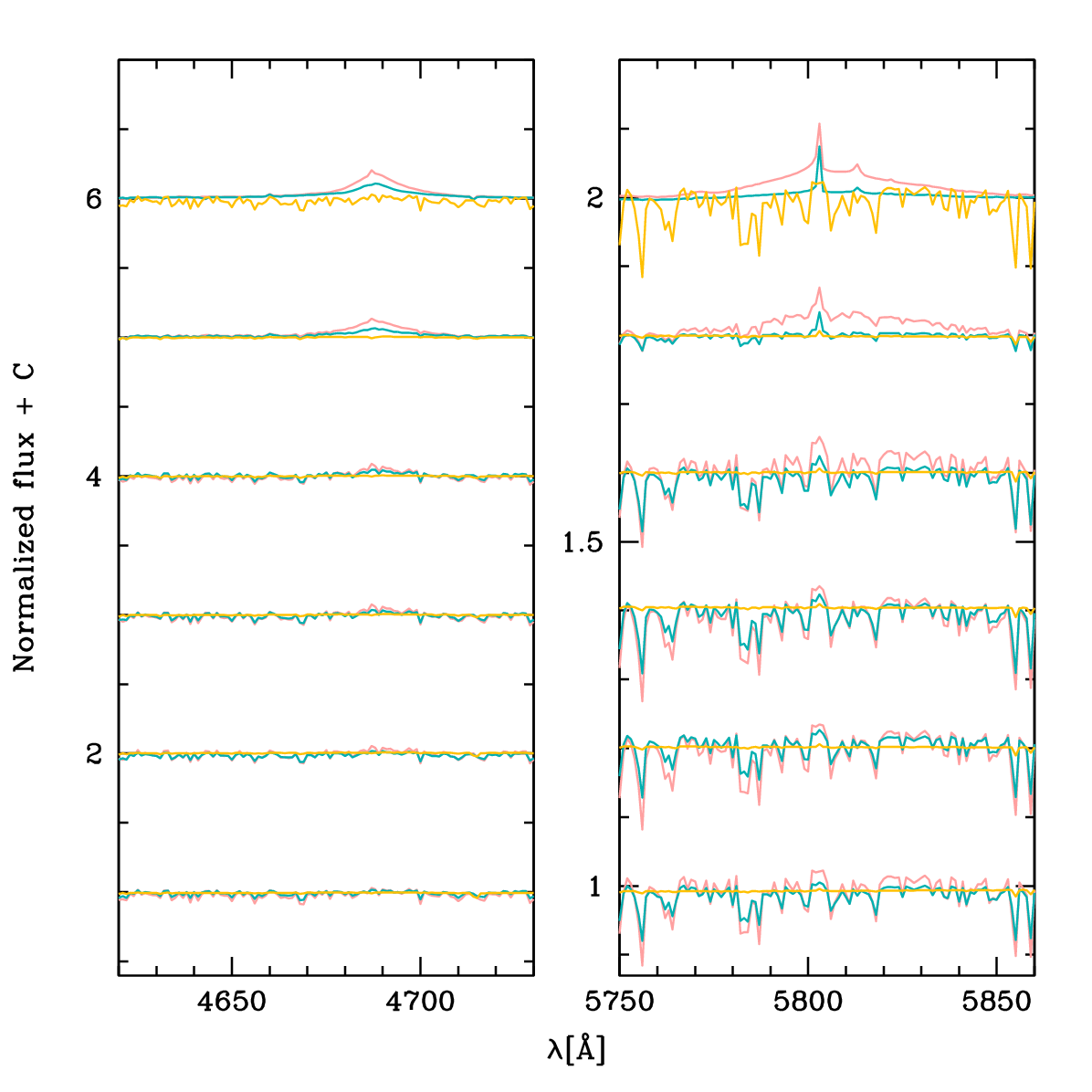} 
\caption{Morphology of the optical blue and red  bumps in burst models.  Different colours correspond to different metallicities (see left panels). The top (bottom) panels correspond to models without (with) Z scaling of the VMS mass loss rates. The left panels show burst models in which age goes from 0 (top) to 2.5~Myr (bottom) with 0.5~Myr increments. The right panels show CSF models with age 3, 5, 8, 10, 20 and 50~Myr from top to bottom.}
\label{comp_bumps}
\end{figure*}

In Fig.~\ref{comp_bumps} we show the morphology of the blue and red WR bumps in our models that include VMS. The blue bump never shows \ion{N}{iii}~4634-42 emission, whatever the star-formation history and the assumption regarding VMS mass loss. \heiiopt\ goes from a weak feature to a strong emission. It is stronger for higher mass loss rates, as expected. The maximum emission is seen in burst models at 2.5~Myr of CSF models at 3~Myr, when the contribution of VMS with respect to other stars is maximum. For models with a metallicity scaling of VMS mass loss rates the \heiiopt\ emission is usually weak and disappears at the lowest metallicity. For CSF models with age larger than $\sim$10~Myr the blue bump is also weak or absent. 

In most models the red bump is characterised by the narrow \ion{C}{iv} doublet emission. A broader and weak component is sometimes observed underneath the doublet. For CSF models the red bump is very weak or absent. 

For the metallicity range considered VMS may not always manifest themselves through the optical bumps. If a strong \heiiopt\ emission is seen,  \ion{N}{iii}~4634-42 is absent. The red bump may or may not be seen. When present it is most of the time dominated by the narrow \ion{C}{iv} doublet.

\subsection{Ionising flux}
\label{Z_Qi}

\begin{figure*}[h]
\centering
\includegraphics[width=0.24\textwidth]{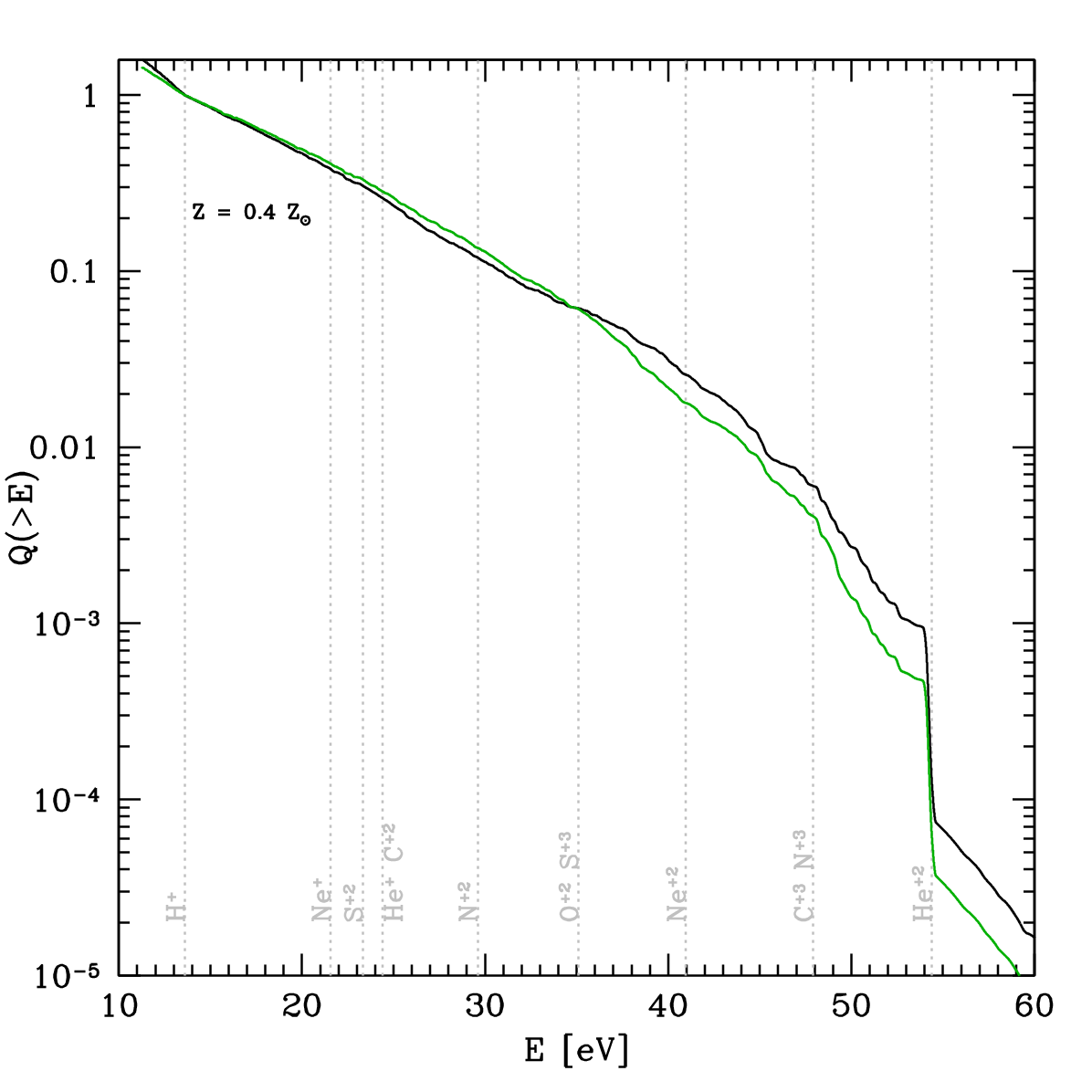}
\includegraphics[width=0.24\textwidth]{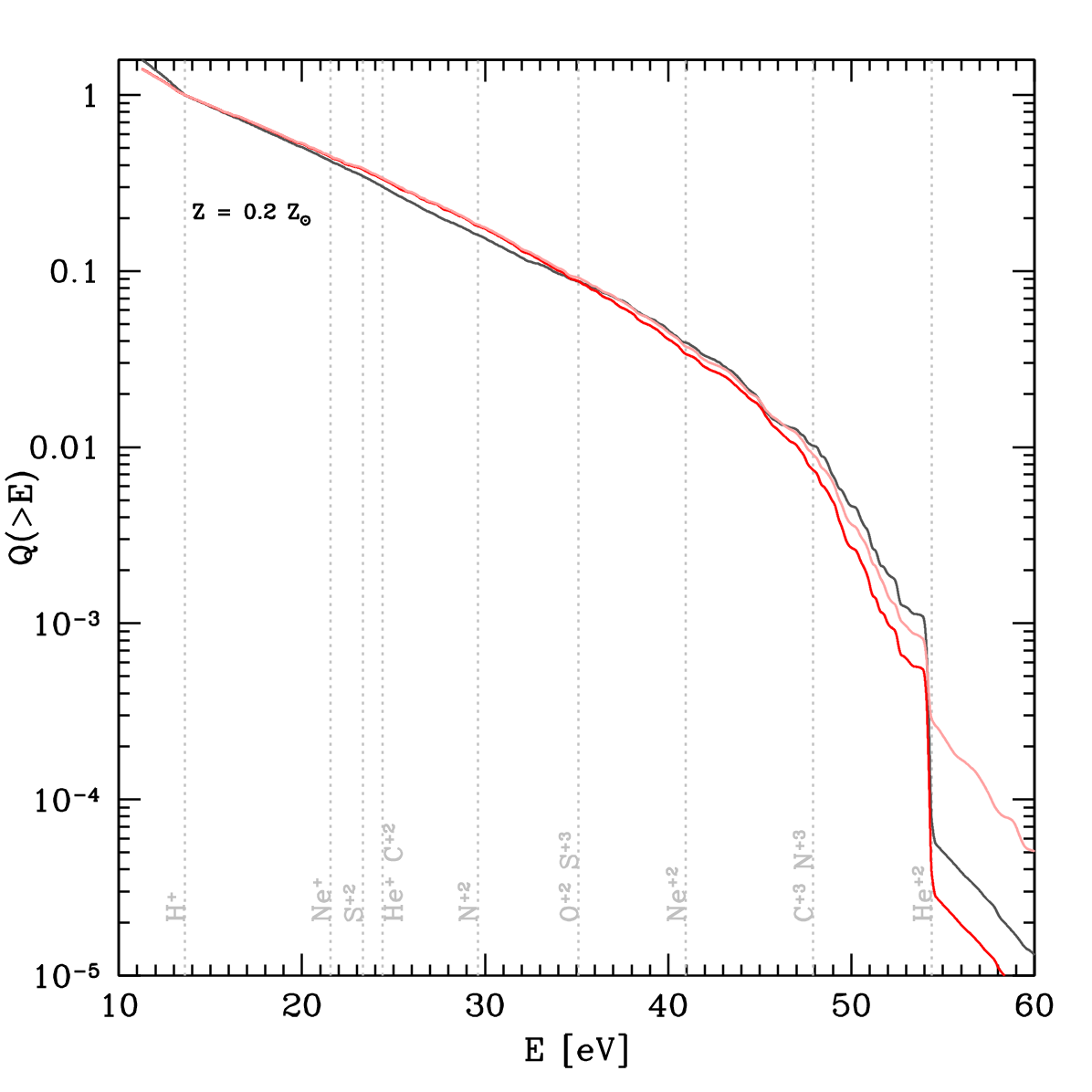}
\includegraphics[width=0.24\textwidth]{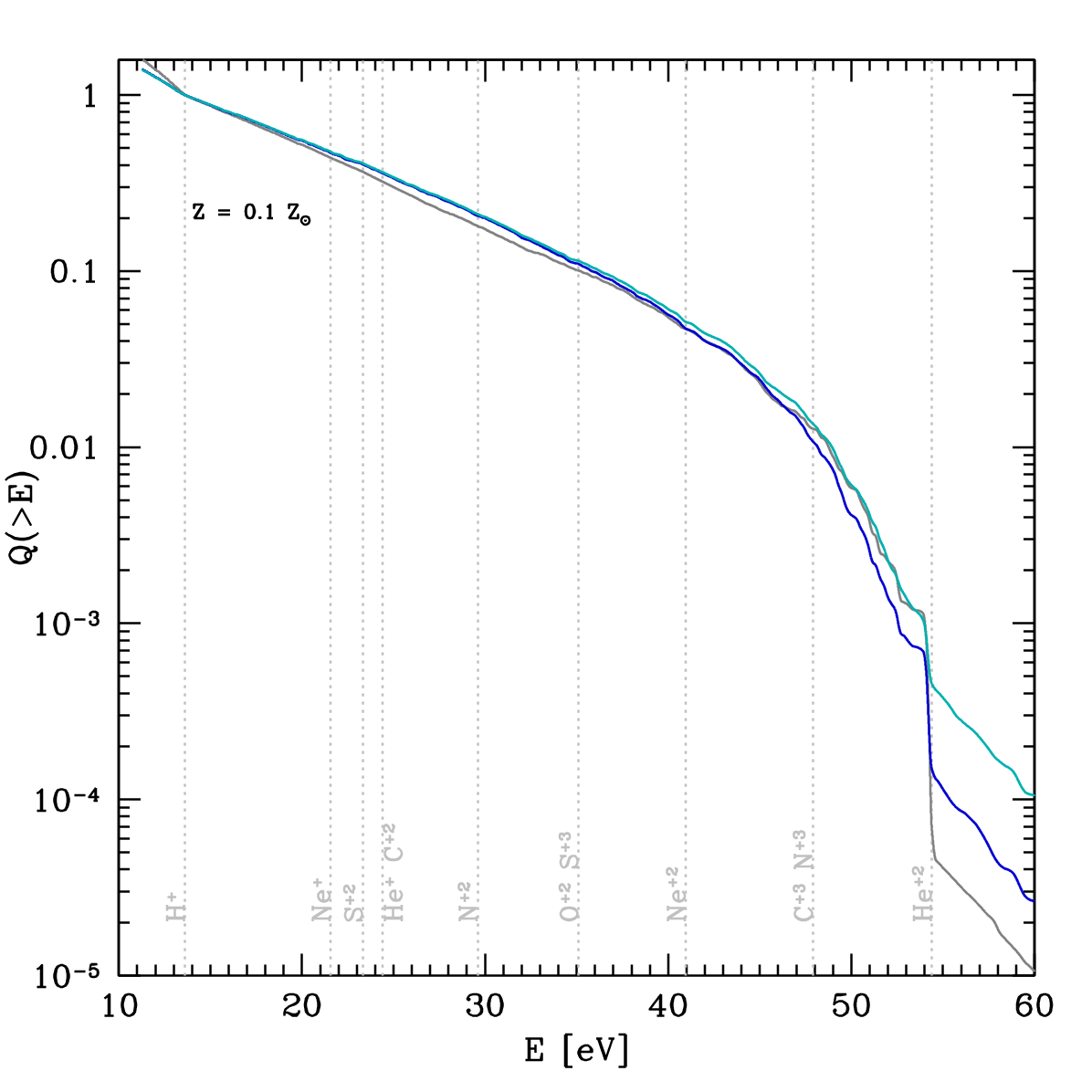}
\includegraphics[width=0.24\textwidth]{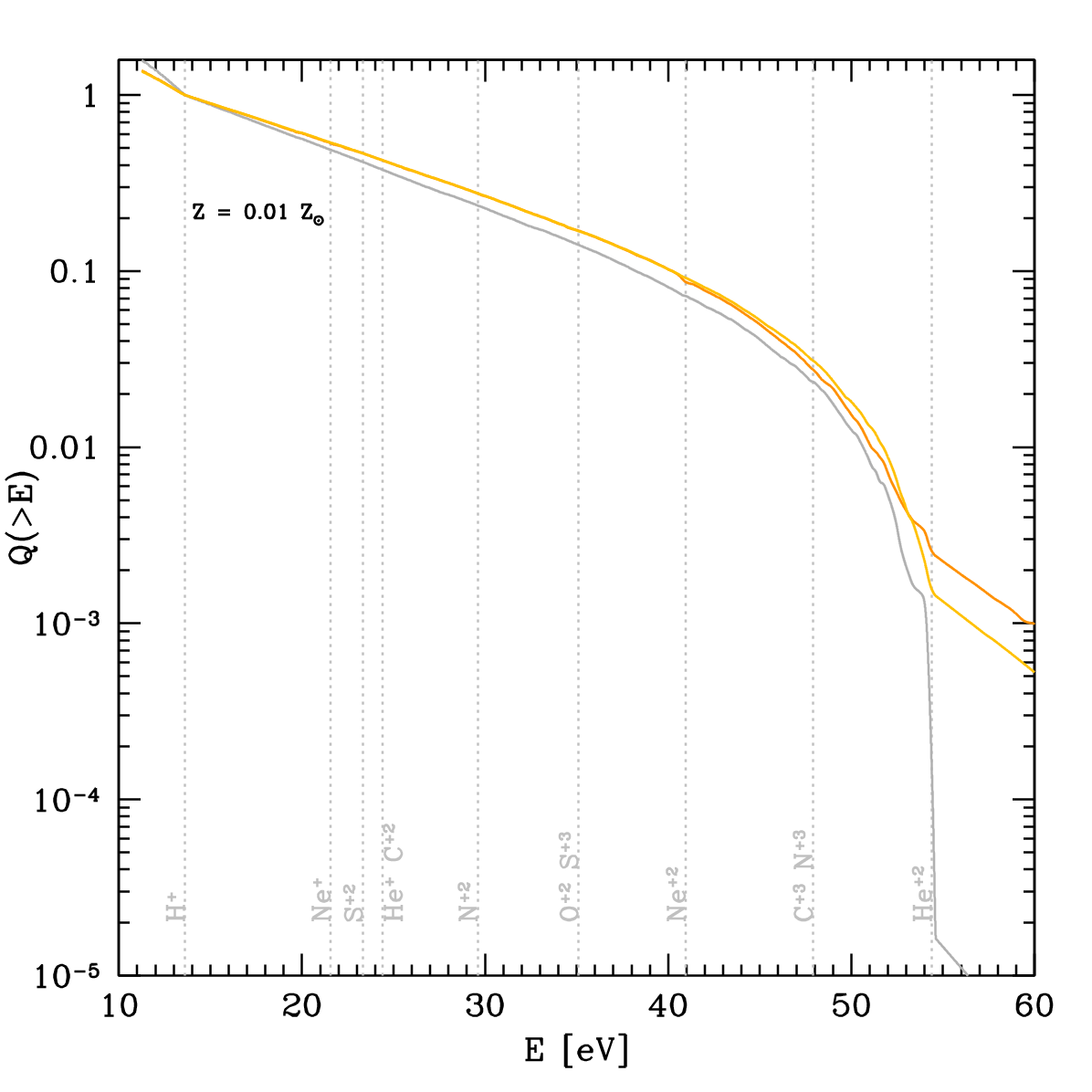}\\
\includegraphics[width=0.24\textwidth]{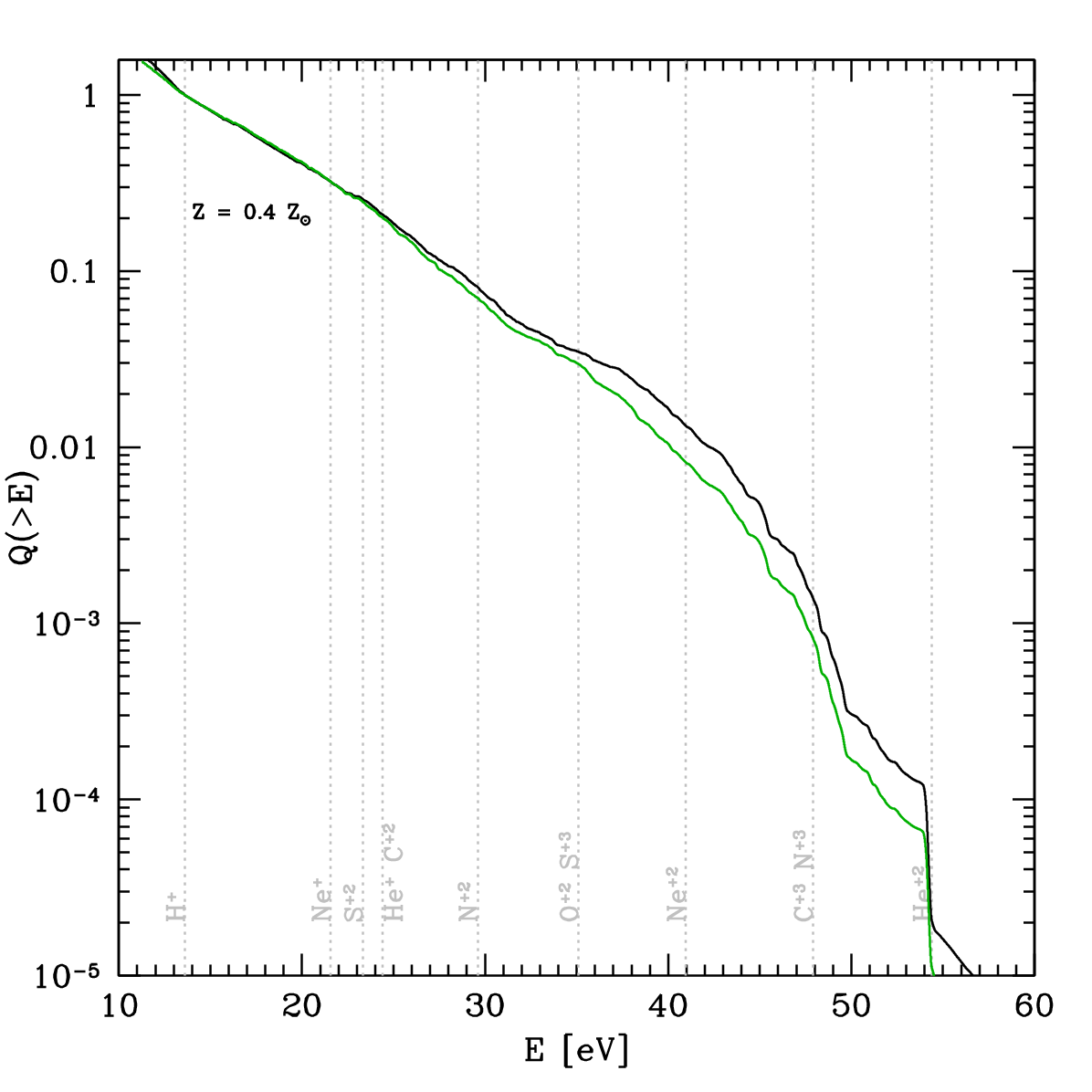}
\includegraphics[width=0.24\textwidth]{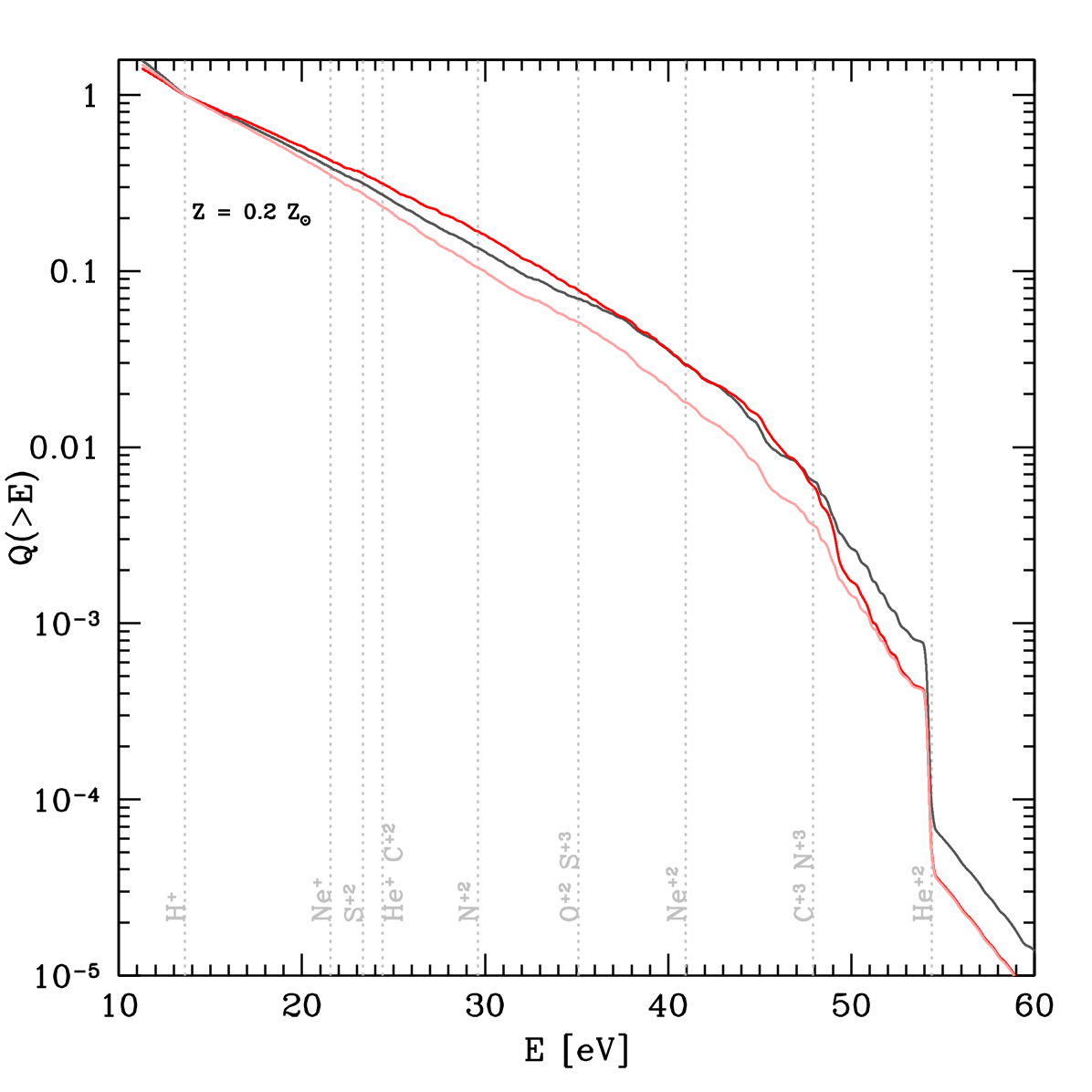}
\includegraphics[width=0.24\textwidth]{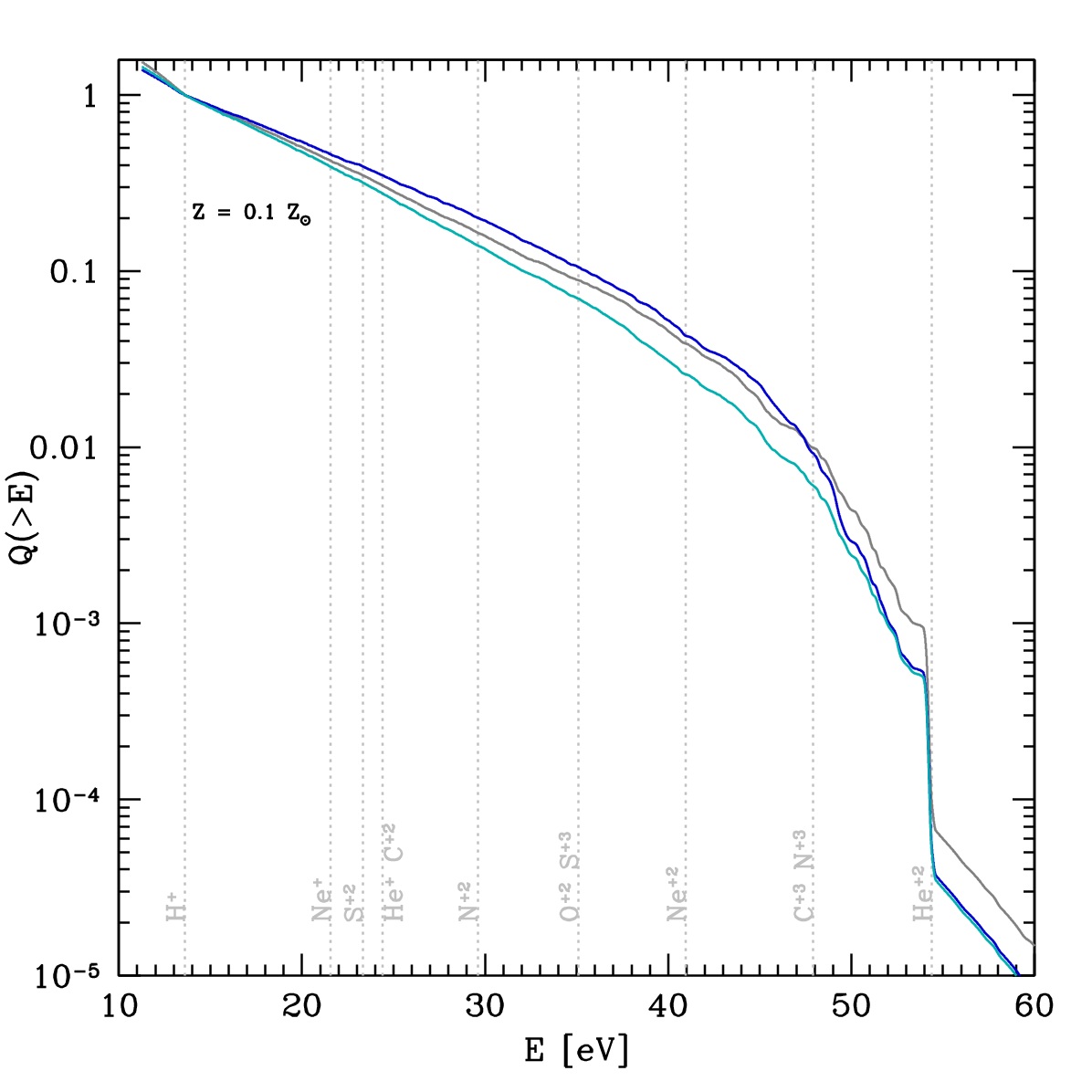}
\includegraphics[width=0.24\textwidth]{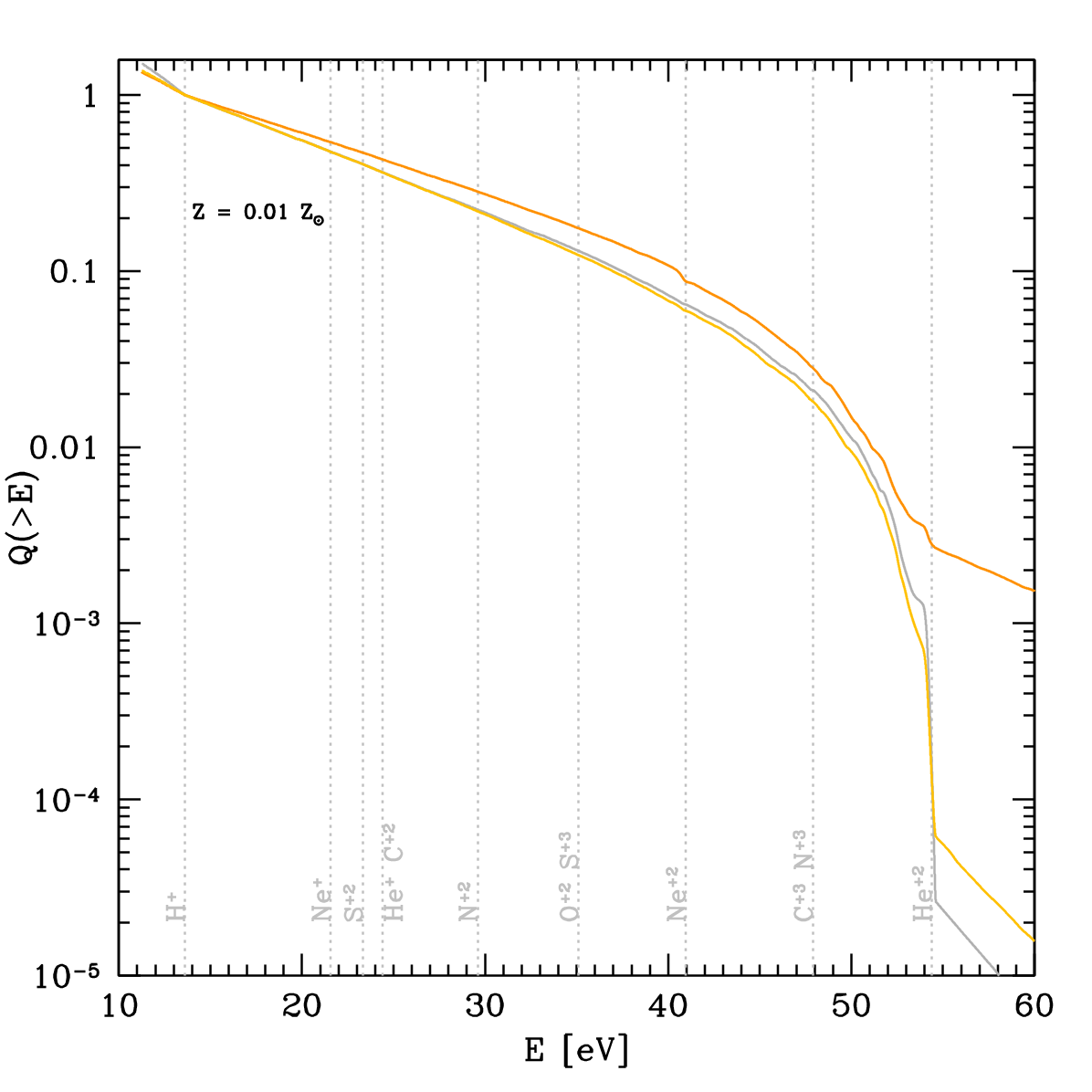}
\caption{Number of photons above energy E for Z=0.4~\zsun, 0.2~\zsun, 0.1~\zsun\ and 0.01~\zsun\ (from left to right). Lighter and heavier colors in each panel correspond to models with (without) Z-scaling of the VMS mass loss rates. The black and grey lines correspond to the BPASS model in each panel. The top (bottom) panels are for an age of 1(2)~Myr. Vertical dotted lines mark the ionisation energy of selected ions.}
\label{QE}
\end{figure*}

\begin{figure*}[h]
\centering
\includegraphics[width=0.24\textwidth]{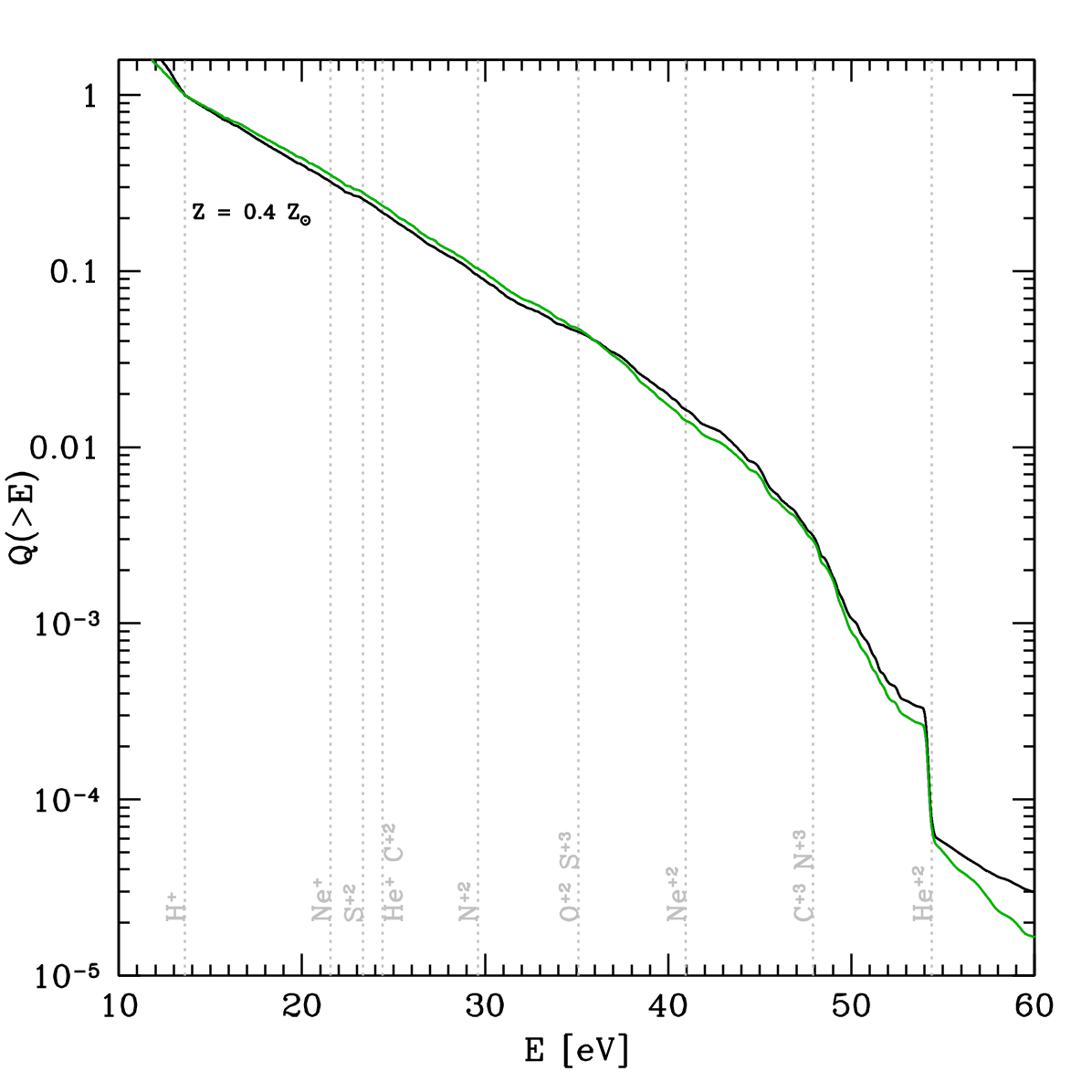}
\includegraphics[width=0.24\textwidth]{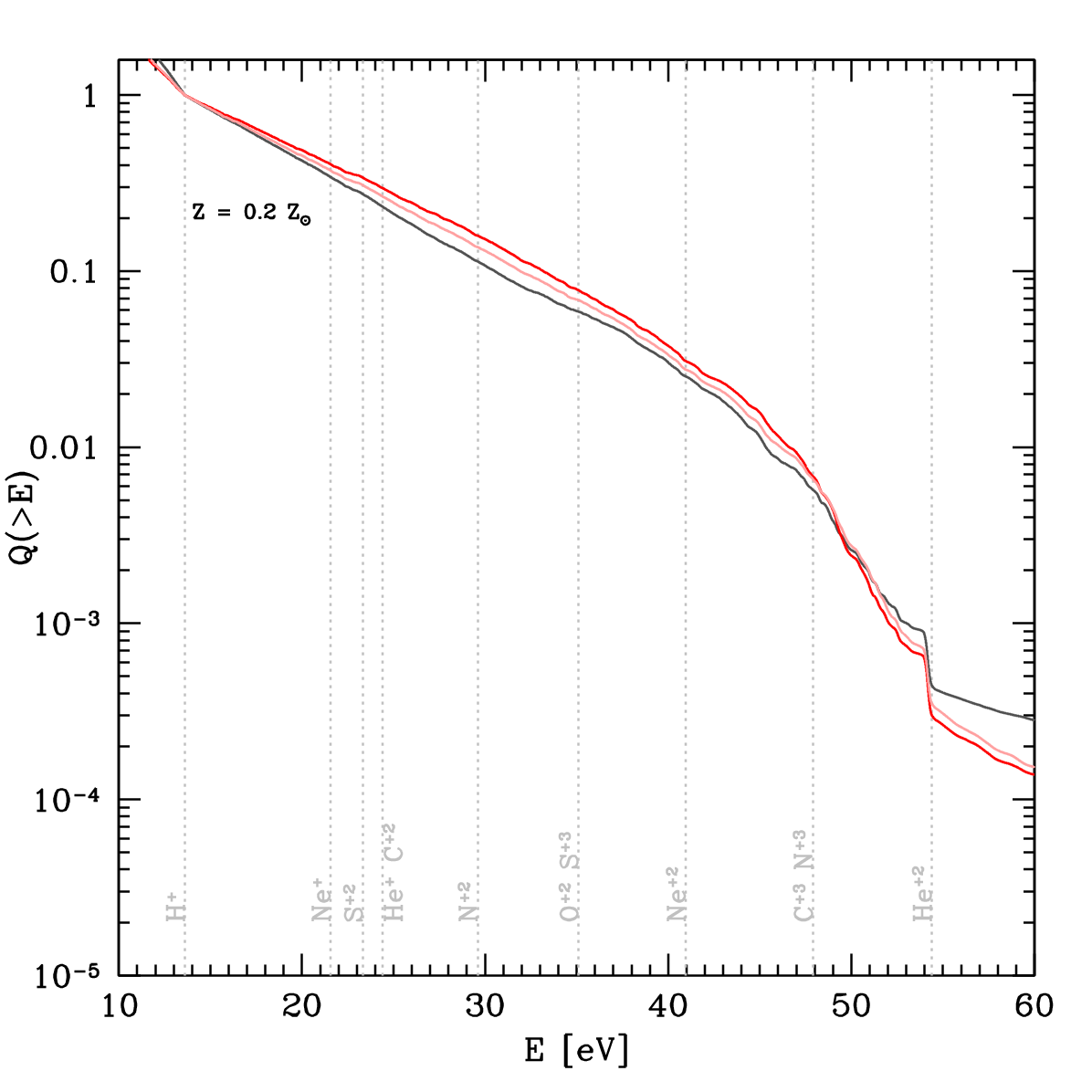}
\includegraphics[width=0.24\textwidth]{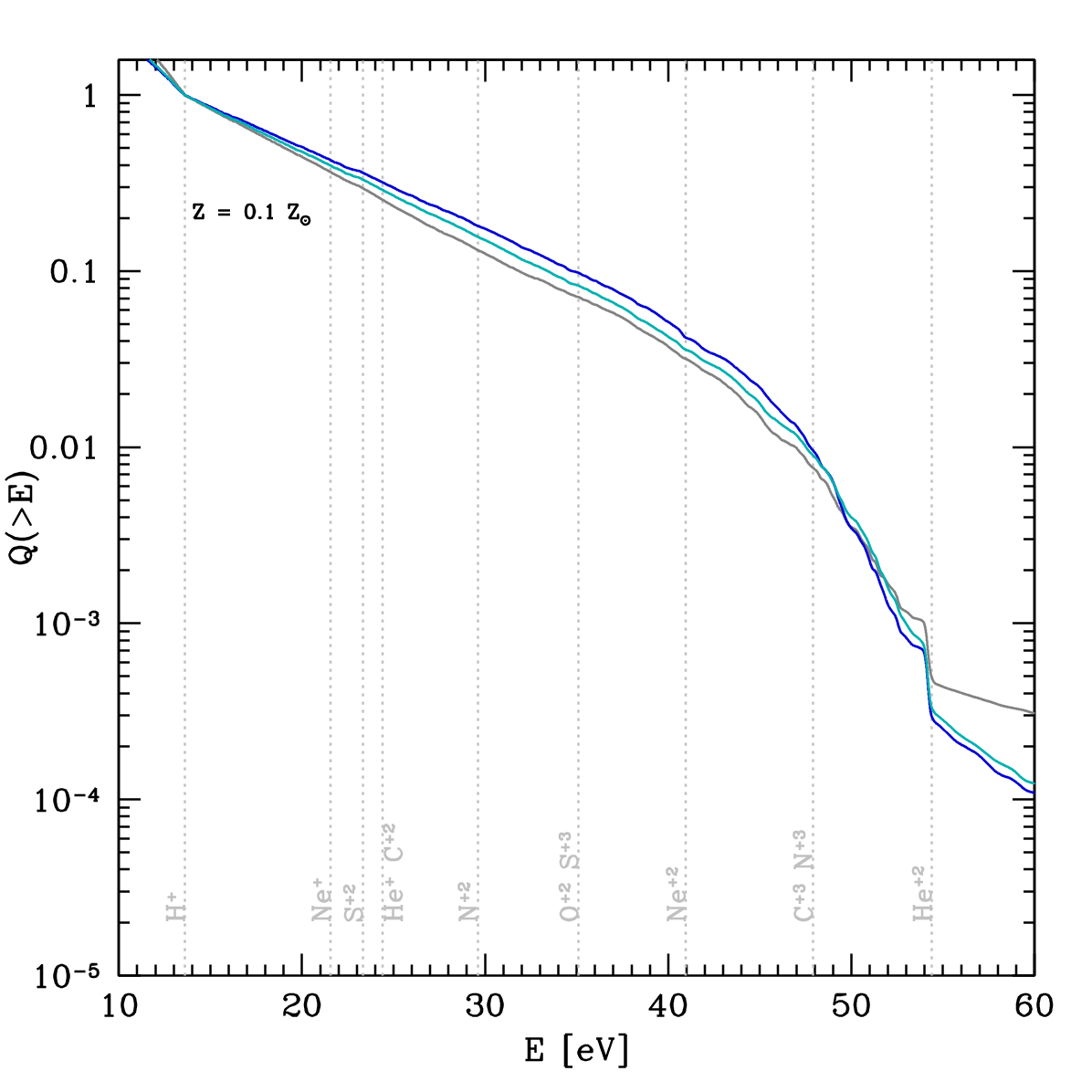}
\includegraphics[width=0.24\textwidth]{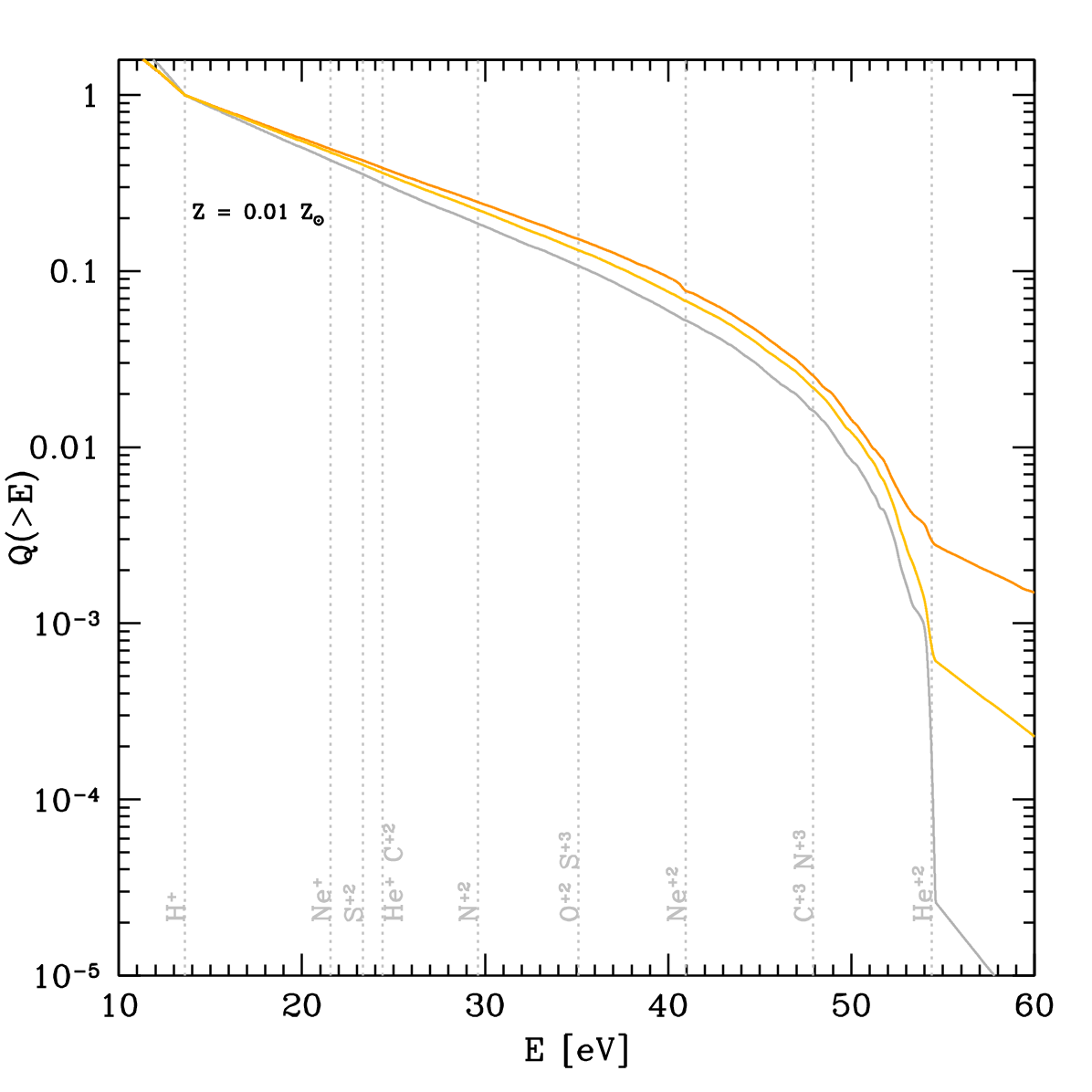}
\caption{Same as Fig.~\ref{QE} for 10~Myr CSF models.}
\label{QEcsf}
\end{figure*}

In this Section we first compare the shape of the ionising spectrum of stellar populations hosting VMS. For this we plot the cumulative number of ionising photons above energy E and normalized to the value at 13.6 eV (which corresponds to the Lyman break). Fig.~\ref{QE} shows this quantity for burst models at 1 and 2~Myr, and for the same metallicities as in the previous Section. The Z=0.4~\zsun\ case at 1~Myr is discussed by \citet{schaerer24} who highlight that the shape of the ionising spectrum is basically unaffected by the presence of VMS below 35~eV. At higher energies VMS lead to a reduction of the ionising flux, mostly because of their dense winds.

The second, third and fourth columns of Fig.~\ref{QE} present the Z = 0.2, 0.1 and 0.01~\zsun\ cases respectively. Compared to the Z=0.4~\zsun\ case there is a global trend of more ionising photons in models with VMS compared to models without them, as metallicity decreases. At 1~Myr the choice of the mass loss rate has little impact on the number of ionising photons below $\sim$47~eV. This is different above that energy, and especially true for \ion{He}{ii} ionising photons. Models without metallicity scaling of the VMS mass loss rates produce less \ion{He}{ii} photons than models with weaker winds. At Z = 0.2~\zsun\ the former have less ionising photons than the BPASS models, while the latter have more ionising photons. Below Z=0.1~\zsun\ all models with VMS produce more \ion{He}{ii} ionising photons than the BPASS models. As described in \citet{schaerer24} the presence of weaker winds leads to a reduced opacity in the \ion{He}{ii} continuum that translates into more flux.

At 2~Myr (bottom panels of Fig.~\ref{QE}) models with VMS and Z = 0.4~\zsun\ have a significantly softer ionising spectrum compared to models without VMS. The reason is that at that age VMS are evolved and cooler than the bulk of less massive stars. VMS thus contribute less flux at high energies, leading to a softer spectrum of the population. 
At 2~Myr the choice of the mass loss rate recipe for low metallicity VMS has an impact on the shape of the ionising continuum. Indeed, depending on the mass loss history VMS will evolve to the red part of the HRD, thus becoming cooler, or will stay in the hot part of the HRD, and will produce ionising photons (see Fig.~\ref{hrd}). The bottom panels of Fig.~\ref{QE} show this effect: below $\sim$45~eV the dark lines, corresponding to the un-scaled VMS mass loss rates, are above the light ones, corresponding to Z-scaled mass loss rates. Regarding the \ion{He}{ii} ionising flux, only the lowest metallicity models produce more photons than models without VMS. 

For completeness we show in Fig.~\ref{QEcsf} Q($>$E) for 10~Myr CSF models. As seen in burst models, a reduction of metallicity translates into a larger number of ionising photons with E$<$45~eV in models with VMS, compared to non-VMS models. Above 54~eV only the Z = 0.01~\zsun\ models produce more ionising photons (right panel). Above that metallicity models with and without VMS produce a rather similar ionising continuum. 

In conclusion, as the shape of the ionising flux is concerned, there is a global trend of more ionising photons below $\sim$45~eV when VMS are included with strong mass loss rates. For higher energies and for models with lower mass loss rates the behaviour is more complex, depends on age, star formation history and energy. As discussed in Sect.~\ref{s_postMS} the treatment of the most advanced phases of evolution also shapes the SED at these energies in a way that remains difficult to quantify at present. 

\smallskip

\begin{figure}[h]
\centering
\includegraphics[width=0.24\textwidth]{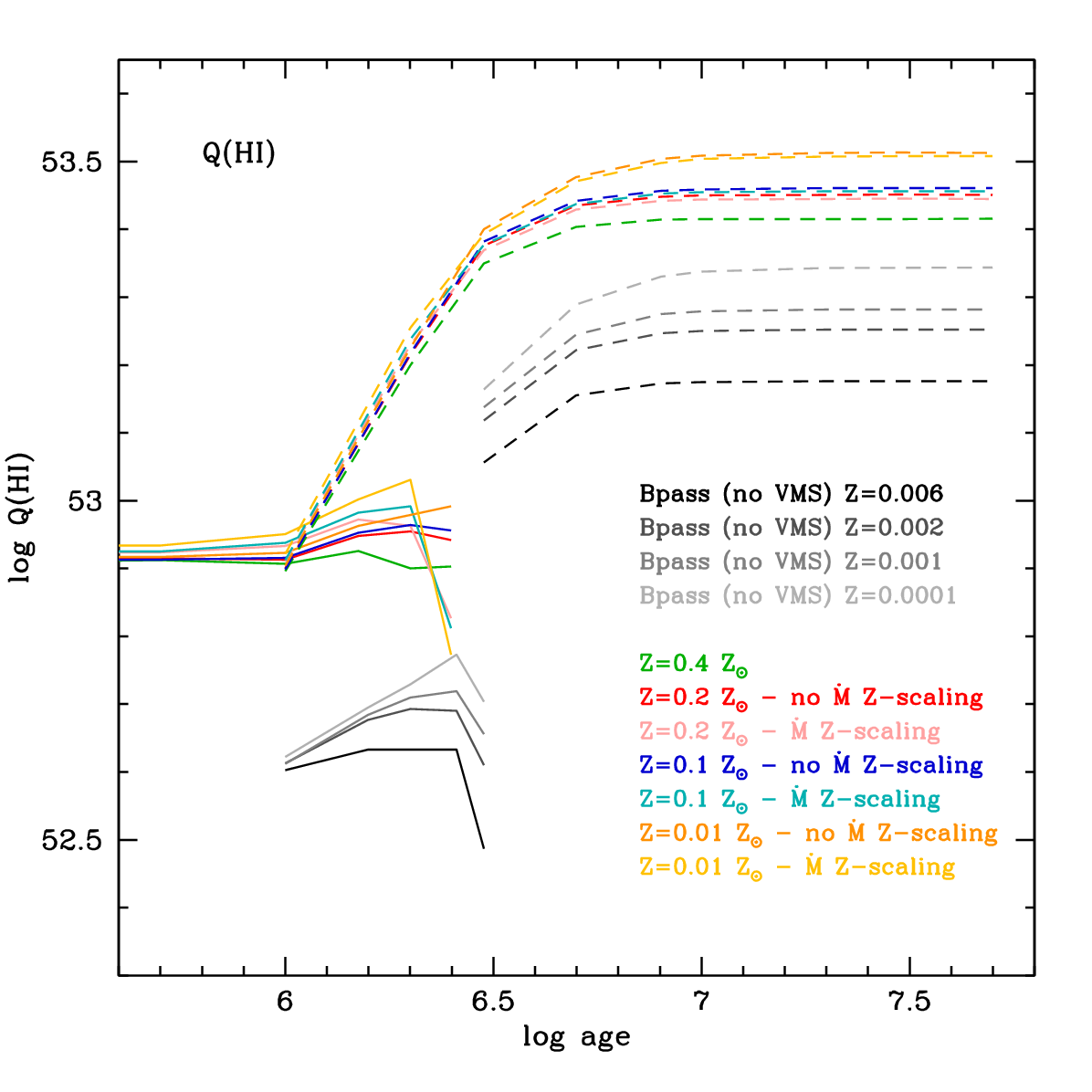}
\includegraphics[width=0.24\textwidth]{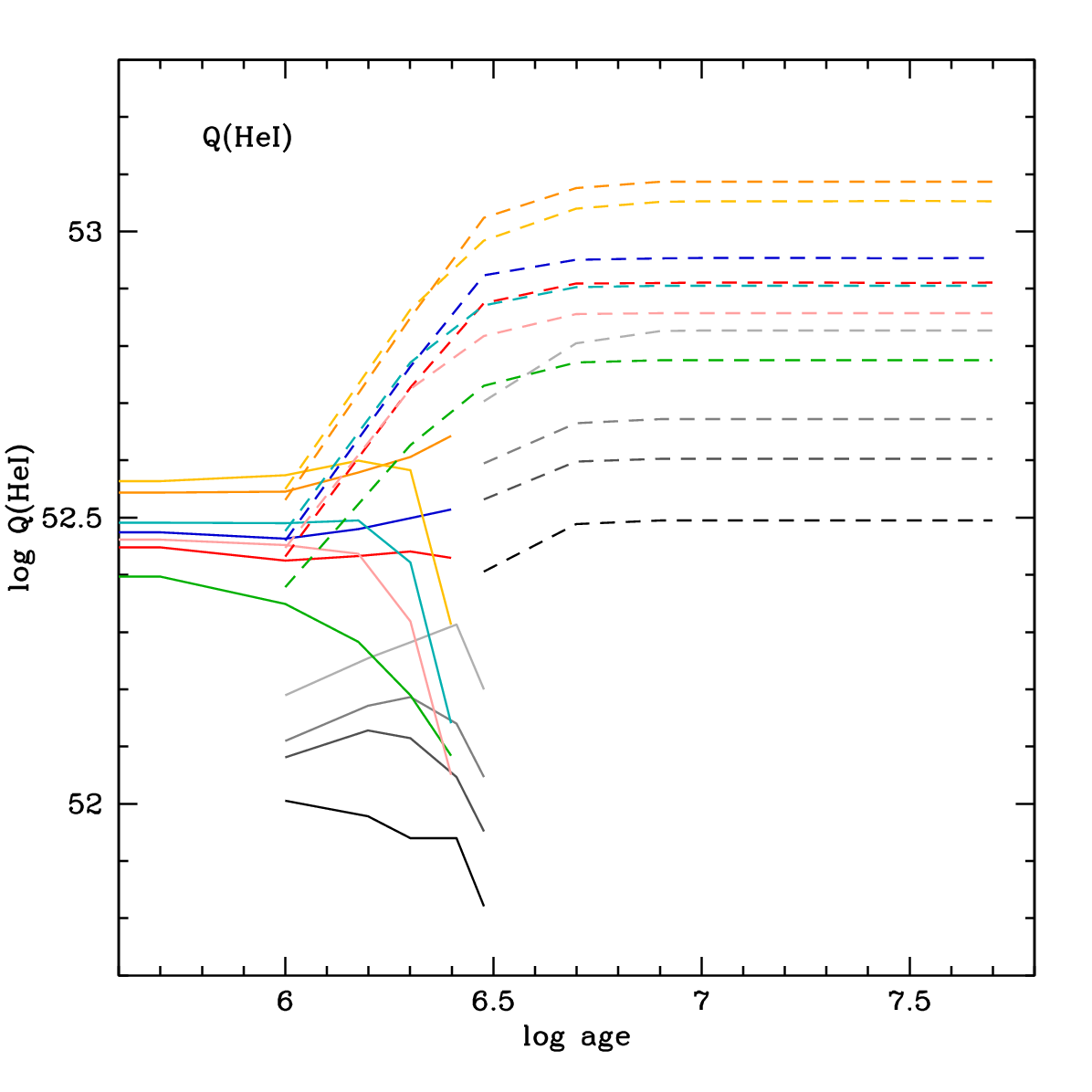}\\
\includegraphics[width=0.24\textwidth]{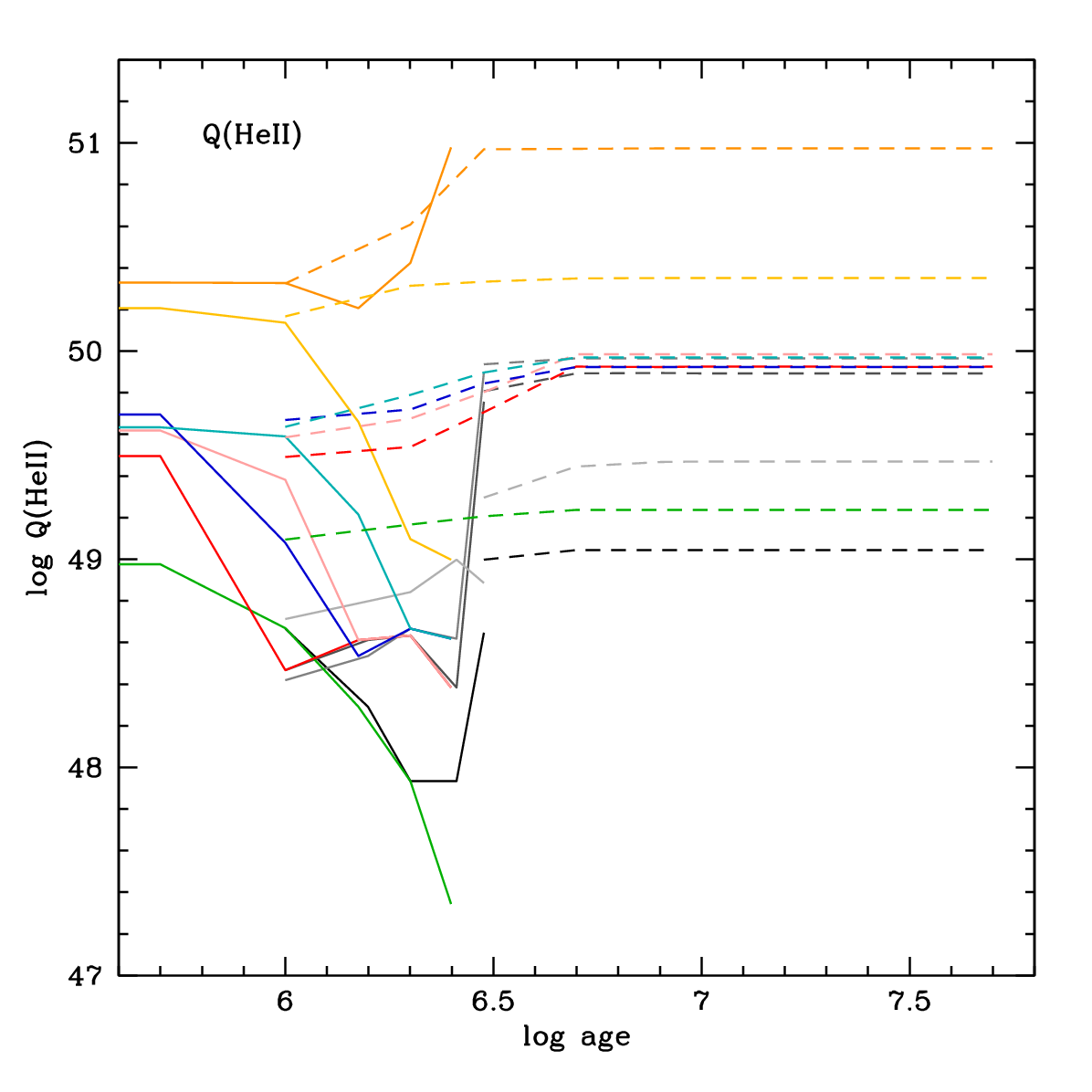}
\includegraphics[width=0.24\textwidth]{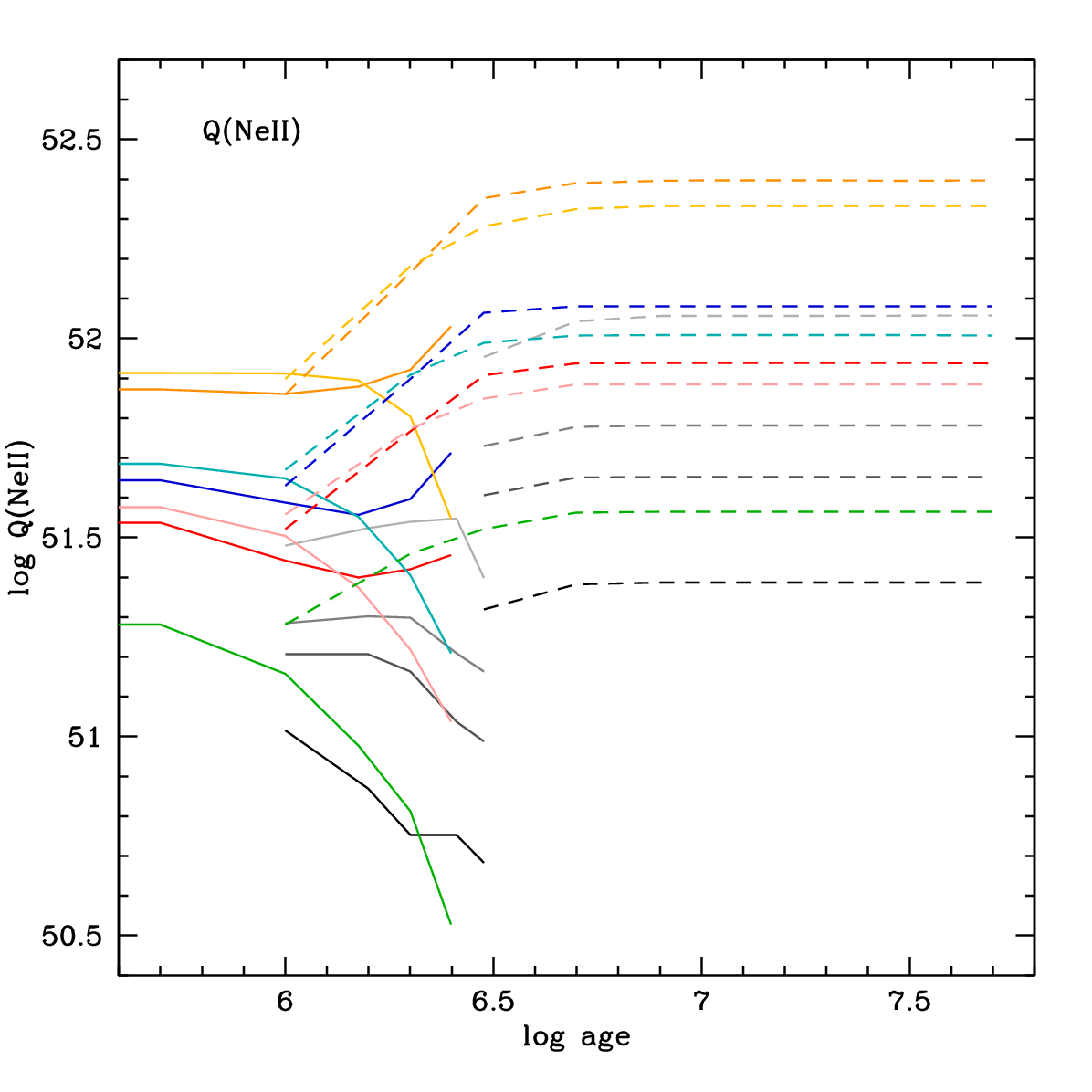}
\caption{Number of ionising photons per second as a function of age for population synthesis models. The upper left (upper right, lower left, lower right) panel is for \ion{H}{i} (\ion{He}{i}, \ion{He}{ii}, \ion{Ne}{ii}). Solid (dashed) lines are burst (CSF) models. Grey lines are for BPASS models that do not include VMS, i.e. the models we use for stars with masses below 100~\msun.}
\label{Qi}
\end{figure}

Fig.~\ref{Qi} shows the effects of metallicity on the number of ionising photons for our models that include VMS. In addition to \ion{H}{i}, \ion{He}{i}, and \ion{He}{ii} we also show the \ion{Ne}{ii} photon flux since it corresponds to an energy of 40.9~eV close to the energy separating the spectral regions where the effects of the assumptions on mass loss rate on the shape of the SED are best seen. When metallicity decreases there is a global trend of more ionising photons. The difference is the smallest for \ion{H}{i} and increases at higher energies as discussed above. We also see that Q(\ion{He}{ii}) has a relatively complex behaviour as already stressed (see also Sect.~\ref{s_postMS}). Fig.~\ref{Qi} also confirms the trend reported by \citet{schaerer25} that VMS increase the number of \ion{H}{i} and \ion{He}{i} ionising photons by a factor$\sim$2. Again, the behaviour of the number of \ion{He}{ii} ionising photons is more complex with no clear trend with metallicity.

\begin{figure}[h]
\centering
\includegraphics[width=0.45\textwidth]{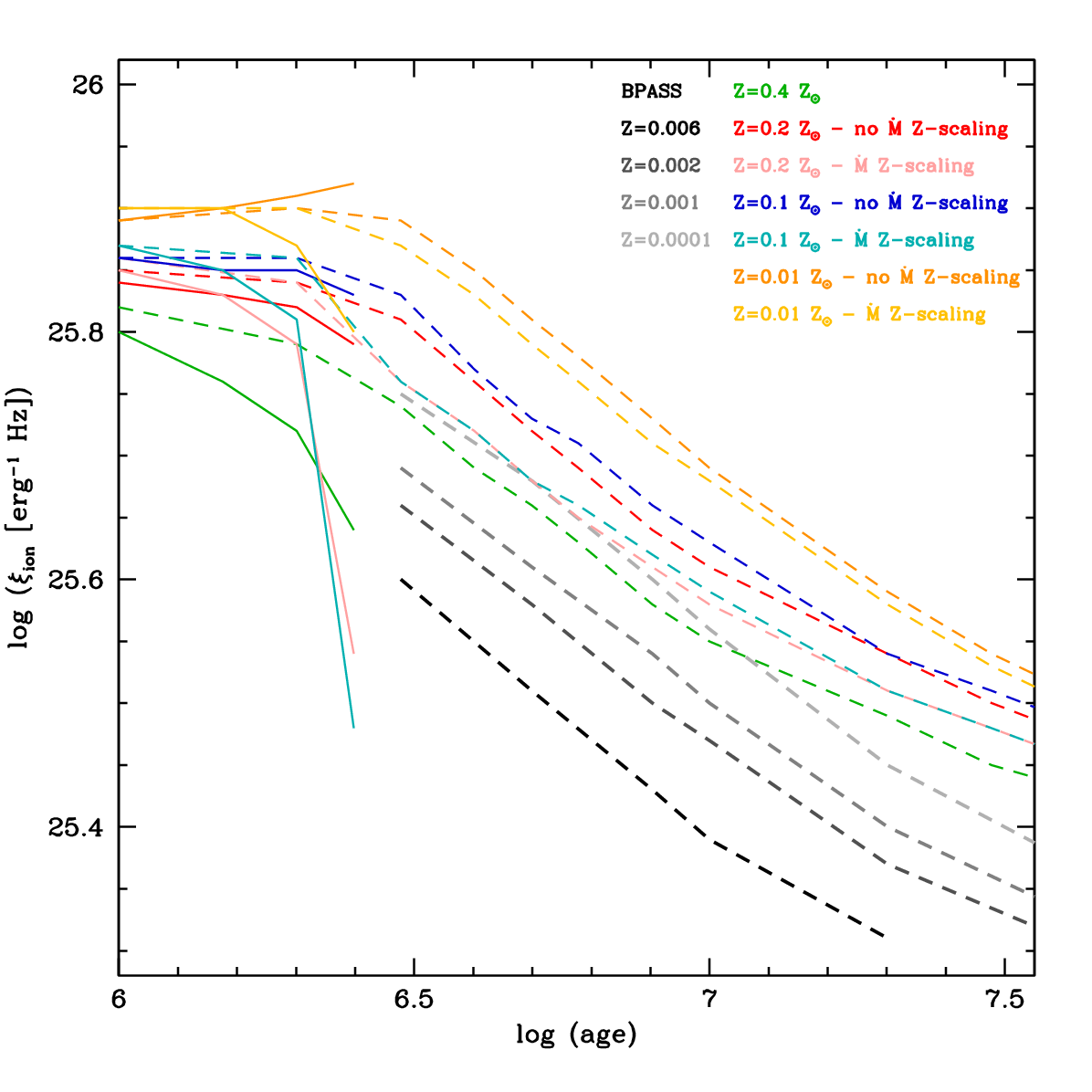}
\caption{ionising photon efficiency as a function of age. Different colours are used for different metallicities and assumption regarding VMS mass loss rate, as explained in the upper right corner of the Figure. Solid (dashed) lines are for burst (CSF) models that include nebular continuum emission.}
\label{Xi_ion}
\end{figure}

\smallskip

Finally Fig.~\ref{Xi_ion} shows the ionising photon efficiency $\xi_{ion}$ defined as the ratio of the number of hydrogen ionising photons to the luminosity at 1500~\AA\ (in units of erg/s/Hz). A reference value for $\log (\xi_{ion})$ is 25.2 erg$^{-1}$ Hz, as defined by \citet{robertson13} based on population synthesis models that do not include VMS \citep{bc03}. The models shown in Fig.~\ref{Xi_ion} include the nebular continuum emission on top of the stellar emission. To add this contribution, we followed the method described in \citet{schaerer02}, with the various coefficients taken from \citet{osterferland}. Nebular emission includes free-free and free-bound emission by H, neutral He and He+. Hydrogen two photons emission is also taken into account. An electron temperature of 10000~K and an electron density of 100~cm$^{-3}$ are adopted.

\citet{schaerer25} show that $\xi_{ion}$ is boosted by the presence of VMS and more generally by a top-heavy initial mass function. The models of 
Fig.~\ref{Xi_ion} confirm this trend: for a given metallicity the models that include VMS have systematically larger $\xi_{ion}$ values compared to the BPASS models (that do not include VMS). 
The novelty seen in Fig.~\ref{Xi_ion} compared to the work of \citet{schaerer25} is the metallicity dependence of $\xi_{ion}$. Taking the CSF models at 10~Myr as an example, $\log (\xi_{ion})$ increases from 25.55 at Z = 0.4~\zsun\ to 25.62 at Z = 0.1~\zsun, and up to 25.69 at Z = 0.01~\zsun. For comparison the BPASS models produce $\log (\xi_{ion})$ = 25.39 to 25.56 for the same metallicity range. The increase of $\xi_{ion}$ also depends on the mass loss assumption for VMS. The models that produce the largest ionising photon efficiency are those with the higher mass loss rates. This is mostly because these models remain hot, leading to a large ionising photon flux. 
Measurements of $\log (\xi_{ion})$ in star-forming galaxies at high redshift often reach values of 25.5 to 26.0, as summarised in Fig.~8 of \citet{seeyave23}. Such values are difficult to explain by standard models (i.e. without VMS) unless bursts of age $\lesssim$5~Myr are invoked. Our models release part of the tension with empirical measurements, offering the possibility to reach large $\xi_{ion}$ at low metallicity.  

\subsection{Comparison to local starbursts}
\label{s_popsynobs}

In this Section we compare our population synthesis models including VMS to spectroscopic observations of low metallicity starburst clusters or galaxies in the Local Universe. The UV data used for the comparisons, all from the Hubble Space Telescope (HST), are described in Table~\ref{tab_data_sbursts} and have been retrieved from the MAST database. We selected all sources that show \heiiuv\ in emission and for which the metallicity estimates are within the range of our models. We found three objects that fulfil these criteria: cluster A in II~Zw~40, cluster A in MrK71 and the starburst region SB-126. 

\begin{table}[h]
\begin{center}
  \caption{HST/COS data for starbursts} \label{tab_data_sbursts}
  \begin{tabular}{lcccc}
\hline
Name &  Grating & PID  & Reference\\    
\hline
II~Zw~40-A  &  G140L & 14102 &  1 \\
MrK71-A &  G140L & 16261 &  2 \\
SB~126 &  G160M+G185M & 15185 & 3 \\
\hline
\end{tabular}
\tablefoot{References: 1- \citet{leitherer18}; 2- \citet{smith23}; 3- \citet{senchyna20}}
\end{center}
\end{table}


\subsubsection{II~Zw~40-A}
\label{s_IIzw40}

\begin{figure*}[h]
\centering
\includegraphics[width=0.47\textwidth]{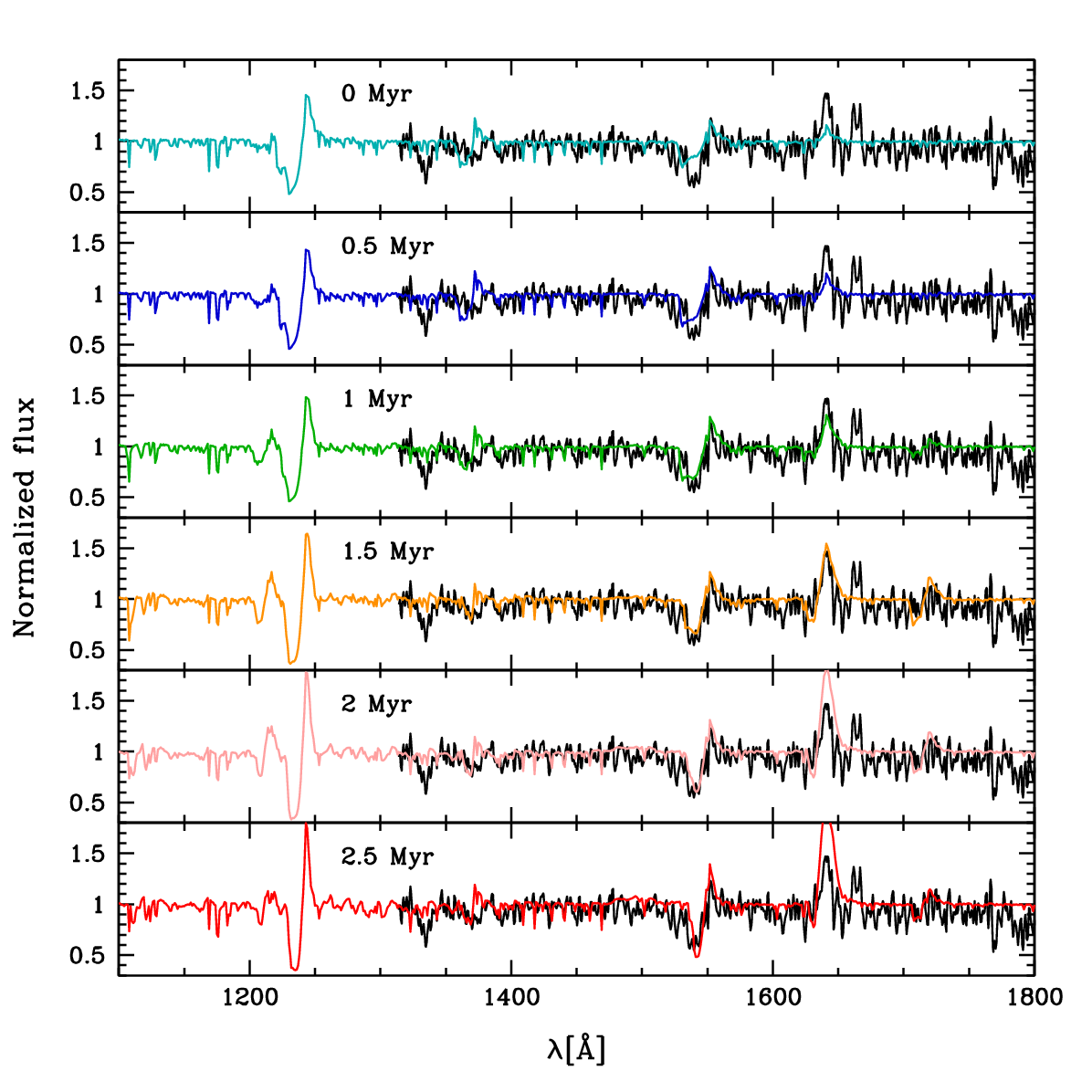}
\includegraphics[width=0.47\textwidth]{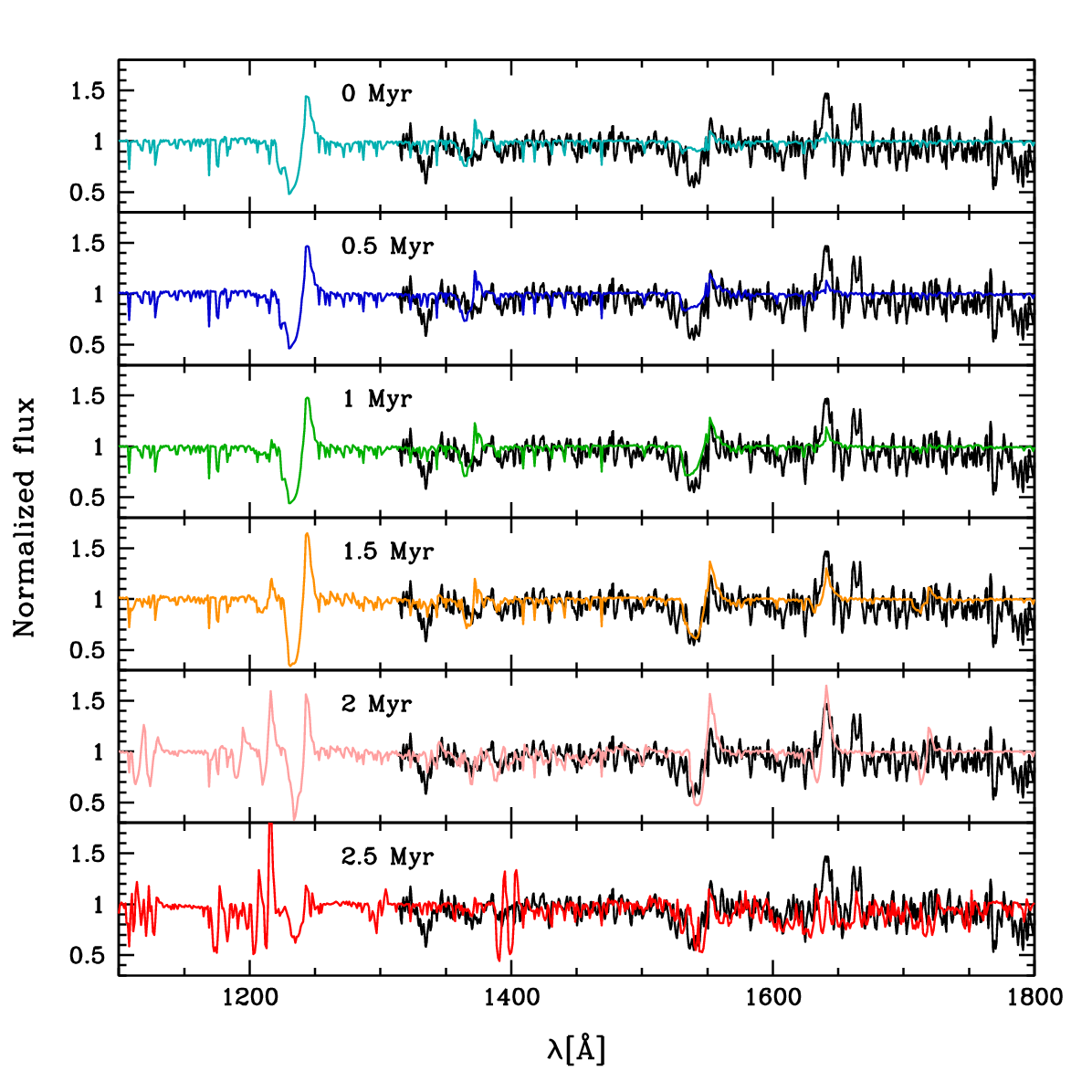}\\
\caption{Comparison between burst population synthesis models including VMS up to 225~\msun\ and the observed UV spectrum of the super star cluster in II~Zw~40. All models are for Z = 0.2~\zsun. Ages range from 0 to 2.5~Myr from top to bottom. The left (right) panels correspond to models without (with) metallicity scaling of the VMS mass loss rates.}
\label{popsyn_zw40}
\end{figure*}

Cluster A in the galaxy II~Zw~40 was studied by \citet{leitherer18} who show that it has one of the strongest \heiiuv\ emission among local starbursts. 
UV spectroscopy reveals the presence of all the strong lines seen in VMS and \citet{leitherer18} suspect that VMS are indeed present in the cluster. In Fig.~\ref{popsyn_zw40} we compare the predictions of our population synthesis models at different ages with the UV spectrum of II~Zw~40-A. The right and left panels correspond to models that have VMS mass loss rates that scale (do not scale) with metallicity. The metallicity of II~Zw~40 is 12+$\log$(O/H)=8.09, close to that of the SMC \citep{guseva00} so we use our models for Z~=~0.2~\zsun in Fig.~\ref{popsyn_zw40}. 
A direct by-eye inspection indicates that the global morphology of the observed spectra is qualitatively reproduced: \heiiuv\ is predicted in emission, \civuv\ is well developed, \nivuv\ is present. \ovuv\ is also predicted but systematically stronger than the observation. Among the two families of models, the 1.5~Myr model with no metallicity scaling of the mass loss rates (orange in the left panel) is particularly remarkable since it reproduces fairly well most observed features. The quality of this fit is not matched by models in which the mass loss rates are scaled with metallicity. In those models \heiiuv\ is too weak compared to the observed spectrum, or is reasonably reproduced but at the cost of a too strong \civuv.

\begin{figure}[h]
\centering
\includegraphics[width=0.47\textwidth]{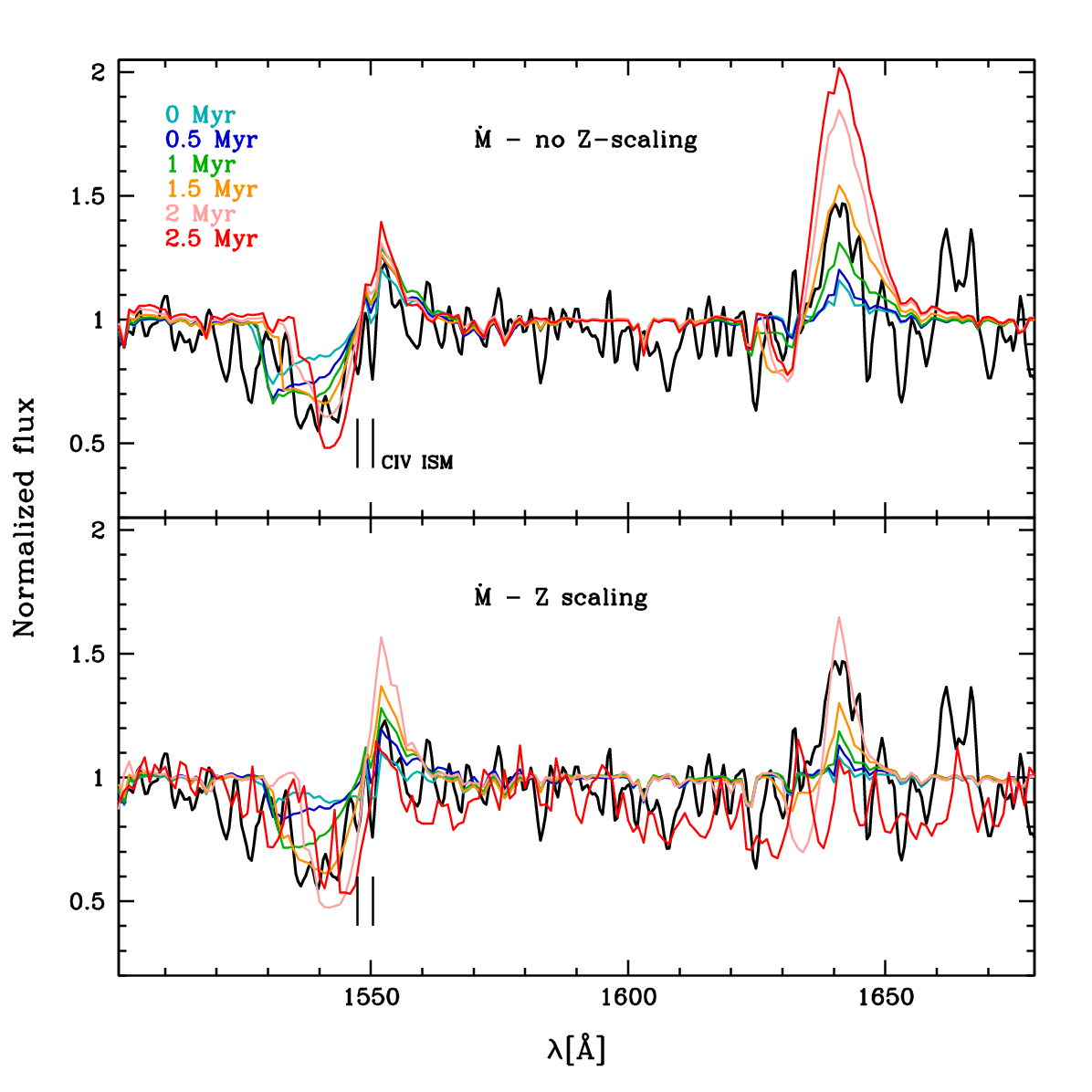}
\caption{Comparison between burst population synthesis models including VMS up to 225~\msun\ to the observed UV spectrum of cluster A in II~Zw~40 (black line). All models are for Z = 0.2~\zsun\ that is close to the metallicity of the cluster. Ages range from 0 to 2.5~Myr and are colour-coded as indicated in the top panel. The top (bottom) panel corresponds to models without (with) metallicity scaling of the VMS mass loss rates. The vertical black lines indicate the Galactic nebular \ion{C}{iv} absorption.}
\label{zoom_zw40}
\end{figure}

To better appreciate the comparisons we show in Fig.~\ref{zoom_zw40} a zoom on the two strongest features, \civuv\ and \heiiuv. The only model of the Z-scaling series that reaches the level of \heiiuv\ emission of II~Zw~40 is that at 2~Myr. But as stressed above the corresponding \civuv\ P-Cygni profile is too strong. For the no Z-scaling mass loss models, we see again that the 1.5~Myr model provides an almost perfect match to \civuv\ and \heiiuv. \citet{leitherer18} determined an age of 2.8$\pm$0.1~Myr for the cluster, from the comparison of the strongest UV features with Starburst99 models restricted to masses below 100~\msun. They could not reproduce \heiiuv\ (see their Fig.~3). Since VMS contribute not only to \heiiuv\ but also to other UV lines, stronger lines are produced at earlier times which explains the slightly younger age we favour. 

So far our population synthesis models include VMS up to 225~\msun. Increasing the upper mass limit to 300~\msun\ slightly boosts the UV emission, especially that of \heiiuv. But this is not sufficient to affect qualitatively the conclusions raised above: the best fit model remains one of the series without Z-scaling of the VMS mass loss rates, and the models with scaled mass loss rates suffer from the same limitations as above. With the \heiiuv\ emission increase when the upper mass limit is extended, the model of the no Z-scaling series at an age of 1~Myr becomes equivalent to that at 1.5~Myr . 

\subsubsection{MrK71-A}
\label{s_mrk71}

\begin{figure*}[h]
\centering
\includegraphics[width=0.47\textwidth]{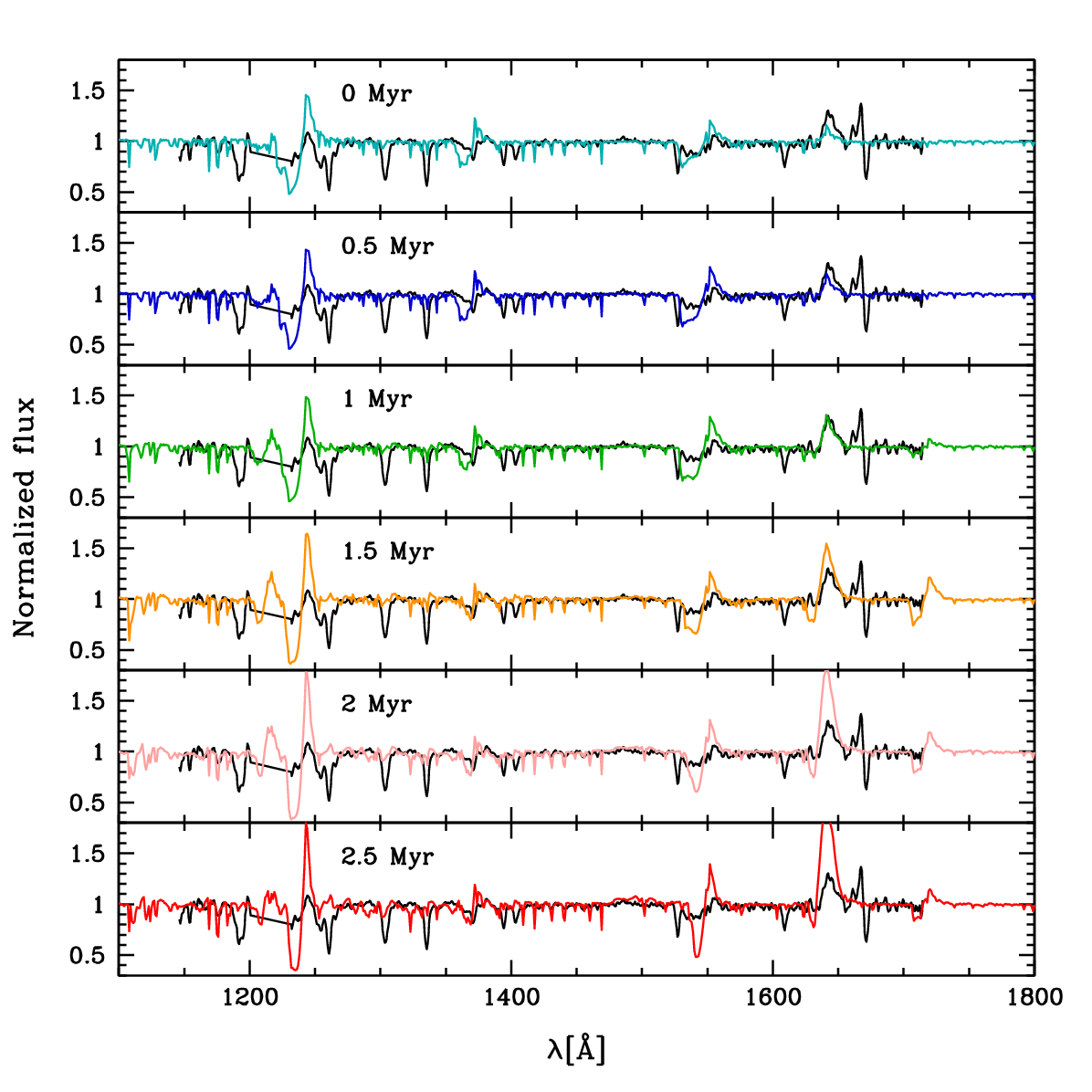}
\includegraphics[width=0.47\textwidth]{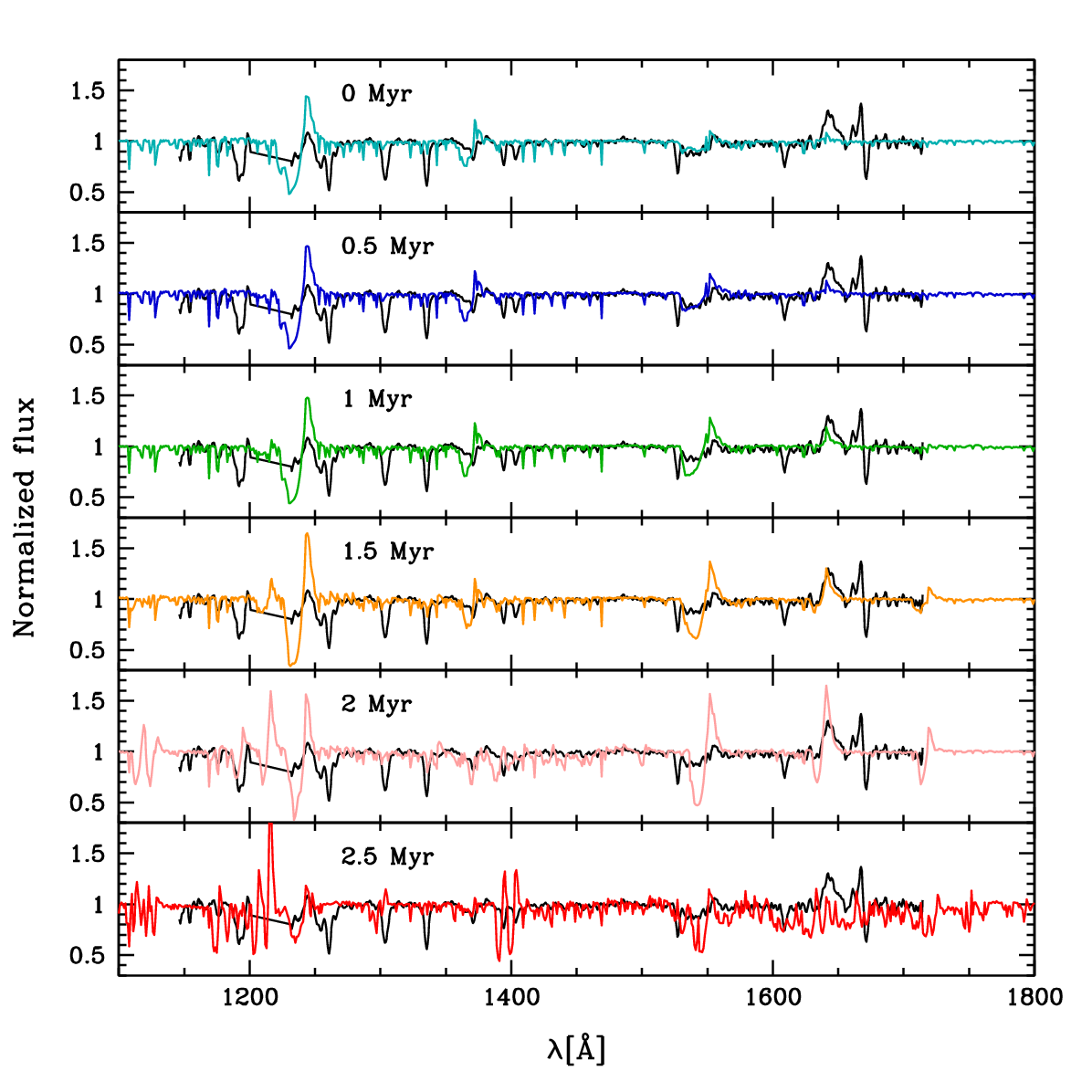}\\
\includegraphics[width=0.47\textwidth]{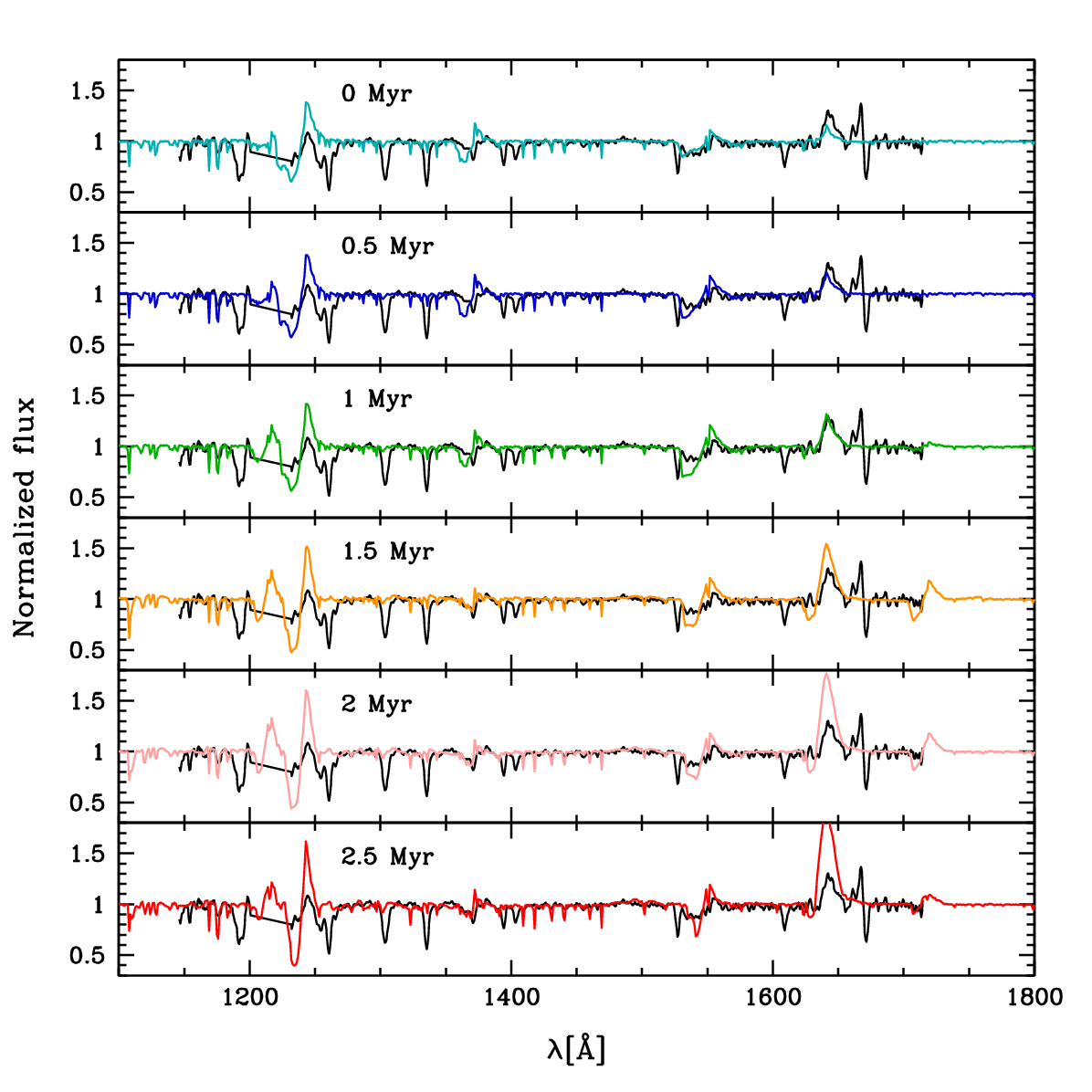}
\includegraphics[width=0.47\textwidth]{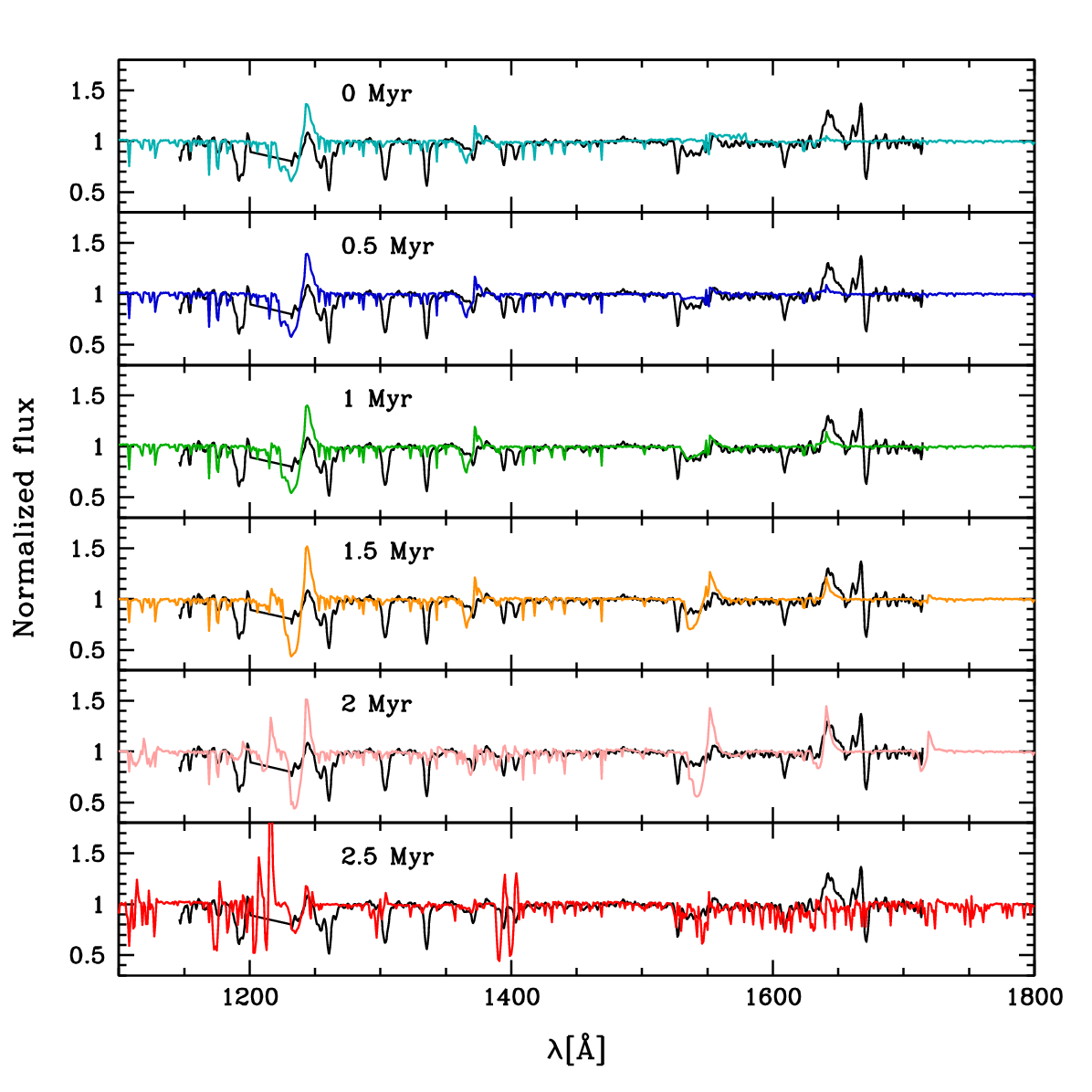}
\caption{Comparison between burst population synthesis models including VMS up to 225~\msun\ and the observed UV spectrum of MrK71-A (black line). Top (bottom) models are for Z=0.2(0.1)~\zsun. In each panel ages range from 0 to 2.5~Myr from top to bottom. The left (right) panels correspond to models without (with) metallicity scaling of the VMS mass loss rates.}
\label{popsyn_mrk71A}
\end{figure*}

The UV spectrum of cluster A in the galaxy MrK71 displays strong lines produced in the winds of massive stars. The \heiiuv\ emission is particularly strong, relative to other lines such as \civuv. As described in Sect.~\ref{s_intro} \citet{smith23} attribute this to the presence of VMS and argue that their mass loss rates do not depend on Z. MrK71 has 12+$\log$(O/H)=7.89 \citep{chen23} slightly below the SMC value. We overplot our Z = 0.2~\zsun\ burst models including VMS on top of the UV spectrum of MrK71-A in Fig.~\ref{popsyn_mrk71A}. Unlike for II~Zw~40, no model reproduces quantitatively all features. In particular the relative strength of \civuv\ and \heiiuv\ is not matched. \civuv\ is weak, which is attributed to the low metallicity of the galaxy \citep{smith23}. All of our models at Z = 0.2~\zsun\ and that have no Z scaling of the VMS mass loss rates overpredict the \civuv\ strength. For the same metallicity, the models with Z scaling of the mass loss rates and age 0 and 0.5 Myr reproduce reasonably the \civuv\ emission, but underpredict \heiiuv. At later ages, the strength of \heiiuv\ increases and matches the observed profile, but then \civuv\ is way too strong. A possible solution to this issue could be that the age of the cluster is lower than 1~Myr. As explained in Sect.~\ref{s_popsynsetup} BPASS does not provide burst models below 1~Myr so we used the 1~Myr BPASS models for the population of normal massive stars at 0 and 0.5~Myr. Since normal O stars contribute to \civuv\ that strengthens with time, we suspect that the use of the 1~Myr BPASS models at 0 and 0.5 Myr leads to an overprediction of the \civuv\ strength. However the strength of \ion{O}{v}~1371 would be even larger in that case, which is opposite to what would be needed to solve the mismatch with \civuv.

We checked whether a lower metallicity would be helpful. The Z = 0.1~\zsun\ models are shown in the lower panels of Fig.~\ref{popsyn_mrk71A}. When no Z scaling of mass loss rates is considered, \heiiuv\ is reproduced at 0.5 Myr, when \civuv\ is slightly over-predicted. The same argument as above for \civuv\ would help to reduce its strength a bit, bringing the model and observed spectra into better agreement. For the case of Z scaling of mass loss rates, \heiiuv\ reaches the observed emission strength between 1.5 and 2~Myr, but \civuv\ is too strong by that age.

Increasing the upper mass limit of our models does not help since all lines are affected. When \heiiuv\ is increased so is \civuv\, so that the discrepancy we describe above remains. Thus there is no model with VMS that quantitatively match the observed UV spectrum of MrK71-A.

\subsubsection{SB~126}
\label{s_sb126}

\begin{figure*}[h]
\centering
\includegraphics[width=0.47\textwidth]{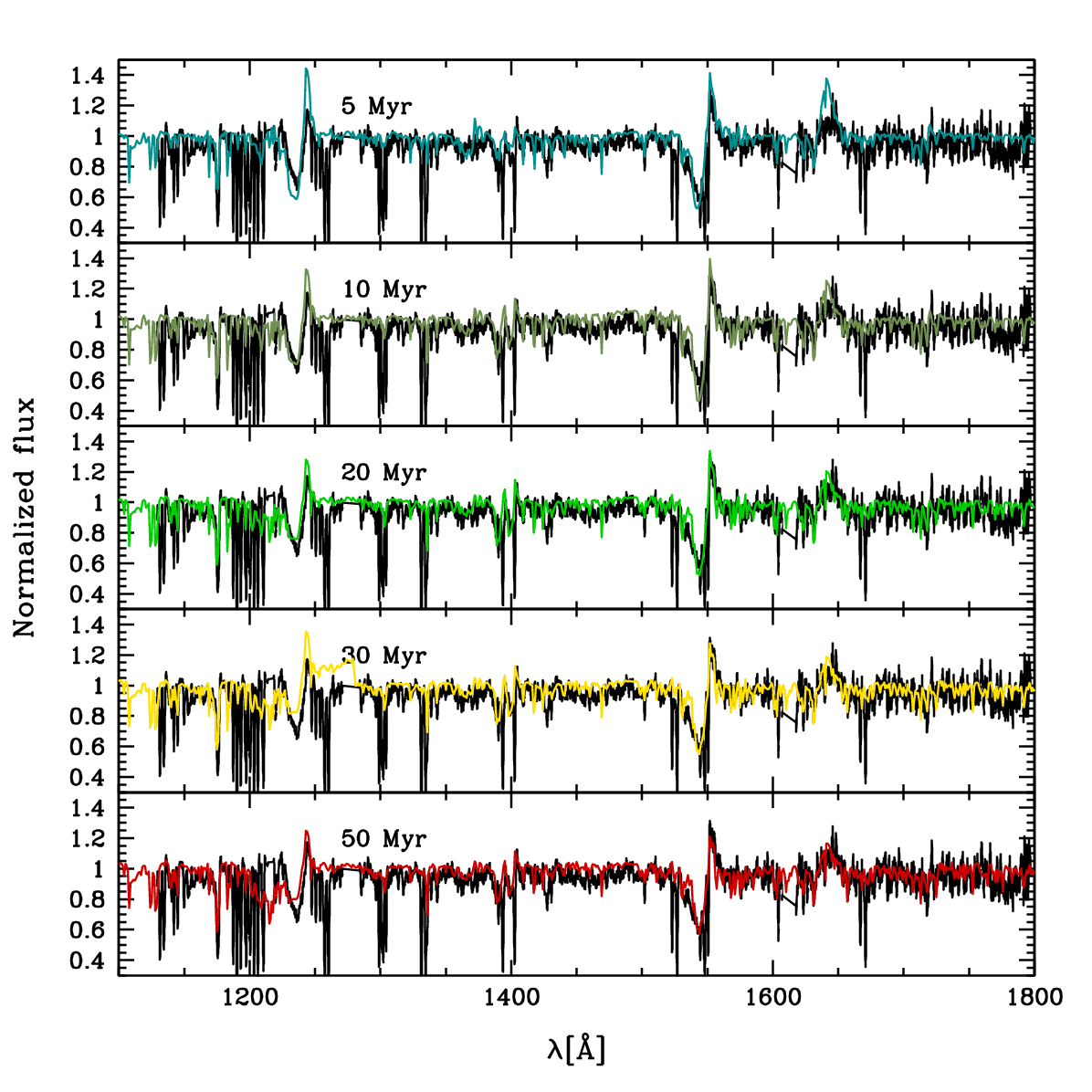}
\includegraphics[width=0.47\textwidth]{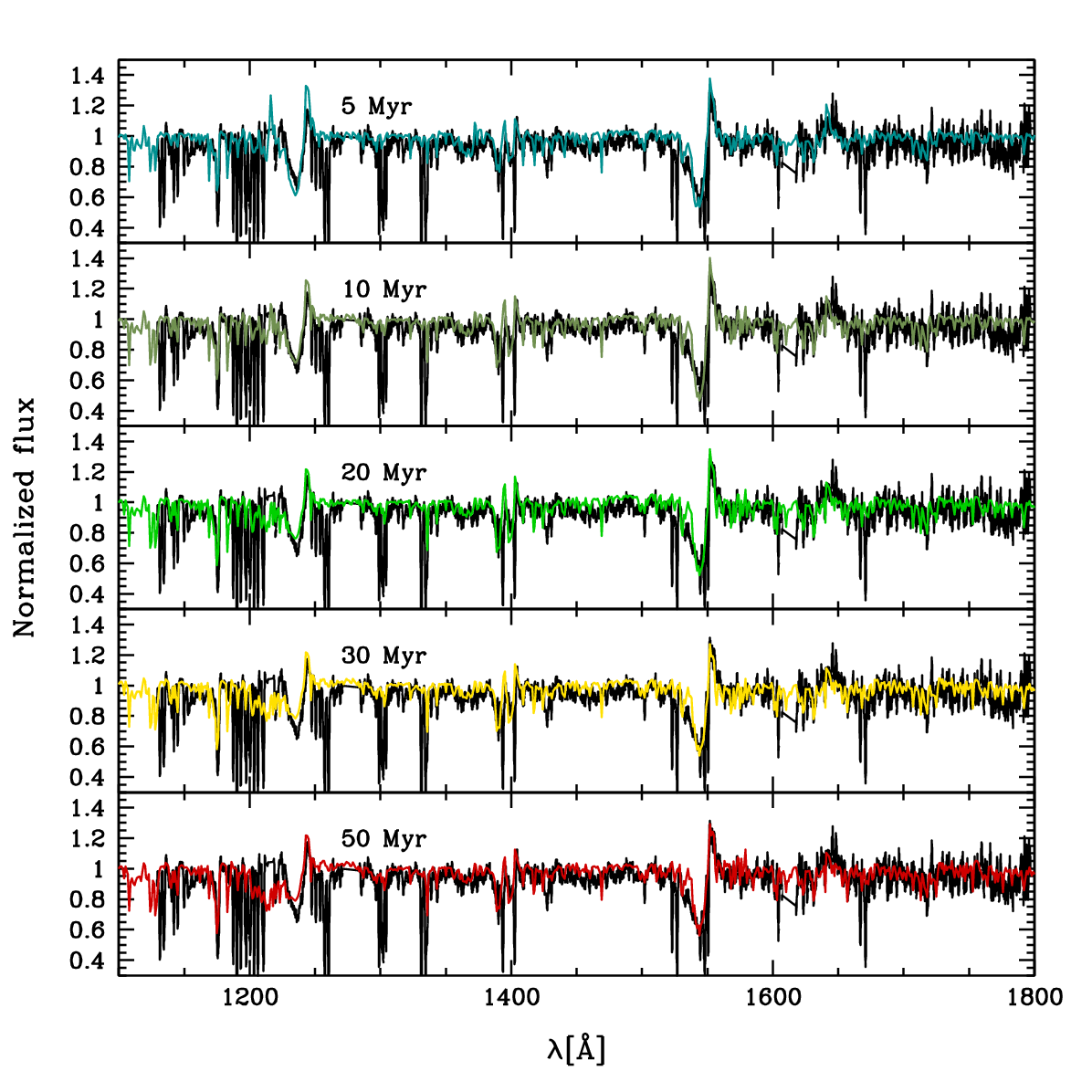}
\caption{Comparison between Z=0.2~\zsun\ CSF population synthesis models including VMS up to 225~\msun\ and the observed UV spectrum of SB126 (black line). Left (right) panels show models without (with) metallicity scaling of the VMS mass loss rates.}
\label{popsyn_sb126}
\end{figure*}

The starburst galaxy SB-126 was studied by \citet{senchyna20}. For this source and others they could not match both the stellar wind features and the nebular emission lines at the same time. To solve this issue they invoke the presence of VMS. SB-126 has a metallicity 12+$\log$(O/H)=8.02$\pm$0.04 according to \citet{senchyna20}. We show how its UV spectrum compares to our Z = 0.2~\zsun\ models in Fig.~\ref{popsyn_sb126}. Since the UV spectrum of SB~126 likely results from a combination of several sources, as stressed by \citet{senchyna20} and visible from their Fig.~2, we use CSF models rather than burst models. The CSF models with ages between 5 and 50~Myr are considered. 

The models with no Z scaling of the mass loss rates of VMS tend to over-predict \heiiuv, although the mismatch is relatively small at ages larger than 20~Myr. Conversely, the strength of this line is slightly underestimated in models with a metallicity scaling of mass loss. The models of this family with ages of 20 to 50~Myr also account reasonably well for \civuv. They also fit nicely both \ion{N}{v}~1240 and the \ion{Si}{iv} doublet near 1400~\AA. 

For SB-126, reasonable fits of the UV spectrum can be obtained. This is an improvement compared to the analysis of \citet{senchyna20} who needed to consider metallicities higher than measured to reproduce the observed lines of \civuv\ and \heiiuv. In our case there is no such tension regarding metallicity. In spite of this relative success a clear conclusion regarding which family of models (mass loss Z-scaling or not) is most appropriate to fit the data is not reached. A metallicity scaling of the VMS mass loss rates that would be shallower than the one adopted here may improve the global fits for this galaxy. However the fact that the spectrum likely results from several sources complicates the analysis. It may be that some sources contain VMS, some not, so the resulting spectrum is a combination of several types of spectra.

\section{Discussion}
\label{s_disc}

In this Section we discuss some of the shortcomings of our modelling regarding stellar evolution, mass loss rates and population synthesis.

\subsection{VMS evolution}
\label{s_discevol}

Models of VMS have been computed by several authors. \citet{yusof13} proposed models for stars up to 500~\msun\ at Z = 0.014, 0.004 and 0.002 which correspond to solar, LMC and SMC metallicity. Mass loss rates are taken from \citet{vink01} as for O-type stars, and thus do not include the specific VMS recipes. The resulting evolutionary tracks are characterised by a vertical evolution with a luminosity decrease, over a wide range of masses and metallicities. This is different from the models of \citet{kohler15} that have a classical redward evolution. These models are also based on the \citet{vink01} mass loss recipe and thus do not include the boosted winds of VMS. 

\citet{martinet23} improve on the Geneva models of \citet{yusof13} by, among other things, using the mass loss recipe of \citet{gh08} for VMS. A global metallicity dependence of the form (Z/\zsun)$^{0.7}$ is assumed. Their models are designed for Z = 0, 10$^{-5}$, 0.006 and 0.014. The tracks show a complex behaviour. At the metallicity of the LMC (Z = 0.006) that is comparable to our computations \citep{mp22}, the models of Martinet et al. start with a classical redward evolution down to \teff$\sim$30000~K before evolving mostly to the blue. Their models show a global trend of evolution that extends more to the red part of the HRD as metallicity decreases. The study of \citet{martinet23} indicates that as mass increases mass loss becomes the dominant ingredient of VMS modelling. Rotation affects the tracks of their 180~\msun\ models but has almost no impact on the 300~\msun\ track.

\citet{sabhahit22} explore two regimes of mass loss for VMS: one based on dependence on the Eddington factor, the other on the luminosity-to-mass ratio. Their computations are for Z = 0.014 and Z = 0.008. For both metallicities the evolution is to the red for the first mass loss scheme, and vertically for the second one. Sabhahit et al. favour the latter evolution because observations of VMS in the Arches cluster and 30~Dor are better accounted for by these models. In a subsequent study they explore the metallicity dependence of the switch from optically thin to optically thick winds \citep{sabhahit23}. Their models cover Z = 0.008, 0.004, and 0.002. Their evolution shifts from a vertical one at Z = 0.008 to a horizontal one (to the red of the HR diagram) at Z = 0.002, because of the reduced mass loss.

All these computations show that mass loss is really a key ingredient of the evolution of VMS. Quite different tracks are obtained even if boosted mass loss rates are included. The exact shape of the dependence of mass loss rates on physical parameters matters. Unfortunately observational constraints remain too sparse to favour one recipe (see also next Sect.~\ref{s_mdotZ}). The number of individual VMS is small and a preferred position in the HR diagram, as advocated by \citet{sabhahit22}, is unclear. Adding more stars from other clusters at different ages and metallicities would be most helpful to provide robust constraints.

\subsection{Advanced phases of evolution}
\label{s_postMS}

Our population synthesis models do not include the very last phases of evolution. The reasons have been stated in \citet{mp22}. The correct modelling of these advanced phases requires a better "coupling" of evolutionary and atmosphere models than what we do in the present study. Indeed in these phases there is a mismatch in the temperature and density structure of evolutionary and atmosphere models. In addition the advanced phases represent no more than $\sim$10\% of the VMS lifetime. 

In spite of these limitations one can still estimate the contribution of the final phases of VMS evolution on the integrated spectra of young starbursts. For this we built test models as follows. We computed new models for the 150 and 200~\msun\ models at Z = 0.1 and 0.01~\zsun. These models are shown by the cross symbols in the middle and right panels of Fig.~\ref{hrd}. Their parameters are listed in Table~\ref{tab_evolved}. The two models on the 150~\msun\ tracks correspond to ages of 2.55~Myr and 2.75~Myr, while the models on the 200~\msun\ tracks have an age of 2.75~Myr. We assume that the 2.55(2.75)~Myr models are representative of the 2.500-2.625 (2.625-2.750)~Myr age range in population synthesis models. 
We recall that stars evolve quickly during their final phases. For instance the Z = 0.1~\zsun\ 150~\msun\ model moves from a position where \teff\ = 70~kK to \teff\ = 150~kK in 0.3~Myr so the above assumptions are made to provide a qualitative rather than quantitative assessment of the effect of these shorts evolutionary phases on integrated spectra.

\begin{figure*}[h]
\centering
\includegraphics[width=0.47\textwidth]{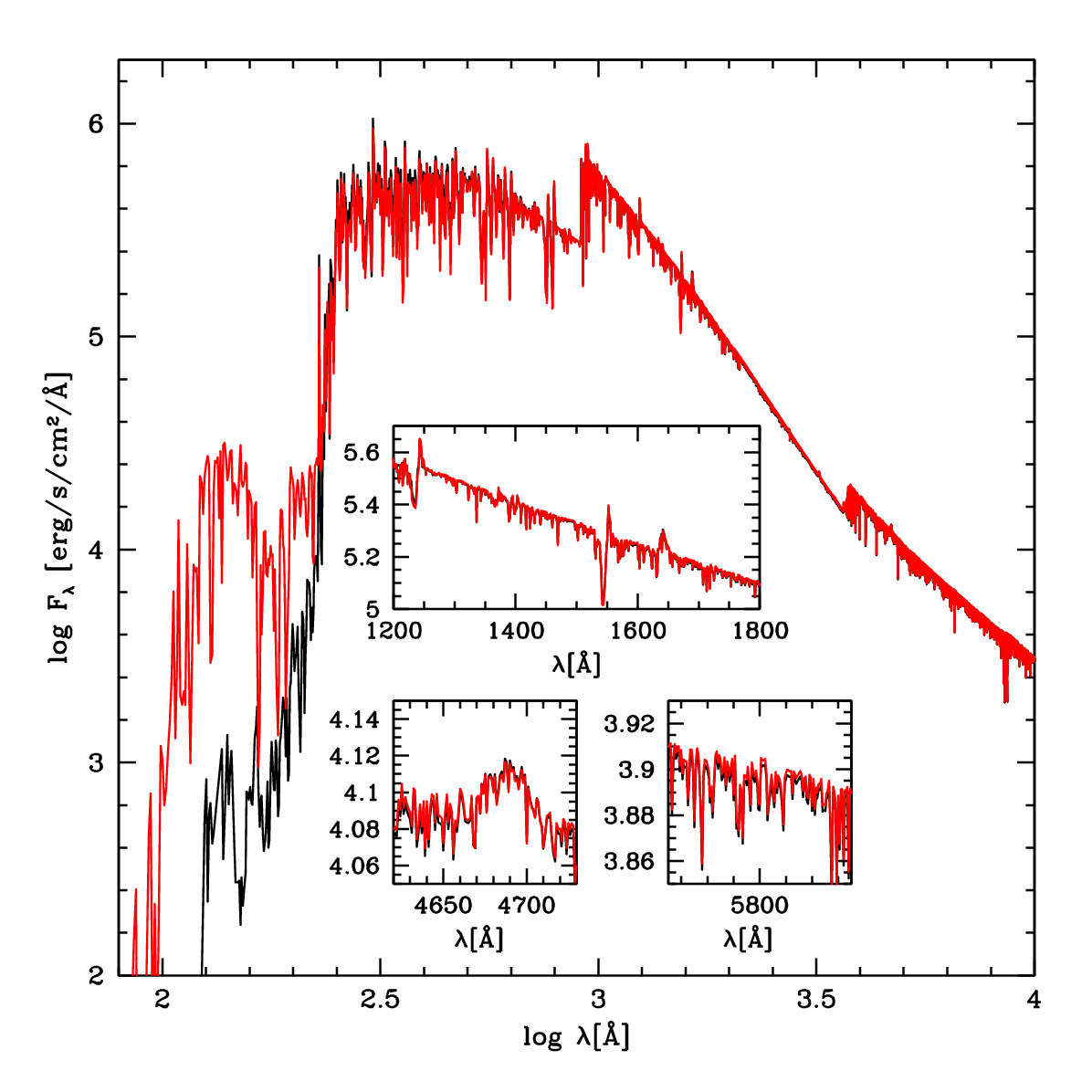}
\includegraphics[width=0.47\textwidth]{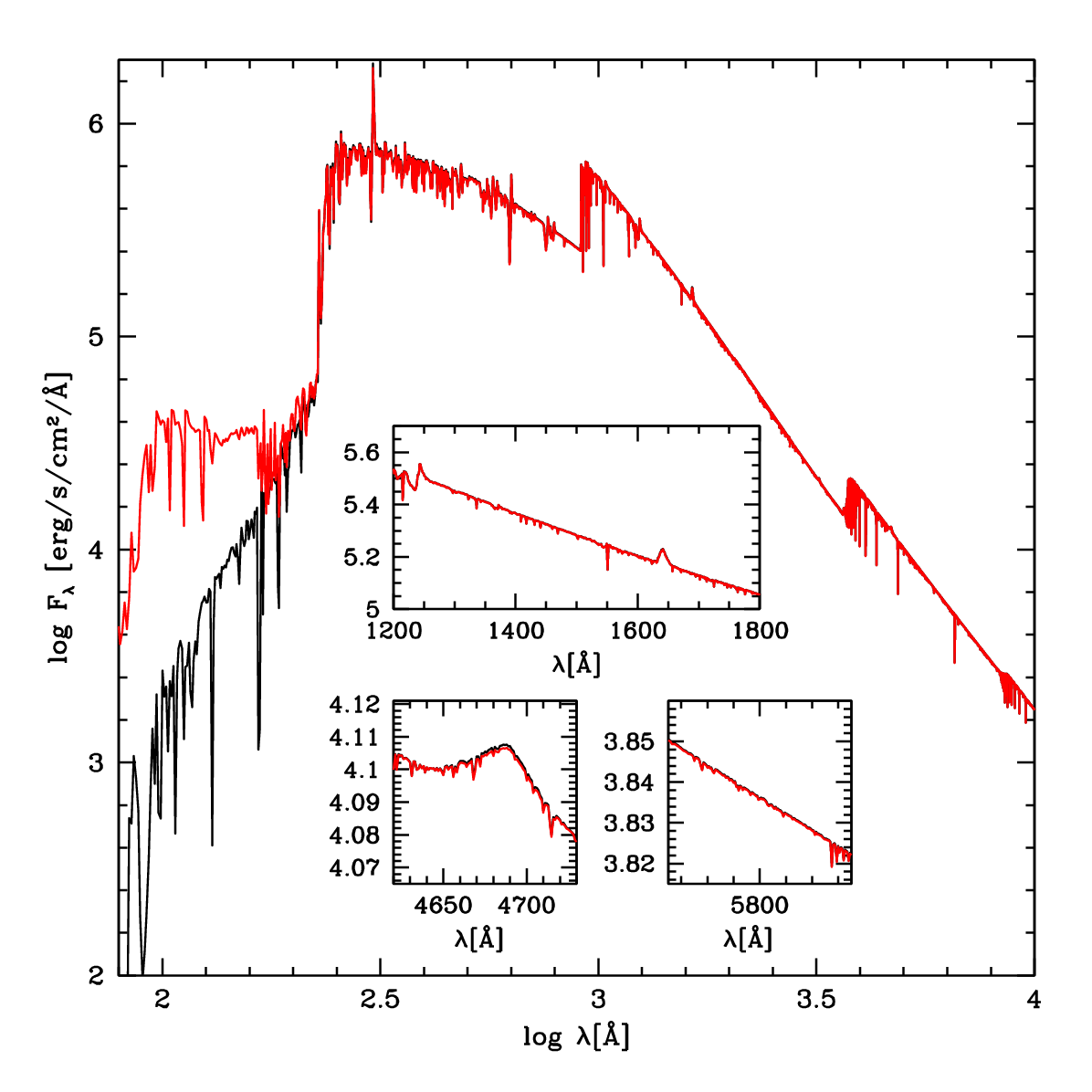}
\caption{Comparison between the spectrum of the initial CSF population synthesis models at 10~Myr (black line) and the spectrum of a test model in which the contribution of evolved phases of evolution is included (red line). The left (right) panel is for Z=0.1 (0.01) \zsun.}
\label{testpostMS}
\end{figure*}

The effect of these advanced phases on the integrated spectrum of starbursts is shown in Fig.~\ref{testpostMS}. The flux is basically unchanged throughout most of the wavelength range. In particular the UV lines, especially \heiiuv, are unaffected. The same conclusion holds for the optical bumps. The most notable difference is the increased flux in the \ion{He}{ii} continuum. The evolved phases correspond to very high effective temperatures not reached in previous phases. Consequently the flux emitted at these wavelengths is boosted. This conclusion should be considered as qualitative rather than quantitative. Indeed as stressed above our test is only a rough estimate of the contribution of evolved phases. A finer sampling of these short phases would be required, together with a better coupling between evolutionary and atmosphere models. Nevertheless our test shows that the UV and optical spectra of CSF models are unaffected by the short advanced phases of VMS evolution. On the other hand the \ion{He}{ii} ionising flux is sensitive to these phases.

\begin{figure}[ht!]
\centering
\includegraphics[width=0.45\textwidth]{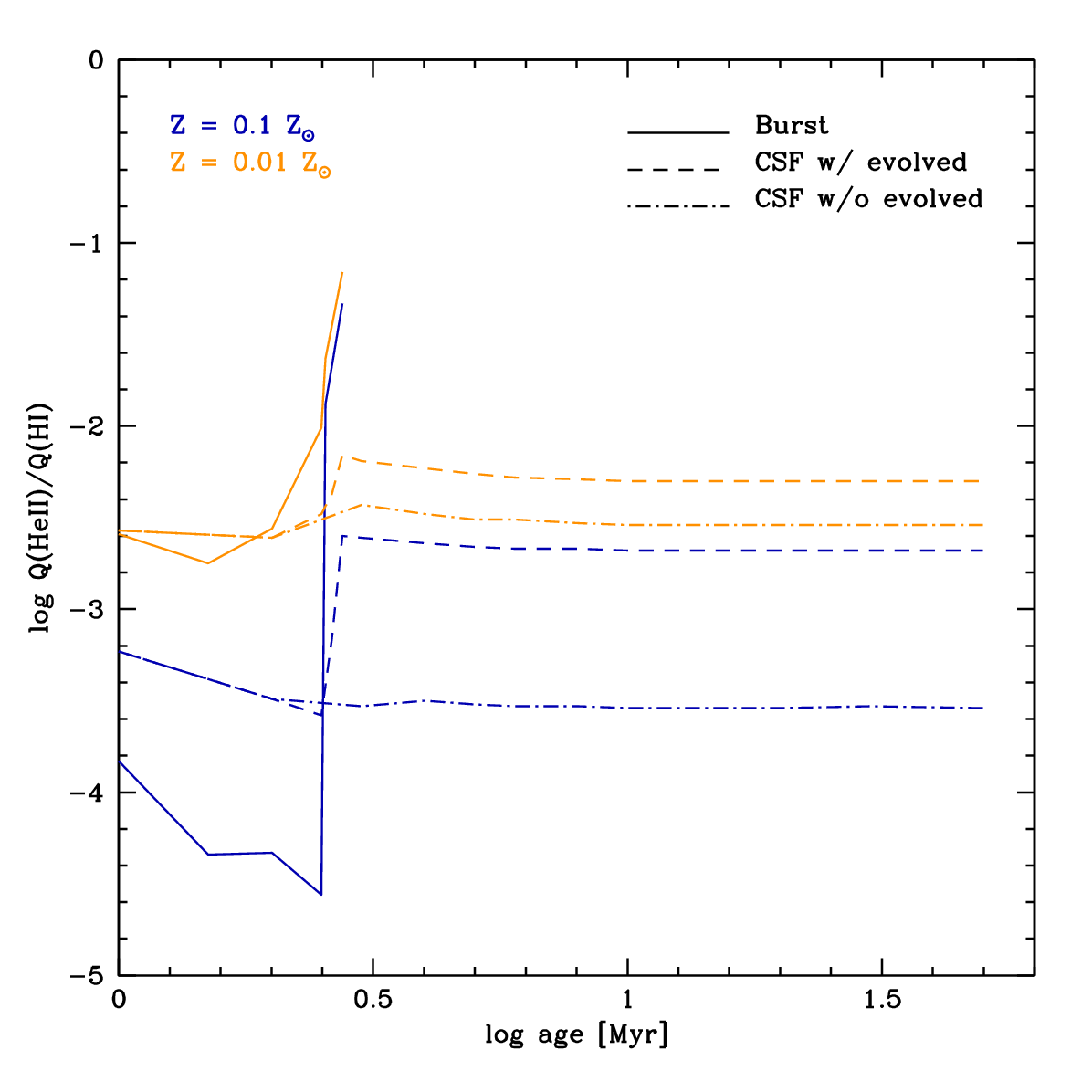}
\caption{Ratio of \ion{He}{ii} to \ion{H}{i} ionising photons fluxes as a function of age. The dashed (dot-dashed) lines correspond to CSF models in which the advanced phases of evolution are included (not included). In burst models these advanced phases are shown at 2.625 and 2.750~Myr. Blue (orange) lines are for Z=0.1 (0.001)~\zsun.}
\label{Q2onQ0}
\end{figure}

We estimate the magnitude of this effect in Fig.~\ref{Q2onQ0}. We show how the ratio of \ion{He}{ii} to \ion{H}{i} ionising photons fluxes changes with time in models that include and do not include these short advanced phases. When VMS evolve into these phases the ratio reaches values as large as 0.1 for a short amount of time ($\sim$0.2~Myr). In CSF models the addition of the advanced phases increases the number of \ion{He}{ii} ionising photons after 2.5~Myr. Q(\ion{He}{ii})/Q(\ion{H}{i}) is multiplied by a factor of 10 at Z = 0.1~\zsun. At lower metallicity the increase is not as large because VMS are already hot during most of their evolution and produce some amount of \ion{He}{ii} ionising flux (see Sect.~\ref{Z_Qi}).

The ratio Q(\ion{He}{ii})/Q(\ion{H}{i}) reaches $2(5)\times 10^{-3}$ for CSF models at Z=0.1(0.01)~\zsun. Under reasonable approximations \citep[see e.g.][]{stasinska15} it can be converted into the ratio of the intensities of \heiiopt\ to H$\beta$: I(\heiiopt)/I(H$\beta$) = 1.74 $\times$ Q(\ion{He}{ii})/Q(\ion{H}{i}). Consequently CSF models that include the short advanced phases of VMS evolution are able to produce I(\heiiopt)/I(H$\beta$) of the order of 1\%. This is sufficient to account for a significant fraction of the observed values in star-forming regions and galaxies where high ionisation lines are observed \citep[e.g. Fig.1 of][]{schaerer19}. However the predicted intensity ratios are too small to explain the highest observed values, up to 0.06, detected in some sources such as II~Zw~18 \citep{izotov97,kehrig15} or SBSS0335-052E \citep{kehrig18,wofford21}. Burst models caught at 2.5-2.7~Myr would produce these kinds of intensity ratios, but their duration is short enough ($\sim$0.2~Myr) to exclude them as a viable explanation for the presence of high ionisation lines.

For the models with scaling of the VMS mass loss rates with metallicity, the evolution is redward so the high \teff\ needed to produce \ion{He}{ii} ionising flux are not reached. In addition the evolution is faster and stars live shorter than in the case of no metallicity scaling of mass loss rates. Consequently no effect of these advanced phases is expected in the UV part of the spectrum. 
These phases have thus a limited impact on the integrated spectral properties of young starbursts that include VMS.

\subsection{VMS mass loss rates at low Z}
\label{s_dizsc_mdotZ}

In this study we have explored the evolution and spectroscopic appearance of VMS and associated populations under two assumptions regarding the VMS mass loss rates. The comparison of population synthesis models with UV spectroscopy of star-forming regions at low metallicity did not allow to favour one assumption over the other. In both cases \heiiuv, the most emblematic feature of VMS, is produced over a wide range of ages and metallicities. Its relative intensity compared to other UV features such as \civuv\ slightly changes depending on the mass loss rates. But confrontation to observations does not help to distinguish which scenario is correct. In fact the truth may be in between, with a metallicity scaling of mass loss rates that is shallower than the assumed linear Z dependence. 

\citet{sander24} present preliminary theoretical modelling of the winds of WN stars. Building on the work of \citet{sander20b} for hydrogen-free WN stars they explore the behaviour of mass loss rates when hydrogen is present. A rather weak dependence is observed. Combined with the effects of metallicity, the scaling of mass loss rates appear to be weaker than a linear Z-scaling (see their Fig.~2). However this exploratory work is for a mass range lower than that of VMS (by a factor of 10). Further investigation in the VMS regime is thus needed. 
\citet{sabhahit23} propose a framework for the transition between optically thin and thick winds, i.e. between normal O-type stars winds and VMS winds. It relies on the concept of transition mass loss developed by \citet{vg12} to parameterise the change between the two winds regimes. Their result indicates that at lower metallicity the boosted mass loss rates of VMS appear at higher luminosity. \citet{sabhahit23} use their formalism to produce evolutionary models of VMS. Different paths are obtained depending on metallicity and mass. There is a global trend of more redward evolution at lower metallicity, because of reduced winds. This is consistent with our results: we have shown that the models with the lowest VMS mass loss rates evolve redward. 

Whatever VMS mass loss recipe we adopt, it appears clear that observational tests are necessary to better understand VMS winds and evolution. 
The observation of individual VMS in different environments is challenging. There are currently very few star-forming regions that are 1) massive enough, 2) young enough and 3) close enough so that VMS can be spectroscopically identified and analysed. As described in Sect.~\ref{s_intro} this leaves no more than a dozen VMS. Future telescopes and instrumentation may be able to break star-forming regions currently unresolved into their stellar components. However, taking R136 as a reference template, spatial resolution of the order 10~mas will be required for the best candidate clusters (such as NGC3125-A, MrK71-A). While this may be possible from an imaging perspective for ELTs and later on HWO, this is a challenge for spatially resolved spectroscopy. 

One may thus look for alternative observational tests to better constrain the mass loss history and the evolution of VMS. The N-emitters presented in Sect.~\ref{s_intro} may be one way. When looking at Tables~\ref{tab_gr_zsmc} to \ref{tab_grz_z0p01} we see that the surface chemistry of models without Z-scaling of the VMS mass loss rates is different from that of the models with a Z dependence. Because of the stronger winds the former models expel more material and expose deeper layers of stars to the surface. More material is thus released in the interstellar environment. These different yields may be probed through the imprint they leave in the surrounding gas. VMS have been listed as one of the potential sources to explain the large nitrogen fraction in the N-emitters \citep{charbonnel23,vink23,senchyna24}. A proper computation of the yields from VMS can in principle be used to see if one of the two mass loss scenarios we considered is favoured to explain these peculiar star-forming regions. This will be the subject of a future study.

\section{Conclusion}
\label{s_conc}

We have presented a study of the evolution and spectroscopic appearance of VMS at metallicities Z=0.2, 0.1 and 0.01~\zsun. We have assumed two mass loss recipes for their winds. The first one assumes no metallicity dependence of VMS winds, and relies on the calibration of \citet{graef21} established for VMS in the LMC. The second recipe assumes a linear scaling with Z, with the \citet{graef21} as an anchor point. We have computed evolutionary models for stars with initial masses 150, 200, 250, and 300~\msun, with the code STAREVOL. We have calculated atmosphere models and synthetic spectra along evolutionary tracks (code CMFGEN), from the ZAMS to ages of about 2.5~Myr (depending on initial mass) with time steps of 0.5~Myr. The resulting spectra have been incorporated in population synthesis models to study the impact of VMS on the integrated light of star-forming regions. 

The main results are summarised as follows:

\begin{itemize}

\item The evolution of VMS critically depends on the assumed mass loss rate recipe. When mass loss rates do not vary with metallicity, VMS at low metallicity spend most of their lifetime close to the ZAMS before evolving to the blue part of the HR diagram. On the contrary stars with reduced mass loss rates at lower metallicity follow a more classical evolution to the red part of the HR diagram. 

\item The UV spectrum of VMS shows a number of emission and P-Cygni lines typical of massive stars, with a significant \heiiuv\ emission that is a specific characteristic of VMS. This line is seen down to Z = 0.1~\zsun\ whatever the assumed VMS mass loss rate recipe. Even at Z = 0.01~\zsun\ the models with the strongest mass loss rates still show this line in emission. 

\item In the optical range the Wolf-Rayet bumps are seen over a wide range of parameters. The blue bump is dominated by \heiiopt\ with no of very weak \ion{N}{iii}~4634-42 emission. When present the red bump shows a narrow \civopt\ doublet emission, sometimes on top of a broader but weak component.  

\item The UV spectrum of population synthesis models have EW(\heiiuv) reaching up to 10~\AA\ for bursts, and 4~\AA\ for constant star-formation history. Low metallicity models predict, on average, larger EW(\heiiuv) than models at Z=0.4~\zsun. We confirm that only models including VMS produce \heiiuv\ in emission. 

\item The ionising spectrum of VMS is harder at lower metallicity, between 13.6 and $\sim$45~eV. It is also harder than that of models without VMS. Models with VMS mass loss rates that do not scale with metallicity, and thus remain hot in the HR diagram, show the hardest spectra. At energies higher than 45~eV the behaviour is more complex, and is also affected by the very final phases of evolution that are only partially included in the present models. \ion{He}{ii} ionising photons are produced at a level that can partially account for observed \ion{He}{ii} nebular lines, although the high intensities sometimes observed are only produced for starbursts at very specific ages. Finally population synthesis models that include nebular emission have higher ionising photon efficiency at lower metallicity. 

\item The UV spectra of three star-forming regions at low metallicity (II~Zw~40-A, MrK71-A and SB126) can be matched qualitatively and sometimes quantitatively by burst or CSF models including VMS. This confirms the importance of including a consistent treatment of VMS evolution and synthetic spectra in population synthesis models of young starbursts. However no clear trend emerges regarding the preferred mass loss recipe at low metallicity (i.e. Z-scaling of mass loss rates or not). 

\end{itemize}

Future spectroscopic identification of individual VMS appears to be difficult, even with the foreseen capabilities. Alternative ways of finding VMS in spatially unresolved populations may be a better way of constraining VMS properties, in particular their winds. Studies of chemistry of young star-forming regions, in particular the N-emitters, may also be relevant. Some of these possibilities will be investigated in subsequent publications.

\vspace{0.5cm}

\noindent \textbf{Data availability}: the synthetic spectra of individual VMS models are distributed on the POLLUX database at \url{https://pollux.oreme.org/}. Evolutionary models and population synthesis models can be shared on reasonable request.

\begin{acknowledgements}

We thank an anonymous referee for useful comments. We thank John Hillier for developing and distributing CMFGEN. 
This research is based on observations made with the NASA/ESA Hubble Space Telescope obtained from the Space Telescope Science Institute, which is operated by the Association of Universities for Research in Astronomy, Inc., under NASA contract NAS 5–26555. These observations are associated with programs 14102, 15185, and 16261. 

\end{acknowledgements}

\bibliographystyle{aa}
\bibliography{vms_lowZ}

\begin{thebibliography}{100}
\expandafter\ifx\csname natexlab\endcsname\relax\def\natexlab#1{#1}\fi

\bibitem[{{Asplund} {et~al.}(2009){Asplund}, {Grevesse}, {Sauval}, \&
  {Scott}}]{asplund09}
{Asplund}, M., {Grevesse}, N., {Sauval}, A.~J., \& {Scott}, P. 2009, \araa, 47,
  481

\bibitem[{{Barniske} {et~al.}(2008){Barniske}, {Oskinova}, \&
  {Hamann}}]{barniske08}
{Barniske}, A., {Oskinova}, L.~M., \& {Hamann}, W.~R. 2008, \aap, 486, 971

\bibitem[{{Bestenlehner}(2020)}]{besten20b}
{Bestenlehner}, J.~M. 2020, \mnras, 493, 3938

\bibitem[{{Bestenlehner} {et~al.}(2020){Bestenlehner}, {Crowther},
  {Caballero-Nieves}, {Schneider}, {Sim{\'o}n-D{\'\i}az}, {Brands}, {de Koter},
  {Gr{\"a}fener}, {Herrero}, {Langer}, {Lennon}, {Ma{\'\i}z Apell{\'a}niz},
  {Puls}, \& {Vink}}]{besten20a}
{Bestenlehner}, J.~M., {Crowther}, P.~A., {Caballero-Nieves}, S.~M., {et~al.}
  2020, \mnras, 499, 1918

\bibitem[{{Bestenlehner} {et~al.}(2014){Bestenlehner}, {Gr{\"a}fener}, {Vink},
  {Najarro}, {de Koter}, {Sana}, {Evans}, {Crowther}, {H{\'e}nault-Brunet},
  {Herrero}, {Langer}, {Schneider}, {Sim{\'o}n-D{\'\i}az}, {Taylor}, \&
  {Walborn}}]{besten14}
{Bestenlehner}, J.~M., {Gr{\"a}fener}, G., {Vink}, J.~S., {et~al.} 2014, \aap,
  570, A38

\bibitem[{{Bestenlehner} {et~al.}(2011){Bestenlehner}, {Vink}, {Gr{\"a}fener},
  {Najarro}, {Evans}, {Bastian}, {Bonanos}, {Bressert}, {Crowther}, {Doran},
  {Friedrich}, {H{\'e}nault-Brunet}, {Herrero}, {de Koter}, {Langer}, {Lennon},
  {Ma{\'\i}z Apell{\'a}niz}, {Sana}, {Soszynski}, \& {Taylor}}]{besten11}
{Bestenlehner}, J.~M., {Vink}, J.~S., {Gr{\"a}fener}, G., {et~al.} 2011, \aap,
  530, L14

\bibitem[{{Brands} {et~al.}(2022){Brands}, {de Koter}, {Bestenlehner},
  {Crowther}, {Sundqvist}, {Puls}, {Caballero-Nieves}, {Abdul-Masih},
  {Driessen}, {Garc{\'\i}a}, {Geen}, {Gr{\"a}fener}, {Hawcroft}, {Kaper},
  {Keszthelyi}, {Langer}, {Sana}, {Schneider}, {Shenar}, \& {Vink}}]{brands22}
{Brands}, S.~A., {de Koter}, A., {Bestenlehner}, J.~M., {et~al.} 2022, \aap,
  663, A36

\bibitem[{{Bruzual} \& {Charlot}(2003)}]{bc03}
{Bruzual}, G. \& {Charlot}, S. 2003, \mnras, 344, 1000

\bibitem[{{Bunker} {et~al.}(2023){Bunker}, {Saxena}, {Cameron}, {Willott},
  {Curtis-Lake}, {Jakobsen}, {Carniani}, {Smit}, {Maiolino}, {Witstok},
  {Curti}, {D'Eugenio}, {Jones}, {Ferruit}, {Arribas}, {Charlot}, {Chevallard},
  {Giardino}, {de Graaff}, {Looser}, {L{\"u}tzgendorf}, {Maseda}, {Rawle},
  {Rix}, {Del Pino}, {Alberts}, {Egami}, {Eisenstein}, {Endsley}, {Hainline},
  {Hausen}, {Johnson}, {Rieke}, {Rieke}, {Robertson}, {Shivaei}, {Stark},
  {Sun}, {Tacchella}, {Tang}, {Williams}, {Willmer}, {Baker}, {Baum},
  {Bhatawdekar}, {Bowler}, {Boyett}, {Chen}, {Circosta}, {Helton}, {Ji},
  {Kumari}, {Lyu}, {Nelson}, {Parlanti}, {Perna}, {Sandles}, {Scholtz},
  {Suess}, {Topping}, {{\"U}bler}, {Wallace}, \& {Whitler}}]{bunker23}
{Bunker}, A.~J., {Saxena}, A., {Cameron}, A.~J., {et~al.} 2023, \aap, 677, A88

\bibitem[{{Cameron} {et~al.}(2023){Cameron}, {Katz}, {Rey}, \&
  {Saxena}}]{cameron23}
{Cameron}, A.~J., {Katz}, H., {Rey}, M.~P., \& {Saxena}, A. 2023, \mnras, 523,
  3516

\bibitem[{{Cantiello} {et~al.}(2009){Cantiello}, {Langer}, {Brott}, {de Koter},
  {Shore}, {Vink}, {Voegler}, {Lennon}, \& {Yoon}}]{cant09}
{Cantiello}, M., {Langer}, N., {Brott}, I., {et~al.} 2009, \aap, 499, 279

\bibitem[{{Cassinelli} {et~al.}(1981){Cassinelli}, {Mathis}, \&
  {Savage}}]{cassi81}
{Cassinelli}, J.~P., {Mathis}, J.~S., \& {Savage}, B.~D. 1981, Science, 212,
  1497

\bibitem[{{Castellano} {et~al.}(2024){Castellano}, {Napolitano}, {Fontana},
  {Roberts-Borsani}, {Treu}, {Vanzella}, {Zavala}, {Arrabal Haro},
  {Calabr{\`o}}, {Llerena}, {Mascia}, {Merlin}, {Paris}, {Pentericci},
  {Santini}, {Bakx}, {Bergamini}, {Cupani}, {Dickinson}, {Filippenko},
  {Glazebrook}, {Grillo}, {Kelly}, {Malkan}, {Mason}, {Morishita},
  {Nanayakkara}, {Rosati}, {Sani}, {Wang}, \& {Yoon}}]{castellano24}
{Castellano}, M., {Napolitano}, L., {Fontana}, A., {et~al.} 2024, \apj, 972,
  143

\bibitem[{{Castor} {et~al.}(1975){Castor}, {Abbott}, \& {Klein}}]{cak}
{Castor}, J.~I., {Abbott}, D.~C., \& {Klein}, R.~I. 1975, \apj, 195, 157

\bibitem[{{Charbonnel} {et~al.}(2023){Charbonnel}, {Schaerer}, {Prantzos},
  {Ram{\'\i}rez-Galeano}, {Fragos}, {Kuruvanthodi}, {Marques-Chaves}, \&
  {Gieles}}]{charbonnel23}
{Charbonnel}, C., {Schaerer}, D., {Prantzos}, N., {et~al.} 2023, \aap, 673, L7

\bibitem[{{Chemerynska} {et~al.}(2024){Chemerynska}, {Atek}, {Dayal}, {Furtak},
  {Feldmann}, {Greene}, {Maseda}, {Nanayakkara}, {Oesch}, {Fujimoto},
  {Labb{\'e}}, {Bezanson}, {Brammer}, {Cutler}, {Leja}, {Pan}, {Price}, {Wang},
  {Weaver}, \& {Whitaker}}]{chemerynska24}
{Chemerynska}, I., {Atek}, H., {Dayal}, P., {et~al.} 2024, \apjl, 976, L15

\bibitem[{{Chen} {et~al.}(2023){Chen}, {Jones}, {Sanders}, {Fadda}, {Sutter},
  {Minchin}, {Huntzinger}, {Senchyna}, {Stark}, {Spilker}, {Weiner}, \&
  {Roberts-Borsani}}]{chen23}
{Chen}, Y., {Jones}, T., {Sanders}, R., {et~al.} 2023, Nature Astronomy, 7, 771

\bibitem[{{Conti} \& {Morris}(1990)}]{cm90}
{Conti}, P.~S. \& {Morris}, P.~W. 1990, \aj, 99, 898

\bibitem[{{Crowther} {et~al.}(2016){Crowther}, {Caballero-Nieves}, {Bostroem},
  {Ma{\'\i}z Apell{\'a}niz}, {Schneider}, {Walborn}, {Angus}, {Brott},
  {Bonanos}, {de Koter}, {de Mink}, {Evans}, {Gr{\"a}fener}, {Herrero},
  {Howarth}, {Langer}, {Lennon}, {Puls}, {Sana}, \& {Vink}}]{crowther16}
{Crowther}, P.~A., {Caballero-Nieves}, S.~M., {Bostroem}, K.~A., {et~al.} 2016,
  \mnras, 458, 624

\bibitem[{{Crowther} \& {Castro}(2024)}]{crowther24}
{Crowther}, P.~A. \& {Castro}, N. 2024, \mnras, 527, 9023

\bibitem[{{Crowther} {et~al.}(2002){Crowther}, {Dessart}, {Hillier}, {Abbott},
  \& {Fullerton}}]{crowther02}
{Crowther}, P.~A., {Dessart}, L., {Hillier}, D.~J., {Abbott}, J.~B., \&
  {Fullerton}, A.~W. 2002, \aap, 392, 653

\bibitem[{{Crowther} {et~al.}(2010){Crowther}, {Schnurr}, {Hirschi}, {Yusof},
  {Parker}, {Goodwin}, \& {Kassim}}]{crowther10}
{Crowther}, P.~A., {Schnurr}, O., {Hirschi}, R., {et~al.} 2010, \mnras, 408,
  731

\bibitem[{{Eldridge} {et~al.}(2017){Eldridge}, {Stanway}, {Xiao}, {McClelland},
  {Taylor}, {Ng}, {Greis}, \& {Bray}}]{bpass}
{Eldridge}, J.~J., {Stanway}, E.~R., {Xiao}, L., {et~al.} 2017, \pasa, 34, e058

\bibitem[{{Farrell} {et~al.}(2020){Farrell}, {Groh}, {Meynet}, {Eldridge},
  {Ekstr{\"o}m}, \& {Georgy}}]{farrell20}
{Farrell}, E.~J., {Groh}, J.~H., {Meynet}, G., {et~al.} 2020, \mnras, 495, 4659

\bibitem[{{Feitzinger} {et~al.}(1980){Feitzinger}, {Schlosser},
  {Schmidt-Kaler}, \& {Winkler}}]{feitzinger80}
{Feitzinger}, J.~V., {Schlosser}, W., {Schmidt-Kaler}, T., \& {Winkler}, C.
  1980, \aap, 84, 50

\bibitem[{{Figer}(2005)}]{figer05}
{Figer}, D.~F. 2005, \nat, 434, 192

\bibitem[{{Fujimoto} {et~al.}(2025){Fujimoto}, {Naidu}, {Chisholm}, {Atek},
  {Endsley}, {Kokorev}, {Furtak}, {Pan}, {Liu}, {Bromm}, {Venditti}, {Visbal},
  {Sarmento}, {Weibel}, {Oesch}, {Brammer}, {Schaerer}, {Adamo}, {Berg},
  {Bezanson}, {Chemerynska}, {Claeyssens}, {Dessauges-Zavadsky}, {Frebel},
  {Korber}, {Labbe}, {Marques-Chaves}, {Matthee}, {McQuinn}, {Mu{\~n}oz},
  {Natarajan}, {Saldana-Lopez}, {Suess}, {Volonteri}, \& {Zitrin}}]{fujimoto25}
{Fujimoto}, S., {Naidu}, R.~P., {Chisholm}, J., {et~al.} 2025, arXiv e-prints
  (ApJ submitted), arXiv:2501.11678

\bibitem[{{Gr{\"a}fener}(2021)}]{graef21}
{Gr{\"a}fener}, G. 2021, \aap, 647, A13

\bibitem[{{Gr{\"a}fener} \& {Hamann}(2008)}]{gh08}
{Gr{\"a}fener}, G. \& {Hamann}, W.~R. 2008, \aap, 482, 945

\bibitem[{{Grassitelli} {et~al.}(2021){Grassitelli}, {Langer}, {Mackey},
  {Gr{\"a}fener}, {Grin}, {Sander}, \& {Vink}}]{grass21}
{Grassitelli}, L., {Langer}, N., {Mackey}, J., {et~al.} 2021, \aap, 647, A99

\bibitem[{{Guseva} {et~al.}(2000){Guseva}, {Izotov}, \& {Thuan}}]{guseva00}
{Guseva}, N.~G., {Izotov}, Y.~I., \& {Thuan}, T.~X. 2000, \apj, 531, 776

\bibitem[{{Hainich} {et~al.}(2015){Hainich}, {Pasemann}, {Todt}, {Shenar},
  {Sander}, \& {Hamann}}]{hainich15}
{Hainich}, R., {Pasemann}, D., {Todt}, H., {et~al.} 2015, \aap, 581, A21

\bibitem[{{Hainich} {et~al.}(2014){Hainich}, {R{\"u}hling}, {Todt}, {Oskinova},
  {Liermann}, {Gr{\"a}fener}, {Foellmi}, {Schnurr}, \& {Hamann}}]{hainich14}
{Hainich}, R., {R{\"u}hling}, U., {Todt}, H., {et~al.} 2014, \aap, 565, A27

\bibitem[{{Hamann} {et~al.}(2006){Hamann}, {Gr{\"a}fener}, \&
  {Liermann}}]{hamann06}
{Hamann}, W.~R., {Gr{\"a}fener}, G., \& {Liermann}, A. 2006, \aap, 457, 1015

\bibitem[{{Hamann} {et~al.}(2019){Hamann}, {Gr{\"a}fener}, {Liermann},
  {Hainich}, {Sander}, {Shenar}, {Ramachandran}, {Todt}, \&
  {Oskinova}}]{hamann19}
{Hamann}, W.~R., {Gr{\"a}fener}, G., {Liermann}, A., {et~al.} 2019, \aap, 625,
  A57

\bibitem[{{Higgins} {et~al.}(2023){Higgins}, {Vink}, {Hirschi}, {Laird}, \&
  {Sabhahit}}]{higgins23}
{Higgins}, E.~R., {Vink}, J.~S., {Hirschi}, R., {Laird}, A.~M., \& {Sabhahit},
  G.~N. 2023, \mnras, 526, 534

\bibitem[{{Hillier} \& {Miller}(1998)}]{hm98}
{Hillier}, D.~J. \& {Miller}, D.~L. 1998, \apj, 496, 407

\bibitem[{{Hummer} \& {Storey}(1987)}]{humstor87}
{Hummer}, D.~G. \& {Storey}, P.~J. 1987, \mnras, 224, 801

\bibitem[{{Isobe} {et~al.}(2023){Isobe}, {Ouchi}, {Tominaga}, {Watanabe},
  {Nakajima}, {Umeda}, {Yajima}, {Harikane}, {Fukushima}, {Xu}, {Ono}, \&
  {Zhang}}]{isobe23}
{Isobe}, Y., {Ouchi}, M., {Tominaga}, N., {et~al.} 2023, \apj, 959, 100

\bibitem[{{Izotov} {et~al.}(1997){Izotov}, {Foltz}, {Green}, {Guseva}, \&
  {Thuan}}]{izotov97}
{Izotov}, Y.~I., {Foltz}, C.~B., {Green}, R.~F., {Guseva}, N.~G., \& {Thuan},
  T.~X. 1997, \apjl, 487, L37

\bibitem[{{James} {et~al.}(2009){James}, {Tsamis}, {Barlow}, {Westmoquette},
  {Walsh}, {Cuisinier}, \& {Exter}}]{james09}
{James}, B.~L., {Tsamis}, Y.~G., {Barlow}, M.~J., {et~al.} 2009, \mnras, 398, 2

\bibitem[{{Ji} {et~al.}(2024){Ji}, {{\"U}bler}, {Maiolino}, {D'Eugenio},
  {Arribas}, {Bunker}, {Charlot}, {Perna}, {Rodr{\'\i}guez Del Pino},
  {B{\"o}ker}, {Cresci}, {Curti}, {Kumari}, \& {Lamperti}}]{ji24}
{Ji}, X., {{\"U}bler}, H., {Maiolino}, R., {et~al.} 2024, \mnras, 535, 881

\bibitem[{{Kalari} {et~al.}(2022){Kalari}, {Horch}, {Salinas}, {Vink},
  {Andersen}, {Bestenlehner}, \& {Rubio}}]{kalari22}
{Kalari}, V.~M., {Horch}, E.~P., {Salinas}, R., {et~al.} 2022, \apj, 935, 162

\bibitem[{{Kehrig} {et~al.}(2018){Kehrig}, {V{\'\i}lchez}, {Guerrero},
  {Iglesias-P{\'a}ramo}, {Hunt}, {Duarte-Puertas}, \&
  {Ramos-Larios}}]{kehrig18}
{Kehrig}, C., {V{\'\i}lchez}, J.~M., {Guerrero}, M.~A., {et~al.} 2018, \mnras,
  480, 1081

\bibitem[{{Kehrig} {et~al.}(2015){Kehrig}, {V{\'\i}lchez}, {P{\'e}rez-Montero},
  {Iglesias-P{\'a}ramo}, {Brinchmann}, {Kunth}, {Durret}, \& {Bayo}}]{kehrig15}
{Kehrig}, C., {V{\'\i}lchez}, J.~M., {P{\'e}rez-Montero}, E., {et~al.} 2015,
  \apjl, 801, L28

\bibitem[{{K{\"o}hler} {et~al.}(2015){K{\"o}hler}, {Langer}, {de Koter}, {de
  Mink}, {Crowther}, {Evans}, {Gr{\"a}fener}, {Sana}, {Sanyal}, {Schneider}, \&
  {Vink}}]{kohler15}
{K{\"o}hler}, K., {Langer}, N., {de Koter}, A., {et~al.} 2015, \aap, 573, A71

\bibitem[{{Kudritzki} {et~al.}(1989){Kudritzki}, {Pauldrach}, {Puls}, \&
  {Abbott}}]{kud}
{Kudritzki}, R.~P., {Pauldrach}, A., {Puls}, J., \& {Abbott}, D.~C. 1989, \aap,
  219, 205

\bibitem[{{Langer}(1992)}]{langer92}
{Langer}, N. 1992, \aap, 265, L17

\bibitem[{{Leitherer} {et~al.}(2018){Leitherer}, {Byler}, {Lee}, \&
  {Levesque}}]{leitherer18}
{Leitherer}, C., {Byler}, N., {Lee}, J.~C., \& {Levesque}, E.~M. 2018, \apj,
  865, 55

\bibitem[{{Leitherer} {et~al.}(2019){Leitherer}, {Lee}, \&
  {Faisst}}]{leitherer19}
{Leitherer}, C., {Lee}, J.~C., \& {Faisst}, A. 2019, \aj, 158, 192

\bibitem[{{Maeder}(1987)}]{maeder87}
{Maeder}, A. 1987, \aap, 178, 159

\bibitem[{{Marques-Chaves} {et~al.}(2020){Marques-Chaves},
  {{\'A}lvarez-M{\'a}rquez}, {Colina}, {P{\'e}rez-Fournon}, {Schaerer}, {Dalla
  Vecchia}, {Hashimoto}, {Jim{\'e}nez-{\'A}ngel}, \& {Shu}}]{marques20}
{Marques-Chaves}, R., {{\'A}lvarez-M{\'a}rquez}, J., {Colina}, L., {et~al.}
  2020, \mnras, 499, L105

\bibitem[{{Marques-Chaves} {et~al.}(2021){Marques-Chaves}, {Schaerer},
  {{\'A}lvarez-M{\'a}rquez}, {Colina}, {Dessauges-Zavadsky},
  {P{\'e}rez-Fournon}, {Saldana-Lopez}, \& {Verhamme}}]{marques21}
{Marques-Chaves}, R., {Schaerer}, D., {{\'A}lvarez-M{\'a}rquez}, J., {et~al.}
  2021, \mnras, 507, 524

\bibitem[{{Marques-Chaves} {et~al.}(2022){Marques-Chaves}, {Schaerer},
  {{\'A}lvarez-M{\'a}rquez}, {Verhamme}, {Ceverino}, {Chisholm}, {Colina},
  {Dessauges-Zavadsky}, {P{\'e}rez-Fournon}, {Saldana-Lopez}, {Upadhyaya}, \&
  {Vanzella}}]{marques22}
{Marques-Chaves}, R., {Schaerer}, D., {{\'A}lvarez-M{\'a}rquez}, J., {et~al.}
  2022, \mnras, 517, 2972

\bibitem[{{Marques-Chaves} {et~al.}(2024){Marques-Chaves}, {Schaerer},
  {Kuruvanthodi}, {Korber}, {Prantzos}, {Charbonnel}, {Weibel}, {Izotov},
  {Messa}, {Brammer}, {Dessauges-Zavadsky}, \& {Oesch}}]{marques24}
{Marques-Chaves}, R., {Schaerer}, D., {Kuruvanthodi}, A., {et~al.} 2024, \aap,
  681, A30

\bibitem[{{Martinet} {et~al.}(2023){Martinet}, {Meynet}, {Ekstr{\"o}m},
  {Georgy}, \& {Hirschi}}]{martinet23}
{Martinet}, S., {Meynet}, G., {Ekstr{\"o}m}, S., {Georgy}, C., \& {Hirschi}, R.
  2023, \aap, 679, A137

\bibitem[{{Martins} {et~al.}(2008){Martins}, {Hillier}, {Paumard},
  {Eisenhauer}, {Ott}, \& {Genzel}}]{arches}
{Martins}, F., {Hillier}, D.~J., {Paumard}, T., {et~al.} 2008, \aap, 478, 219

\bibitem[{{Martins} \& {Palacios}(2022)}]{mp22}
{Martins}, F. \& {Palacios}, A. 2022, \aap, 659, A163

\bibitem[{{Martins} {et~al.}(2023){Martins}, {Schaerer}, {Marques-Chaves}, \&
  {Upadhyaya}}]{martins23}
{Martins}, F., {Schaerer}, D., {Marques-Chaves}, R., \& {Upadhyaya}, A. 2023,
  \aap, 678, A159

\bibitem[{{Maschmann} {et~al.}(2024){Maschmann}, {Leitherer}, {Faisst}, {Lee},
  \& {Minsley}}]{masch24}
{Maschmann}, D., {Leitherer}, C., {Faisst}, A.~L., {Lee}, J.~C., \& {Minsley},
  R. 2024, \apj, 961, 159

\bibitem[{{Nakajima} {et~al.}(2023){Nakajima}, {Ouchi}, {Isobe}, {Harikane},
  {Zhang}, {Ono}, {Umeda}, \& {Oguri}}]{nakajima23}
{Nakajima}, K., {Ouchi}, M., {Isobe}, Y., {et~al.} 2023, \apjs, 269, 33

\bibitem[{{Osterbrock} \& {Ferland}(2006)}]{osterferland}
{Osterbrock}, D.~E. \& {Ferland}, G.~J. 2006, {Astrophysics of gaseous nebulae
  and active galactic nuclei}

\bibitem[{{Pascale} {et~al.}(2023){Pascale}, {Dai}, {McKee}, \&
  {Tsang}}]{pascale23}
{Pascale}, M., {Dai}, L., {McKee}, C.~F., \& {Tsang}, B. T.~H. 2023, \apj, 957,
  77

\bibitem[{{Patr{\'\i}cio} {et~al.}(2016){Patr{\'\i}cio}, {Richard}, {Verhamme},
  {Wisotzki}, {Brinchmann}, {Turner}, {Christensen}, {Weilbacher}, {Blaizot},
  {Bacon}, {Contini}, {Lagattuta}, {Cantalupo}, {Cl{\'e}ment}, \&
  {Soucail}}]{patricio16}
{Patr{\'\i}cio}, V., {Richard}, J., {Verhamme}, A., {et~al.} 2016, \mnras, 456,
  4191

\bibitem[{{Puls} {et~al.}(2000){Puls}, {Springmann}, \& {Lennon}}]{puls00}
{Puls}, J., {Springmann}, U., \& {Lennon}, M. 2000, \aaps, 141, 23

\bibitem[{{Robertson} {et~al.}(2013){Robertson}, {Furlanetto}, {Schneider},
  {Charlot}, {Ellis}, {Stark}, {McLure}, {Dunlop}, {Koekemoer}, {Schenker},
  {Ouchi}, {Ono}, {Curtis-Lake}, {Rogers}, {Bowler}, \&
  {Cirasuolo}}]{robertson13}
{Robertson}, B.~E., {Furlanetto}, S.~R., {Schneider}, E., {et~al.} 2013, \apj,
  768, 71

\bibitem[{{Sabhahit} {et~al.}(2022){Sabhahit}, {Vink}, {Higgins}, \&
  {Sander}}]{sabhahit22}
{Sabhahit}, G.~N., {Vink}, J.~S., {Higgins}, E.~R., \& {Sander}, A. A.~C. 2022,
  \mnras, 514, 3736

\bibitem[{{Sabhahit} {et~al.}(2025){Sabhahit}, {Vink}, {Sander},
  {Bernini-Peron}, {Crowther}, {Lefever}, \& {Shenar}}]{sabhahit25}
{Sabhahit}, G.~N., {Vink}, J.~S., {Sander}, A. A.~C., {et~al.} 2025, \aap, 696,
  A200

\bibitem[{{Sabhahit} {et~al.}(2023){Sabhahit}, {Vink}, {Sander}, \&
  {Higgins}}]{sabhahit23}
{Sabhahit}, G.~N., {Vink}, J.~S., {Sander}, A. A.~C., \& {Higgins}, E.~R. 2023,
  \mnras, 524, 1529

\bibitem[{{Sander} {et~al.}(2012){Sander}, {Hamann}, \& {Todt}}]{sander12}
{Sander}, A., {Hamann}, W.~R., \& {Todt}, H. 2012, \aap, 540, A144

\bibitem[{{Sander} {et~al.}(2014){Sander}, {Todt}, {Hainich}, \&
  {Hamann}}]{sander14}
{Sander}, A., {Todt}, H., {Hainich}, R., \& {Hamann}, W.~R. 2014, \aap, 563,
  A89

\bibitem[{{Sander}(2024)}]{sander24}
{Sander}, A. A.~C. 2024, in IAU Symposium, Vol. 361, Massive Stars Near and
  Far, ed. J.~{Mackey}, J.~S. {Vink}, \& N.~{St-Louis}, 473--478

\bibitem[{{Sander} \& {Vink}(2020)}]{sander20b}
{Sander}, A. A.~C. \& {Vink}, J.~S. 2020, \mnras, 499, 873

\bibitem[{{Schaerer}(2002)}]{schaerer02}
{Schaerer}, D. 2002, \aap, 382, 28

\bibitem[{{Schaerer} {et~al.}(2019){Schaerer}, {Fragos}, \&
  {Izotov}}]{schaerer19}
{Schaerer}, D., {Fragos}, T., \& {Izotov}, Y.~I. 2019, \aap, 622, L10

\bibitem[{{Schaerer} {et~al.}(2025){Schaerer}, {Guibert}, {Marques-Chaves}, \&
  {Martins}}]{schaerer25}
{Schaerer}, D., {Guibert}, J., {Marques-Chaves}, R., \& {Martins}, F. 2025,
  \aap, 693, A271

\bibitem[{{Schaerer} {et~al.}(2024){Schaerer}, {Marques-Chaves}, {Xiao}, \&
  {Korber}}]{schaerer24}
{Schaerer}, D., {Marques-Chaves}, R., {Xiao}, M., \& {Korber}, D. 2024, \aap,
  687, L11

\bibitem[{{Schnurr} {et~al.}(2008){Schnurr}, {Casoli}, {Chen{\'e}}, {Moffat},
  \& {St-Louis}}]{schnurr08}
{Schnurr}, O., {Casoli}, J., {Chen{\'e}}, A.~N., {Moffat}, A.~F.~J., \&
  {St-Louis}, N. 2008, \mnras, 389, L38

\bibitem[{{Seaton}(1978)}]{seaton78}
{Seaton}, M.~J. 1978, \mnras, 185, 5P

\bibitem[{{Seeyave} {et~al.}(2023){Seeyave}, {Wilkins}, {Kuusisto}, {Lovell},
  {Irodotou}, {Simmonds}, {Vijayan}, {Thomas}, {Roper}, {Byrne}, {Jones},
  {Turner}, \& {Conselice}}]{seeyave23}
{Seeyave}, L. T.~C., {Wilkins}, S.~M., {Kuusisto}, J.~K., {et~al.} 2023,
  \mnras, 525, 2422

\bibitem[{{Senchyna} {et~al.}(2024){Senchyna}, {Plat}, {Stark}, {Rudie},
  {Berg}, {Charlot}, {James}, \& {Mingozzi}}]{senchyna24}
{Senchyna}, P., {Plat}, A., {Stark}, D.~P., {et~al.} 2024, \apj, 966, 92

\bibitem[{{Senchyna} {et~al.}(2021){Senchyna}, {Stark}, {Charlot},
  {Chevallard}, {Bruzual}, \& {Vidal-Garc{\'\i}a}}]{senchyna20}
{Senchyna}, P., {Stark}, D.~P., {Charlot}, S., {et~al.} 2021, \mnras, 503, 6112

\bibitem[{{Shenar} {et~al.}(2023){Shenar}, {Sana}, {Crowther}, {Bostroem},
  {Mahy}, {Najarro}, {Oskinova}, \& {Sander}}]{shenar23}
{Shenar}, T., {Sana}, H., {Crowther}, P.~A., {et~al.} 2023, \aap, 679, A36

\bibitem[{{Smith} {et~al.}(2016){Smith}, {Crowther}, {Calzetti}, \&
  {Sidoli}}]{smith16}
{Smith}, L.~J., {Crowther}, P.~A., {Calzetti}, D., \& {Sidoli}, F. 2016, \apj,
  823, 38

\bibitem[{{Smith} {et~al.}(2023){Smith}, {Oey}, {Hernandez}, {Ryon},
  {Leitherer}, {Charlot}, {Bruzual}, {Calzetti}, {Chu}, {Hayes}, {James},
  {Jaskot}, \& {{\"O}stlin}}]{smith23}
{Smith}, L.~J., {Oey}, M.~S., {Hernandez}, S., {et~al.} 2023, \apj, 958, 194

\bibitem[{{Stasi{\'n}ska} {et~al.}(2015){Stasi{\'n}ska}, {Izotov}, {Morisset},
  \& {Guseva}}]{stasinska15}
{Stasi{\'n}ska}, G., {Izotov}, Y., {Morisset}, C., \& {Guseva}, N. 2015, \aap,
  576, A83

\bibitem[{{Stolte} {et~al.}(2006){Stolte}, {Brandner}, {Brandl}, \&
  {Zinnecker}}]{stolte06}
{Stolte}, A., {Brandner}, W., {Brandl}, B., \& {Zinnecker}, H. 2006, \aj, 132,
  253

\bibitem[{{Upadhyaya} {et~al.}(2024){Upadhyaya}, {Marques-Chaves}, {Schaerer},
  {Martins}, {P{\'e}rez-Fournon}, {Palacios}, \& {Stanway}}]{upad}
{Upadhyaya}, A., {Marques-Chaves}, R., {Schaerer}, D., {et~al.} 2024, \aap,
  686, A185

\bibitem[{{Vanzella} {et~al.}(2023){Vanzella}, {Loiacono}, {Bergamini},
  {Me{\v{s}}tri{\'c}}, {Castellano}, {Rosati}, {Meneghetti}, {Grillo},
  {Calura}, {Mignoli}, {Brada{\v{c}}}, {Adamo}, {Rihtar{\v{s}}i{\v{c}}},
  {Dickinson}, {Gronke}, {Zanella}, {Annibali}, {Willott}, {Messa}, {Sani},
  {Acebron}, {Bolamperti}, {Comastri}, {Gilli}, {Caputi}, {Ricotti},
  {Gruppioni}, {Ravindranath}, {Mercurio}, {Strait}, {Martis}, {Pascale},
  {Caminha}, {Annunziatella}, \& {Nonino}}]{vanzella23}
{Vanzella}, E., {Loiacono}, F., {Bergamini}, P., {et~al.} 2023, \aap, 678, A173

\bibitem[{{Villar-Mart{\'\i}n} {et~al.}(2004){Villar-Mart{\'\i}n},
  {Cervi{\~n}o}, \& {Gonz{\'a}lez Delgado}}]{villar04}
{Villar-Mart{\'\i}n}, M., {Cervi{\~n}o}, M., \& {Gonz{\'a}lez Delgado}, R.~M.
  2004, \mnras, 355, 1132

\bibitem[{{Vink}(2015)}]{vink15}
{Vink}, J.~S., ed. 2015, Astrophysics and Space Science Library, Vol. 412,
  {Very Massive Stars in the Local Universe}, ed. J.~S. {Vink}

\bibitem[{{Vink}(2023)}]{vink23}
{Vink}, J.~S. 2023, \aap, 679, L9

\bibitem[{{Vink} {et~al.}(2001){Vink}, {de Koter}, \& {Lamers}}]{vink01}
{Vink}, J.~S., {de Koter}, A., \& {Lamers}, H.~J.~G.~L.~M. 2001, \aap, 369, 574

\bibitem[{{Vink} \& {Gr{\"a}fener}(2012)}]{vg12}
{Vink}, J.~S. \& {Gr{\"a}fener}, G. 2012, \apjl, 751, L34

\bibitem[{{Vink} {et~al.}(2011){Vink}, {Muijres}, {Anthonisse}, {de Koter},
  {Gr{\"a}fener}, \& {Langer}}]{vink11}
{Vink}, J.~S., {Muijres}, L.~E., {Anthonisse}, B., {et~al.} 2011, \aap, 531,
  A132

\bibitem[{{Wofford} {et~al.}(2014){Wofford}, {Leitherer}, {Chandar}, \&
  {Bouret}}]{wofford14}
{Wofford}, A., {Leitherer}, C., {Chandar}, R., \& {Bouret}, J.-C. 2014, \apj,
  781, 122

\bibitem[{{Wofford} {et~al.}(2023){Wofford}, {Sixtos}, {Charlot}, {Bruzual},
  {Cullen}, {Stanton}, {Hern{\'a}ndez}, {Smith}, \& {Hayes}}]{wofford23}
{Wofford}, A., {Sixtos}, A., {Charlot}, S., {et~al.} 2023, \mnras, 523, 3949

\bibitem[{{Wofford} {et~al.}(2021){Wofford}, {Vidal-Garc{\'\i}a}, {Feltre},
  {Chevallard}, {Charlot}, {Stark}, {Herenz}, \& {Hayes}}]{wofford21}
{Wofford}, A., {Vidal-Garc{\'\i}a}, A., {Feltre}, A., {et~al.} 2021, \mnras,
  500, 2908

\bibitem[{{Xu} {et~al.}(2013){Xu}, {Goriely}, {Jorissen}, {Chen}, \&
  {Arnould}}]{xu13}
{Xu}, Y., {Goriely}, S., {Jorissen}, A., {Chen}, G.~L., \& {Arnould}, M. 2013,
  \aap, 549, A106

\bibitem[{{Yusof} {et~al.}(2013){Yusof}, {Hirschi}, {Meynet}, {Crowther},
  {Ekstr{\"o}m}, {Frischknecht}, {Georgy}, {Abu Kassim}, \&
  {Schnurr}}]{yusof13}
{Yusof}, N., {Hirschi}, R., {Meynet}, G., {et~al.} 2013, \mnras, 433, 1114

\end{thebibliography}

\onecolumn

\begin{appendix}

\section{Model parameters}
\label{s_tabparam}

Tables~\ref{tab_gr_zsmc} to \ref{tab_grz_z0p01} gather the parameters for our sets of models at the three metallicities explored, and for each metallicity with the two assumptions regarding VMS mass loss rate. 

\begin{table*}[ht]
\begin{center}
  \caption{Parameters of the atmosphere models with mass loss rates not scaled with metallicity and at Z~=~0.2~\zsun.} \label{tab_gr_zsmc}
  \small
\begin{tabular}{lcccccccccccc}
\hline
 M & $T_{\rm eff}$  &  $\log g$ & $\log(L/L_{\odot})$ & R& $\log \dot{M}$ & $\varv_{\infty}$ & age & H & He & C & N & O \\    
  \msun & K          &                   &          &  R$\odot$ &            & \kms       & Myr  &  & & & & \\
\hline
150 \msun & & & & & & & & & \\
\hline
149.9 & 60415 & 4.31 & 6.379 &   14.20 & -5.01 &   3815 &  0.01 & 0.746 & 0.252 & 3.921~$10^{-4}$ & 1.148~$10^{-4}$ & 9.508~$10^{-4}$ \\
144.6 & 57642 & 4.20 & 6.386 &   15.73 & -4.90 &   3447 &  0.50 & 0.746 & 0.252 & 3.921~$10^{-4}$ & 1.148~$10^{-4}$ & 9.508~$10^{-4}$ \\
137 & 54852 & 4.09 & 6.396 &   17.57 & -4.73 &   3132 &  1.00 & 0.746 & 0.251 & 3.921~$10^{-4}$ & 1.148~$10^{-4}$ & 9.508~$10^{-4}$ \\
125 & 51673 & 3.93 & 6.404 &   19.98 & -4.53 &   2626 &  1.50 & 0.709 & 0.289 & 2.156~$10^{-5}$ & 1.371~$10^{-3}$ & 9.002~$10^{-6}$ \\
122.7 & 51490 & 3.85 & 6.407 &   20.19 & -4.35 &   2316 &  2.00 & 0.534 & 0.464 & 2.171~$10^{-5}$ & 1.371~$10^{-3}$ & 8.290~$10^{-6}$ \\
82.5 & 54443 & 3.84 & 6.401 &  17.93 & -4.25 &   2081 &  2.50 & 0.263 & 0.735 & 2.359~$10^{-5}$ & 1.370~$10^{-3}$ & 7.592~$10^{-6}$ \\
\hline                                                
200 \msun &  & & & & & & & & \\                       
\hline  
201.0  & 61745 & 4.29 & 6.565 &   16.84 & -4.69 &   3837 & 0.00 & 0.746 & 0.252 & 3.921~$10^{-4}$ & 1.148~$10^{-4}$ & 9.508~$10^{-4}$ \\
188.1 & 58232 & 4.16 & 6.563 &   18.88 & -4.59 &   3384 & 0.50 & 0.746 & 0.252 & 3.895~$10^{-4}$ & 1.200~$10^{-4}$ & 9.504~$10^{-4}$ \\
173.8  & 55451 & 4.04 & 6.563 &   20.87 & -4.42 &   2985 & 1.00 & 0.716 & 0.282 & 2.215~$10^{-5}$ & 1.370~$10^{-3}$ & 9.574~$10^{-6}$ \\
151.9  & 54421 & 3.95 & 6.562 &   21.69 & -4.30 &   2646 & 1.50 & 0.585 & 0.413 & 2.256~$10^{-5}$ & 1.370~$10^{-3}$ & 8.137~$10^{-6}$ \\
123.0  & 54862 & 3.88 & 6.555 &   21.23 & -4.19 &   1948 & 2.00 & 0.371 & 0.627 & 2.358~$10^{-5}$ & 1.370~$10^{-3}$ & 7.692~$10^{-6}$ \\
86.4  & 57821 & 3.85 & 6.523 &   18.37 & -4.09 &   2003 & 2.50 & 0.091 & 0.906 & 2.658~$10^{-5}$ & 1.367~$10^{-3}$ & 6.638~$10^{-6}$ \\
\hline                                                
250 \msun & & & & & & & & & \\                       
\hline                                                                              
250 &  61207 &  4.23 &  6.698 & 19.98 & -4.51 &    3693 &  0.00 & 0.746 & 0.252 & 3.921~$10^{-4}$ & 1.148~$10^{-4}$ & 9.508~$10^{-4}$ \\
232.5 &  58286 &  4.12 &  6.693 & 21.89 & -4.38 &    3284 &  0.49 & 0.746 & 0.252 & 3.921~$10^{-4}$ & 1.148~$10^{-4}$ & 9.508~$10^{-4}$ \\
208 &  56272 &  4.02 &  6.688 & 23.37 & -4.26 &    2929 &  1.00 & 0.669 & 0.329 & 2.273~$10^{-5}$ & 1.370~$10^{-3}$ & 8.088~$10^{-6}$ \\
176.9 &  55599 &  3.94 &  6.680 & 23.71 & -4.16 &    2622 &  1.50 & 0.506 & 0.492 & 2.333~$10^{-5}$ & 1.370~$10^{-3}$ & 7.820~$10^{-6}$ \\
138.6 &  55194 &  3.84 &  6.661 & 23.54 & -4.07 &    2253 &  2.00 & 0.281 & 0.717 & 2.462~$10^{-5}$ & 1.369~$10^{-3}$ & 7.311~$10^{-6}$ \\
111.9 &  54955 &  3.76 &  6.636 & 23.07 & -4.02 &    1968 &  2.30 & 0.121 & 0.877 & 2.635~$10^{-5}$ & 1.367~$10^{-3}$ & 6.708~$10^{-6}$\\
\hline                                                
300 \msun & & & & & & & & & \\                       
\hline                                                                              
299.9 & 62395 & 4.24 & 6.807 &  21.78 & -4.35 &   3306  & 0.01  & 0.746 & 0.252 & 3.921~$10^{-4}$ & 1.148~$10^{-4}$ & 9.508~$10^{-4}$ \\
274.7 & 58129 & 4.09 & 6.796 &  24.79 & -4.23 &   3249  & 0.50  & 0.738 & 0.260 & 2.155~$10^{-5}$ & 1.282~$10^{-3}$ & 1.109~$10^{-4}$ \\
241.4 & 56384 & 3.99 & 6.787 &  26.08 & -4.14 &   2867  & 1.00  & 0.629 & 0.369 & 2.318~$10^{-5}$ & 1.370~$10^{-3}$ & 7.887~$10^{-6}$ \\
200.6 & 55562 & 3.90 & 6.774 &  26.46 & -4.06 &   2528  & 1.50  & 0.451 & 0.547 & 2.390~$10^{-5}$ & 1.369~$10^{-3}$ & 7.586~$10^{-6}$ \\
152 & 53836 & 3.75 & 6.746 &  27.28 & -3.98 &   2062  & 2.00  & 0.220 & 0.778 & 2.541~$10^{-5}$ & 1.368~$10^{-3}$ & 7.018~$10^{-6}$ \\
104.5 & 70779 & 4.10 & 6.707 &  15.09 & -3.75 &   2049  & 2.40  & 0.000 & 0.998 & 3.365~$10^{-5}$ & 1.361~$10^{-3}$ & 4.386~$10^{-6}$  \\
\hline
\end{tabular}
\tablefoot{Columns are mass of the corresponding model according to stellar evolution rounded up to the first decimal, effective temperature, surface gravity, luminosity, radius, mass-loss rate, wind terminal velocity, age, and surface abundances of H, He, C, N, and O given in mass fraction. For each mass, the lines of this table correspond to the characteristics of the models marked as open squares in Fig.~\ref{hrd}).}
\end{center}
\end{table*}

\normalsize

\begin{table*}[ht]
\begin{center}
  \caption{Parameters of the atmosphere models with mass loss rates scaled linearly with Z/Z$_{LMC}$ and at Z~=~0.2~\zsun.} \label{tab_grz_zsmc}
  \small
\begin{tabular}{lccccccccccccc}
\hline
M & $T_{\rm eff}$  &  $\log g$ & $\log(L/L_{\odot})$ & R& $\log \dot{M}$ & $\varv_{\infty}$ & age & H & He & C & N & O \\    
  \msun & K          &                   &          &  R$\odot$ &            & \kms       & Myr  &  & & & & \\
\hline
150 \msun  & & & & & & & & & \\
\hline
150 &  60268 & 4.31 & 6.378 & 14.25 & -5.42  &  3835 &  0.01 & 0.746 & 0.252 & 3.921~$10^{-4}$ & 1.148~$10^{-4}$ & 9.508~$10^{-4}$ \\
147.8 &  57750 & 4.21 & 6.398 & 15.89 & -5.28  &  3527 &  0.50 & 0.746 & 0.252 & 3.921~$10^{-4}$ & 1.148~$10^{-4}$ & 9.508~$10^{-4}$ \\
144.6 &  55054 & 4.09 & 6.421 & 17.94 & -5.12  &  3144 &  1.01 & 0.746 & 0.251 & 3.921~$10^{-4}$ & 1.148~$10^{-4}$ & 9.508~$10^{-4}$ \\
139.9 &  50126 & 3.89 & 6.444 & 22.22 & -4.92  &  2657 &  1.50 & 0.746 & 0.251 & 3.921~$10^{-4}$ & 1.148~$10^{-4}$ & 9.508~$10^{-4}$ \\
131.7 &  39811 & 3.44 & 6.470 & 36.30 & -4.65  &  1878 &  2.00 & 0.746 & 0.251 & 3.921~$10^{-4}$ & 1.148~$10^{-4}$ & 9.508~$10^{-4}$ \\
116.1 &  24788 & 2.53 & 6.502 & 97.18 & -4.38  &  970. &  2.50 & 0.611 & 0.387 & 2.127~$10^{-5}$ & 1.372~$10^{-3}$ & 8.424~$10^{-6}$ \\
\hline                                                
200 \msun & & & & & & & & & \\                       
\hline 
200 &  61317 & 4.28 & 6.562 & 17.01 & -5.18  &  3821 &  0.00 & 0.746 & 0.252 & 3.921~$10^{-4}$ & 1.148~$10^{-4}$ & 9.508~$10^{-4}$ \\
195.4 &  58330 & 4.16 & 6.580 & 19.19 & -4.97  &  3399 &  0.50 & 0.746 & 0.252 & 2.734~$10^{-4}$ & 2.596~$10^{-4}$ & 9.481~$10^{-4}$ \\
188.9 &  54318 & 4.01 & 6.598 & 22.60 & -4.80  &  2995 &  1.00 & 0.746 & 0.252 & 9.444~$10^{-5}$ & 4.664~$10^{-4}$ & 9.469~$10^{-4}$ \\
178.5 &  46579 & 3.69 & 6.618 & 31.45 & -4.57  &  2244 &  1.51 & 0.739 & 0.259 & 2.035~$10^{-5}$ & 1.245~$10^{-3}$ & 1.541~$10^{-4}$ \\
161.6 &  36072 & 3.18 & 6.643 & 53.98 & -4.37  &  1504 &  2.00 & 0.629 & 0.368 & 2.236~$10^{-5}$ & 1.370~$10^{-3}$ & 8.189~$10^{-6}$ \\
140.1 &  22404 & 2.27 & 6.670 & 144.34& -4.17  &  791. &  2.40 & 0.446 & 0.552 & 2.230~$10^{-5}$ & 1.371~$10^{-3}$ & 7.815~$10^{-6}$ \\
\hline                                                
250 \msun & & & & & & & & & \\                       
\hline                                                                              
249.3 &  61883 &  4.25 &  6.699  & 19.57 & -4.90 & 3745  & 0.03 & 0.746 & 0.252 & 3.921~$10^{-4}$ & 1.148~$10^{-4}$ & 9.508~$10^{-4}$ \\
242.9 &  58368 &  4.13 &  6.712  & 22.33 & -4.77 & 3394  & 0.50 & 0.746 & 0.252 & 3.921~$10^{-4}$ & 1.148~$10^{-4}$ & 9.508~$10^{-4}$  \\
232.4 &  52944 &  3.92 &  6.728  & 27.64 & -4.59 & 2771  & 1.00 & 0.746 & 0.252 & 3.921~$10^{-4}$ & 1.148~$10^{-4}$ & 9.508~$10^{-4}$ \\
216.2 &  45091 &  3.59 &  6.747  & 38.94 & -4.40 & 2083  & 1.50 & 0.688 & 0.310 & 2.270~$10^{-5}$ & 1.370~$10^{-3}$ & 8.107~$10^{-6}$ \\
191.5 &  32932 &  2.97 &  6.771  & 75.04 & -4.21 & 1294  & 2.00 & 0.537 & 0.461 & 2.311~$10^{-5}$ & 1.370~$10^{-3}$ & 7.836~$10^{-6}$ \\
170.1 &  20580 &  2.08 &  6.792  & 196.86 & -4.07 & 681  & 2.30 & 0.386 & 0.612 & 2.348~$10^{-5}$ & 1.370~$10^{-3}$ & 7.516~$10^{-6}$ \\
\hline                                                
300 \msun & & & & & & & & & \\                       
\hline                                                                              
299.8 &  62390 & 4.24 & 6.807 &  21.80 & -4.75 &   3774  & 0.02  & 0.746 & 0.252 & 3.921~$10^{-4}$ & 1.148~$10^{-4}$ & 9.506~$10^{-4}$ \\
289.8 &  57903 & 4.08 & 6.818 &  25.62 & -4.62 &   3234  & 0.50  & 0.746 & 0.252 & 3.921~$10^{-4}$ & 1.148~$10^{-4}$ & 9.506~$10^{-4}$  \\
275 &  51520 & 3.84 & 6.832 &  32.89 & -4.44 &   2596  & 1.00  & 0.733 & 0.265 & 2.238~$10^{-5}$ & 1.336~$10^{-3}$ & 4.750~$10^{-5}$ \\
253.2 &  42901 & 3.47 & 6.850 &  48.43 & -4.28 &   1910  & 1.50  & 0.645 & 0.353 & 2.317~$10^{-5}$ & 1.370~$10^{-3}$ & 7.888~$10^{-6}$ \\
221.1 &  28522 & 2.68 & 6.874 & 112.64 & -4.10 &   1058  & 2.00  & 0.470 & 0.528 & 2.375~$10^{-5}$ & 1.369~$10^{-3}$ & 7.575~$10^{-6}$ \\
\hline
\end{tabular}
\tablefoot{Columns are mass of the corresponding model according to stellar evolution rounded up to the first decimal, effective temperature, surface gravity, luminosity, radius, mass-loss rate, wind terminal velocity, age, and surface abundances of H, He, C, N, and O given in mass fraction. For each mass, the lines of this table correspond to the characteristics of the models marked as open squares in Fig.~\ref{hrd}).}
\end{center}
\end{table*}

\normalsize

\begin{table*}[ht]
\begin{center}
  \caption{Parameters of the atmosphere models with mass loss rates not scaled with metallicity and at Z~=~0.1~\zsun.} \label{tab_gr_z0p1}
  \small
\begin{tabular}{lccccccccccccc}
\hline
  M & $T_{\rm eff}$  &  $\log g$ & $\log(L/L_{\odot})$ & R& $\log \dot{M}$ & $\varv_{\infty}$ & age & H & He & C & N & O \\    
  \msun & K          &                   &          &  R$\odot$ &            & \kms       & Myr  &  & & & & \\
\hline
150 \msun  & & & & & & & & & \\
\hline
149.9  & 61551 & 4.34 & 6.378 & 13.66 & -5.02 & 3863  & 0.01 & 0.748 & 0.251 & 2.379~$10^{-4}$ & 6.967~$10^{-5}$ & 5.769~$10^{-4}$ \\
144.6  & 59081 & 4.25 & 6.386 & 14.97 & -4.90 & 3584  & 0.50 & 0.748 & 0.251 & 2.379~$10^{-4}$ & 6.967~$10^{-5}$ & 5.769~$10^{-4}$ \\
136.8  & 56530 & 4.14 & 6.396 & 16.53 & -4.73 & 3209  & 1.00 & 0.748 & 0.251 & 2.379~$10^{-4}$ & 6.967~$10^{-5}$ & 5.769~$10^{-4}$ \\
124.9  & 54025 & 4.01 & 6.403 & 18.26 & -4.53 & 2762  & 1.50 & 0.710 & 0.289 & 1.356~$10^{-5}$ & 8.317~$10^{-4}$ & 4.985~$10^{-6}$ \\
107  & 54190 & 3.94 & 6.409 & 18.27 & -4.37 & 2433  & 2.00 & 0.549 & 0.450 & 1.342~$10^{-5}$ & 8.320~$10^{-4}$ & 4.807~$10^{-6}$ \\
82.3  & 59346 & 3.99 & 6.405 & 15.16 & -4.24 & 2261  & 2.50 & 0.273 & 0.726 & 1.453~$10^{-5}$ & 8.311~$10^{-4}$ & 4.410~$10^{-6}$ \\
\hline                                                
200 \msun & & & & & & & & & \\                       
\hline                                                                              
199.9 &  63308 &  4.33 &  6.564 &  16.00 &  -4.70 &    3900  & 0.01 & 0.748 & 0.251 & 2.379~$10^{-4}$ & 6.967~$10^{-5}$ & 5.769~$10^{-4}$ \\
188.8 &  59991 &  4.21 &  6.563 &  17.79 &  -4.58 &    3470  & 0.50 & 0.748 & 0.251 & 1.358~$10^{-4}$ & 1.924~$10^{-4}$ & 5.752~$10^{-4}$ \\
172.8 &  57830 &  4.11 &  6.563 &  19.17 &  -4.42 &    3087  & 1.00 & 0.717 & 0.282 & 1.397~$10^{-5}$ & 8.314~$10^{-4}$ & 4.837~$10^{-6}$ \\
150.8 &  57746 &  4.05 &  6.562 &  19.19 &  -4.30 &    2782  & 1.50 & 0.413 & 0.586 & 1.418~$10^{-5}$ & 8.312~$10^{-4}$ & 4.704~$10^{-6}$ \\
122.2 &  59953 &  4.03 &  6.554 &  17.64 &  -4.19 &    2505  & 2.00 & 0.372 & 0.626 & 1.480~$10^{-5}$ & 8.307~$10^{-4}$ & 4.453~$10^{-6}$ \\
86.3 &  67074 &  4.11 &  6.521 &  13.57 &  -4.09 &    2334  & 2.50 & 0.093 & 0.906 & 1.670~$10^{-5}$ & 8.291~$10^{-4}$ & 3.841~$10^{-6}$ \\

\hline                                                
250 \msun & & & & & & & & & \\                       
\hline                                                                              
249.5 &  63479 &  4.30 &  6.698 & 18.58 & -4.50 & 3892 &  0.02 & 0.748 & 0.251 & 2.379~$10^{-4}$ & 6.967~$10^{-5}$ & 5.769~$10^{-4}$\\
232.5 &  60447 &  4.19 &  6.692 & 20.34 & -4.38 & 3465 &  0.50 & 0.748 & 0.251 & 2.379~$10^{-4}$ & 6.967~$10^{-5}$ & 5.769~$10^{-4}$ \\
207.9 &  59311 &  4.11 &  6.687 & 21.00 & -4.26 & 3073 &  1.00 & 0.670 & 0.329 & 1.431~$10^{-5}$ & 8.311~$10^{-4}$ & 4.679~$10^{-6}$\\
176.7 & 59969 &  4.07 &  6.679 & 20.36 & -4.16 & 2811 &  1.50 & 0.508 & 0.491 & 1.467~$10^{-5}$ & 8.308~$10^{-4}$ & 4.526~$10^{-6}$\\
138.6 & 62020 &  4.04 &  6.659 & 18.60 & -4.07 & 2500 &  2.00 & 0.285 & 0.714 & 1.546~$10^{-5}$ & 8.302~$10^{-4}$ & 4.237~$10^{-6}$\\
102 & 66146 &  4.06 &  6.621 & 15.65 & -3.99 & 2276 &  2.40 & 0.073 & 0.926 & 1.725~$10^{-5}$ & 8.286~$10^{-4}$ & 3.684~$10^{-6}$\\
\hline                                                
300 \msun & &  & & & & & & & \\                       
\hline                                                            
300 &  64627 &  4.30 &  6.807 & 20.32 & -4.34  &  3893 & 0.00 & 0.748 & 0.251 & 2.379~$10^{-4}$ & 6.967~$10^{-5}$ & 5.769~$10^{-4}$ \\
275.2 &  60740 & 4.17 & 6.796 &  22.70 & -4.23 &  3425  & 0.50  & 
0.741 & 0.258 & 1.401~$10^{-5}$ & 8.062~$10^{-4}$  & 3.365~$10^{-5}$ \\
241.8 &  60174 & 4.10 & 6.786 &  22.87 & -4.13 &  3027  & 1.00  & 0.630 & 0.369 & 1.459~$10^{-5}$ & 8.309~$10^{-4}$ & 4.565~$10^{-6}$ \\
201.9 &  60897 & 4.06 & 6.773 &  21.99 & -4.06 &  2270  & 1.50  & 0.453 & 0.546 & 1.503~$10^{-5}$ & 8.305~$10^{-4}$ & 4.394~$10^{-6}$ \\
153.8 &  61992 & 4.00 & 6.744 &  20.53 & -3.98 &  2405  & 2.00  & 0.224 & 0.775 & 1.597~$10^{-5}$ & 8.297~$10^{-4}$ & 4.069~$10^{-6}$ \\
115.8 &  65720 & 4.02 & 6.705 &  17.46 & -3.90 & 2187  & 2.33  & 0.049 & 0.950 & 1.796~$10^{-5}$ & 8.279~$10^{-4}$ & 3.498~$10^{-6}$  \\
\hline
\end{tabular}
\tablefoot{Columns are mass of the corresponding model according to stellar evolution rounded up to the first decimal, effective temperature, surface gravity, luminosity, radius, mass-loss rate, wind terminal velocity, age, and surface abundances of H, He, C, N, and O given in mass fraction. For each mass, the lines of this table correspond to the characteristics of the models marked as filled squares in Fig.~\ref{hrd}).}
\end{center}
\end{table*}

\normalsize

\begin{table*}[ht]
\begin{center}
  \caption{Parameters of the atmosphere models with mass loss rates scaled linearly with Z and at Z~=~0.1~\zsun.} \label{tab_grz_z0p1}
  \small
\begin{tabular}{lcccccccccccc}
\hline
M & $T_{\rm eff}$  &  $\log g$ & $\log(L/L_{\odot})$ & R& $\log \dot{M}$ & $\varv_{\infty}$ & age & H & He & C & N & O \\    
 \msun &  K          &                   &          &  R$\odot$ &            & \kms       & Myr  &  & & & & \\
\hline
150 \msun & & & & & & & & & \\
\hline
150 & 61566 & 4.34 & 6.378 &  13.66 &  -5.63 &   3862  & 0.01 & 0.748 & 0.251 & 2.379~$10^{-4}$ & 6.967~$10^{-5}$ & 5.769~$10^{-4}$ \\
148.6 & 59236 & 4.25 & 6.402 &  15.16 &  -5.49 &   3591  & 0.50 & 0.748 & 0.251 & 2.379~$10^{-4}$ & 6.967~$10^{-5}$ & 5.769~$10^{-4}$ \\
146.7 & 56821 & 4.14 & 6.427 &  16.97 &  -5.33 &   3219  & 1.00 & 0.748 & 0.251 & 2.379~$10^{-4}$ & 6.967~$10^{-5}$ & 5.769~$10^{-4}$ \\
143.8 & 52448 & 3.97 & 6.454 &  20.54 &  -5.15 &   2804  & 1.50 & 0.748 & 0.251 & 2.379~$10^{-4}$ & 6.967~$10^{-5}$ & 5.769~$10^{-4}$ \\
139.2 & 43724 & 3.61 & 6.483 &  30.56 &  -4.92 &   2119  & 2.00 & 0.748 & 0.251 & 2.379~$10^{-4}$ & 6.967~$10^{-5}$ & 5.769~$10^{-4}$ \\
130.4 & 22079 & 2.36 & 6.518 & 124.76 &  -4.59 &    896  & 2.50 & 0.745 & 0.253 & 7.566~$10^{-5}$ & 5.361~$10^{-4}$ & 2.600~$10^{-4}$ \\
\hline                                                
200 \msun & & & & & & & & & \\                       
\hline                                                                              
200 & 63321 & 4.33 & 6.564 &  15.99 &  -5.32 &   3897  &  0.00 & 0.748 & 0.251 & 2.379~$10^{-4}$ & 6.967~$10^{-5}$ & 5.769~$10^{-4}$ \\
197.2 & 60169 & 4.22 & 6.584 &  18.13 &  -5.18 &   3565  &  0.51 & 0.748 & 0.251 & 2.167~$10^{-4}$ & 9.790~$10^{-5}$ & 5.757~$10^{-4}$ \\
193.2 & 56682 & 4.08 & 6.607 &  20.97 &  -5.02 &   3107  &  1.01 & 0.748 & 0.251 & 1.269~$10^{-4}$ & 2.027~$10^{-4}$ & 5.752~$10^{-4}$ \\
187.2 & 49803 & 3.82 & 6.631 &  27.93 &  -4.82 &   2509  &  1.50 & 0.748 & 0.251 & 3.687~$10^{-5}$ & 3.069~$10^{-4}$ & 5.743~$10^{-4}$ \\
176.9 & 35628 & 3.18 & 6.660 &  56.41 &  -4.56 &   1502  &  2.00 & 0.731 & 0.268 & 1.393~$10^{-5}$ & 8.305~$10^{-4}$ & 5.919~$10^{-6}$ \\
170.7 & 26821 & 2.66 & 6.673 & 101.06 &  -4.46 &   1067  &  2.20 & 0.691 & 0.307 & 1.399~$10^{-5}$ & 8.314~$10^{-4}$ & 4.795~$10^{-6}$ \\
\hline                                                
250 \msun & & & & & & & & & \\                       
\hline                                                                              
249.8 & 63530 & 4.30 & 6.699  & 18.57 & -5.11 & 3884 & 0.03 & 0.748 & 0.251 & 2.379~$10^{-4}$ & 6.967~$10^{-5}$ & 5.769~$10^{-4}$ \\
245.6 & 60640 & 4.19 & 6.717  & 20.80 & -4.98 & 3474 & 0.50 & 0.748 & 0.251 & 2.379~$10^{-4}$ & 6.967~$10^{-5}$ & 5.769~$10^{-4}$ \\
239.3 & 56146 & 4.03 & 6.738  & 24.86 & -4.82 & 3024 & 1.00 & 0.748 & 0.251 & 2.379~$10^{-4}$ & 6.967~$10^{-5}$ & 5.769~$10^{-4}$ \\
229.6 & 46858 & 3.67 & 6.761  & 36.65 & -4.60 & 2206 & 1.50 & 0.748 & 0.251 & 2.068~$10^{-4}$ & 1.546~$10^{-4}$ & 5.214~$10^{-4}$ \\
213.7 & 30564 & 2.87 & 6.789  & 88.95 & -4.40 & 1231 & 2.00 & 0.670 & 0.328 & 1.432~$10^{-5}$ & 8.311~$10^{-4}$ & 4.665~$10^{-6}$ \\
\hline                                                
300 \msun & & & & & & & & & \\                       
\hline                                                          
300 & 64263 & 4.29 & 6.807 &  20.54 & -4.96 &  3868  & 0.00  & 0.748 & 0.251 & 2.379~$10^{-4}$ & 6.967~$10^{-5}$ & 5.769~$10^{-4}$ \\
293.9 & 60705 & 4.17 & 6.823 & 23.45 & -4.84 &  3477  & 0.50  & 0.748 & 0.251 & 2.379~$10^{-4}$ & 6.967~$10^{-5}$ & 5.769~$10^{-4}$ \\
285.1 & 55047 & 3.96 & 6.842 & 29.16 & -4.67 &  2813  & 1.00  & 0.748 & 0.251 & 2.379~$10^{-4}$ & 6.967~$10^{-5}$ & 5.769~$10^{-4}$ \\
271.5 & 44384 & 3.55 & 6.865 &  46.02 & -4.47 & 2046  & 1.50  & 0.724 & 0.275 & 1.443~$10^{-5}$ & 8.309~$10^{-4}$ & 4.760~$10^{-6}$ \\
250.3 & 23967 & 2.41 & 6.892 & 162.86 & -4.28 & 893  & 2.00  & 0.624 & 0.375 & 1.460~$10^{-5}$ & 8.309~$10^{-4}$ & 4.541~$10^{-6}$ \\
\hline
\end{tabular}
\tablefoot{Columns are mass of the corresponding model according to stellar evolution rounded up to the first decimal, effective temperature, surface gravity, luminosity, radius, mass-loss rate, wind terminal velocity, age, and surface abundances of H, He, C, N, and O given in mass fraction. For each mass, the lines of this table correspond to the characteristics of the models marked as filled squares in Fig.~\ref{hrd}).}
\end{center}
\end{table*}

\normalsize

\begin{table*}[ht]
\begin{center}
  \caption{Parameters of the atmosphere models with mass loss rates not scaled with metallicity and at Z~=~0.01~\zsun.} \label{tab_gr_z0p01}
  \small
\begin{tabular}{lccccccccccccc}
\hline
  M & $T_{\rm eff}$  &  $\log g$ & $\log(L/L_{\odot})$ & R& $\log \dot{M}$ & $\varv_{\infty}$ & age & H & He & C & N & O \\    
  \msun & K          &                   &          &  R$\odot$ &            & \kms       & Myr  &  & & & & \\
\hline
150 \msun  & & & & & & & & & \\
\hline
149.5  & 65950 & 4.46 & 6.379 & 11.92 & -5.00 & 4141  & 0.01 & 0.751 & 0.249 & 2.491~$10^{-5}$ & 7.295~$10^{-6}$ & 6.040~$10^{-5}$ \\
145.3  & 64470 & 4.40 & 6.387 & 12.59 & -4.89 & 3898  & 0.50 & 0.751 & 0.249 & 2.491~$10^{-5}$ & 7.295~$10^{-6}$ & 6.040~$10^{-5}$ \\
136.9  & 62095 & 4.30 & 6.396 & 13.71 & -4.71 & 3498  & 1.00 & 0.751 & 0.249 & 2.491~$10^{-5}$ & 7.295~$10^{-6}$ & 6.040~$10^{-5}$ \\
122.5  & 60486 & 4.20 & 6.402 & 14.55 & -4.45 & 3051  & 1.50 & 0.702 & 0.298 & 1.545~$10^{-6}$ & 8.703~$10^{-5}$ & 4.203~$10^{-7}$ \\
103.6  & 62714 & 4.19 & 6.403 & 13.55 & -4.38 & 2804  & 2.00 & 0.527 & 0.473 & 9.491~$10^{-7}$ & 8.773~$10^{-5}$ & 4.042~$10^{-7}$ \\
81.0   & 70365 & 4.29 & 6.396 & 10.67 & -4.25 & 2714  & 2.50 & 0.263 & 0.737 & 1.328~$10^{-6}$ & 8.732~$10^{-5}$ & 3.708~$10^{-7}$ \\
\hline                                                
200 \msun & & & & & & & & & \\                       
\hline                                                                              
197.9  & 67602 & 4.44 & 6.564 & 14.03 & -4.70 & 4122  & 0.01 & 0.751 & 0.249 & 2.491~$10^{-5}$ & 7.295~$10^{-6}$ & 6.040~$10^{-5}$ \\
187.4  & 65799 & 4.37 & 6.563 & 14.80 & -4.57 & 3799  & 0.50 & 0.751 & 0.249 & 1.422~$10^{-5}$ & 2.014~$10^{-5}$ & 6.024~$10^{-5}$ \\
170.9  & 64269 & 4.29 & 6.563 & 15.51 & -4.36 & 3402  & 1.00 & 0.715 & 0.285 & 1.709~$10^{-6}$ & 8.685~$10^{-5}$ & 4.078~$10^{-7}$ \\
146.4  & 65803 & 4.27 & 6.556 & 14.67 & -4.30 & 3119  & 1.50 & 0.572 & 0.428 & 1.624~$10^{-6}$ & 8.696~$10^{-5}$ & 3.963~$10^{-7}$ \\
120.0  & 69940 & 4.30 & 6.546 & 12.85 & -4.20 & 2942  & 2.00 & 0.365 & 0.635 & 1.730~$10^{-6}$ & 8.685~$10^{-5}$ & 3.752~$10^{-7}$ \\
83.4   & 82277 & 4.46 & 6.510 & 8.90 & -4.10 & 2797  & 2.50 & 0.094 & 0.906 & 2.051~$10^{-6}$ & 8.653~$10^{-5}$ & 3.233~$10^{-7}$ \\
\hline                                                
250 \msun & & & & & & & & & \\                       
\hline                                                                              
247.4  & 68733 & 4.43 & 6.700 & 15.87 & -4.49 & 4128  & 0.03 & 0.751 & 0.249 & 2.491~$10^{-5}$ & 7.295~$10^{-6}$ & 6.040~$10^{-5}$ \\
232.3  & 66802 & 4.36 & 6.693 & 16.68 & -4.37 & 3788  & 0.50 & 0.751 & 0.249 & 2.491~$10^{-5}$ & 7.295~$10^{-6}$ & 6.040~$10^{-5}$ \\
207.6  & 66804 & 4.32 & 6.684 & 16.50 & -4.25 & 3498  & 1.00 & 0.664 & 0.336 & 1.753~$10^{-6}$ & 8.681~$10^{-5}$ & 3.950~$10^{-7}$ \\
175.8  & 68824 & 4.31 & 6.674 & 15.36 & -4.16 & 3231  & 1.50 & 0.504 & 0.496 & 1.783~$10^{-6}$ & 8.679~$10^{-5}$ & 3.818~$10^{-7}$ \\
137.6  & 73957 & 4.35 & 6.652 & 12.97 & -4.07 & 3012  & 2.00 & 0.284 & 0.716 & 1.897~$10^{-6}$ & 8.668~$10^{-5}$ & 3.570~$10^{-7}$ \\
99.4   & 84863 & 4.49 & 6.610 & 9.38 & -3.99 & 2874  & 2.40 & 0.072 & 0.928 & 2.128~$10^{-6}$ & 8.645~$10^{-5}$ & 3.094~$10^{-7}$ \\
\hline                                                
300 \msun & &  & & & & & & & \\
\hline                                                            
299.5  & 69809 & 4.43 & 6.810 & 17.47 & -4.33 & 4160  & 0.02 & 0.751 & 0.249 & 2.491~$10^{-5}$ & 7.295~$10^{-6}$ & 6.040~$10^{-5}$ \\
273.6  & 67652 & 4.35 & 6.796 & 18.30 & -4.22 & 3731  & 0.50 & 0.744 & 0.256 & 1.777~$10^{-6}$ & 8.678~$10^{-5}$ & 3.931~$10^{-7}$ \\
241.1  & 68494 & 4.33 & 6.783 & 17.59 & -4.14 & 3502  & 1.00 & 0.625 & 0.375 & 1.796~$10^{-6}$ & 8.677~$10^{-5}$ & 3.852~$10^{-7}$ \\
197.4  & 70950 & 4.32 & 6.767 & 16.09 & -4.06 & 3158  & 1.50 & 0.453 & 0.547 & 1.847~$10^{-6}$ & 8.672~$10^{-5}$ & 3.708~$10^{-7}$ \\
151.1  & 76698 & 4.37 & 6.736 & 13.29 & -3.98 & 2960  & 2.00 & 0.226 & 0.774 & 1.969~$10^{-6}$ & 8.661~$10^{-5}$ & 3.429~$10^{-7}$ \\
117.8  & 85051 & 4.48 & 6.698 & 10.34    & -3.92 & 2888  & 2.30 & -     & 0.931 & 2.157~$10^{-6}$ & 8.642~$10^{-5}$ & 3.040~$10^{-7}$ \\
\hline
\end{tabular}
\tablefoot{Columns are mass of the corresponding model according to stellar evolution rounded up to the first decimal, effective temperature, surface gravity, luminosity, radius, mass-loss rate, wind terminal velocity, age, and surface abundances of H, He, C, N, and O given in mass fraction. For each mass, the lines of this table correspond to the characteristics of the models marked as filled circles in Fig.~\ref{hrd}).}
\end{center}
\end{table*}

\normalsize

\begin{table*}[ht]
\begin{center}
  \caption{Parameters of the atmosphere models with mass loss rates scaled linearly with Z and at Z~=~0.01~\zsun.} \label{tab_grz_z0p01}
  \small
\begin{tabular}{lccccccccccccc}
\hline
  M & $T_{\rm eff}$  &  $\log g$ & $\log(L/L_{\odot})$ & R& $\log \dot{M}$ & $\varv_{\infty}$ & age & H & He & C & N & O \\    
  \msun & K          &                   &          &  R$\odot$ &            & \kms       & Myr  &  & & & & \\
\hline
150 \msun  & & & & & & & & & \\
\hline
150.5  & 66051 & 4.46 & 6.385 & 11.96 & -6.59 & 4138  & 0.01 & 0.751 & 0.249 & 2.491~$10^{-5}$ & 7.295~$10^{-6}$ & 6.040~$10^{-5}$ \\
149.9  & 64726 & 4.40 & 6.408 & 12.79 & -6.48 & 3903 & 0.50 & 0.751 & 0.249 & 2.491~$10^{-5}$ & 7.295~$10^{-6}$ & 6.040~$10^{-5}$ \\
150.2  & 62531 & 4.31 & 6.409 & 14.20 & -6.31 & 3593 & 1.00 & 0.751 & 0.249 & 2.491~$10^{-5}$ & 7.295~$10^{-6}$ & 6.040~$10^{-5}$ \\
150.5  & 58739 & 4.17 & 6.471 & 16.70 & -6.14 & 3194 & 1.50 & 0.751 & 0.249 & 2.491~$10^{-5}$ & 7.295~$10^{-6}$ & 6.040~$10^{-5}$ \\
148.7 & 51769 & 3.91 & 6.506 & 22.39 & -5.95 & 2586 & 2.00 & 0.751 & 0.249 & 2.491~$10^{-5}$ & 7.295~$10^{-6}$ & 6.040~$10^{-5}$ \\
149.7 & 37868 & 3.33 & 6.546 & 43.71 & -5.73 & 1738	& 2.50 & 0.249 & 2.491~$10^{-5}$ & 7.295~$10^{-6}$ & 6.040~$10^{-5}$ \\
\hline                                                
200 \msun & & & & & & & & & \\                       
\hline                                                                              
201.9  & 67673 & 4.45 & 6.564 & 14.01 & -6.31 & 4206  & 0.01 & 0.751 & 0.249 & 2.491~$10^{-5}$ & 7.295~$10^{-6}$ & 6.040~$10^{-5}$ \\
200.7  & 66125 & 4.38 & 6.592 & 15.14 & -6.17 & 3894  & 0.50 & 0.751 & 0.249 & 2.491~$10^{-5}$ & 7.295~$10^{-6}$ & 6.040~$10^{-5}$ \\
199.4  & 63209 & 4.27 & 6.620 & 17.13 & -6.01 & 3494  & 1.00 & 0.751 & 0.249 & 2.471~$10^{-5}$ & 7.698~$10^{-6}$ & 6.035~$10^{-5}$ \\
197.9  & 57802 & 4.08 & 6.652 & 21.24 & -5.84 & 2954  & 1.50 & 0.751 & 0.249 & 2.471~$10^{-5}$ & 7.698~$10^{-6}$ & 6.035~$10^{-5}$ \\
197.0  & 47388 & 3.70 & 6.685 & 32.82 & -5.66 & 2210  & 2.00 & 0.751 & 0.249 & 2.471~$10^{-5}$ & 7.698~$10^{-6}$ & 6.035~$10^{-5}$ \\
198.3  & 35520 & 3.18 & 6.707 & 59.91 & -5.53 & 2756  & 2.30 & 0.751 & 0.249 & 2.467~$10^{-5}$ & 7.750~$10^{-6}$ & 6.035~$10^{-5}$ \\
\hline                                                
250 \msun & & & & & & & & & \\                       
\hline                                                                              
251.6  & 68911 & 4.44 & 6.701 & 15.82 & -6.10 & 4200  & 0.02 & 0.751 & 0.249 & 2.491~$10^{-5}$ & 7.295~$10^{-6}$ & 6.040~$10^{-5}$ \\
251.9  & 67088 & 4.37 & 6.725 & 17.16 & -5.97 & 3904  & 0.50 & 0.751 & 0.249 & 2.491~$10^{-5}$ & 7.295~$10^{-6}$ & 6.040~$10^{-5}$ \\
247.6  & 63509 & 4.24 & 6.753 & 19.76 & -5.83 & 3403  & 1.00 & 0.751 & 0.249 & 2.491~$10^{-5}$ & 7.295~$10^{-6}$ & 6.040~$10^{-5}$ \\
249.0  & 56482 & 4.01 & 6.781 & 25.80 & -5.67 & 2832  & 1.50 & 0.751 & 0.249 & 2.491~$10^{-5}$ & 7.295~$10^{-6}$ & 6.040~$10^{-5}$ \\
249.3  & 41849 & 3.46 & 6.811 & 48.67 & -5.50 & 1924  & 2.00 & 0.751 & 0.249 & 2.491~$10^{-5}$ & 7.295~$10^{-6}$ & 6.040~$10^{-5}$ \\
\hline                                                
300 \msun & &  & & & & & & & \\
\hline                                                            
298.1  & 69895 & 4.43 & 6.810 & 17.42 & -5.94 & 4141  & 0.02 & 0.751 & 0.249 & 2.491~$10^{-5}$ & 7.295~$10^{-6}$ & 6.040~$10^{-5}$ \\
297.1  & 67656 & 4.35 & 6.832 & 19.07 & -5.83 & 3796  & 0.50 & 0.751 & 0.249 & 2.491~$10^{-5}$ & 7.295~$10^{-6}$ & 6.040~$10^{-5}$ \\
301.7  & 63487 & 4.22 & 6.858 & 22.32 & -5.69 & 3403  & 1.00 & 0.751 & 0.249 & 2.491~$10^{-5}$ & 7.295~$10^{-6}$ & 6.040~$10^{-5}$ \\
296.9  & 55111 & 3.94 & 6.885 & 30.56 & -5.54 & 2663  & 1.50 & 0.751 & 0.249 & 2.491~$10^{-5}$ & 7.295~$10^{-6}$ & 6.040~$10^{-5}$ \\
294.1  & 36252 & 3.18 & 6.914 & 73.01 & -5.37 & 1555  & 2.00 & 0.751 & 0.249 & 2.491~$10^{-5}$ & 7.295~$10^{-6}$ & 6.040~$10^{-5}$ \\
\hline
\end{tabular}
\tablefoot{Columns are mass of the corresponding model according to stellar evolution rounded up to the first decimal, effective temperature, surface gravity, luminosity, radius, mass-loss rate, wind terminal velocity, age, and surface abundances of H, He, C, N, and O given in mass fraction. For each mass, the lines of this table correspond to the characteristics of the models marked as filled circles in Fig.~\ref{hrd}).}
\end{center}
\end{table*}

\normalsize

\begin{table*}[ht]
\begin{center}
  \caption{Parameters of the atmosphere models for the advanced phases tested in Sect.~\ref{s_postMS}.} \label{tab_evolved}
  \small
\begin{tabular}{lccccccccccccc}
\hline
  M & $T_{\rm eff}$  &  $\log g$ & $\log(L/L_{\odot})$ & R& $\log \dot{M}$ & $\varv_{\infty}$ & age & H & He & C & N & O \\    
  \msun & K          &                   &          &  R$\odot$ &            & \kms       & Myr  &  & & & & \\
\hline
 & & & Z = 0.1~\zsun\ & & & & & & \\
\hline
78.7  & 60571 & 4.01 & 6.403 & 15.20 & -4.23 & 2254  & 2.55 & 0.236 & 0.763 & 1.484~$10^{-5}$ & 8.308~$10^{-4}$ & 4.337~$10^{-6}$ \\
66.7  & 103323 & 4.83 & 6.439 & 5.20 & -3.94 & 4546 & 2.75 & 0.097 & 0.902 & 1.629~$10^{-5}$ & 8.295~$10^{-4}$ & 3.945~$10^{-6}$ \\
81.8  & 102201 & 4.78 & 6.558 & 6.10 & -3.84 & 3013 & 2.55 & 0.057 & 0.942 & 1.735~$10^{-5}$ & 8.285~$10^{-4}$ & 3.659~$10^{-6}$ \\
\hline         
 & & & Z = 0.01~\zsun\ & & & & & & \\
\hline
77.0  & 72384 & 4.32 & 6.393 & 10.10 & -4.23 & 2707  & 2.55 & 0.220 & 0.780 & 1.474~$10^{-6}$ & 8.716~$10^{-5}$ & 3.631~$10^{-7}$ \\
61.8  & 134223 & 5.28 & 6.410 & 3.00 & -3.99 & 3993 & 2.75 & 0.064 & 0.936 & 9.807~$10^{-7}$ & 8.776~$10^{-5}$ & 3.215~$10^{-7}$ \\
79.0  & 135932 & 5.28 & 6.538 & 3.37 & -3.88 & 4060 & 2.55 & 0.052 & 0.948 & 2.148~$10^{-6}$ & 8.643~$10^{-5}$ & 3.047~$10^{-7}$ \\
\end{tabular}
\tablefoot{Columns are mass of the corresponding model according to stellar evolution rounded up to the first decimal, effective temperature, surface gravity, luminosity, radius, mass-loss rate, wind terminal velocity, age, and surface abundances of H, He, C, N, and O given in mass fraction. For each mass, the lines of this table correspond to the characteristics of the models marked as crosses in Fig.~\ref{hrd}).}
\end{center}
\end{table*}

\normalsize

\FloatBarrier
\newpage

\section{Synthetic spectra for additional 200~\msun\ models}
\label{s_ap}

In this Section we show the UV and optical spectra of all the individual models of VMS with an initial mass of 200~\msun. They are complementary to those displayed in Fig.~\ref{sv200_z0p1}. The spectra for all masses are available at the POLLUX database (\url{https://pollux.oreme.org/}).

\begin{figure*}[h]
\centering
\includegraphics[width=0.49\textwidth]{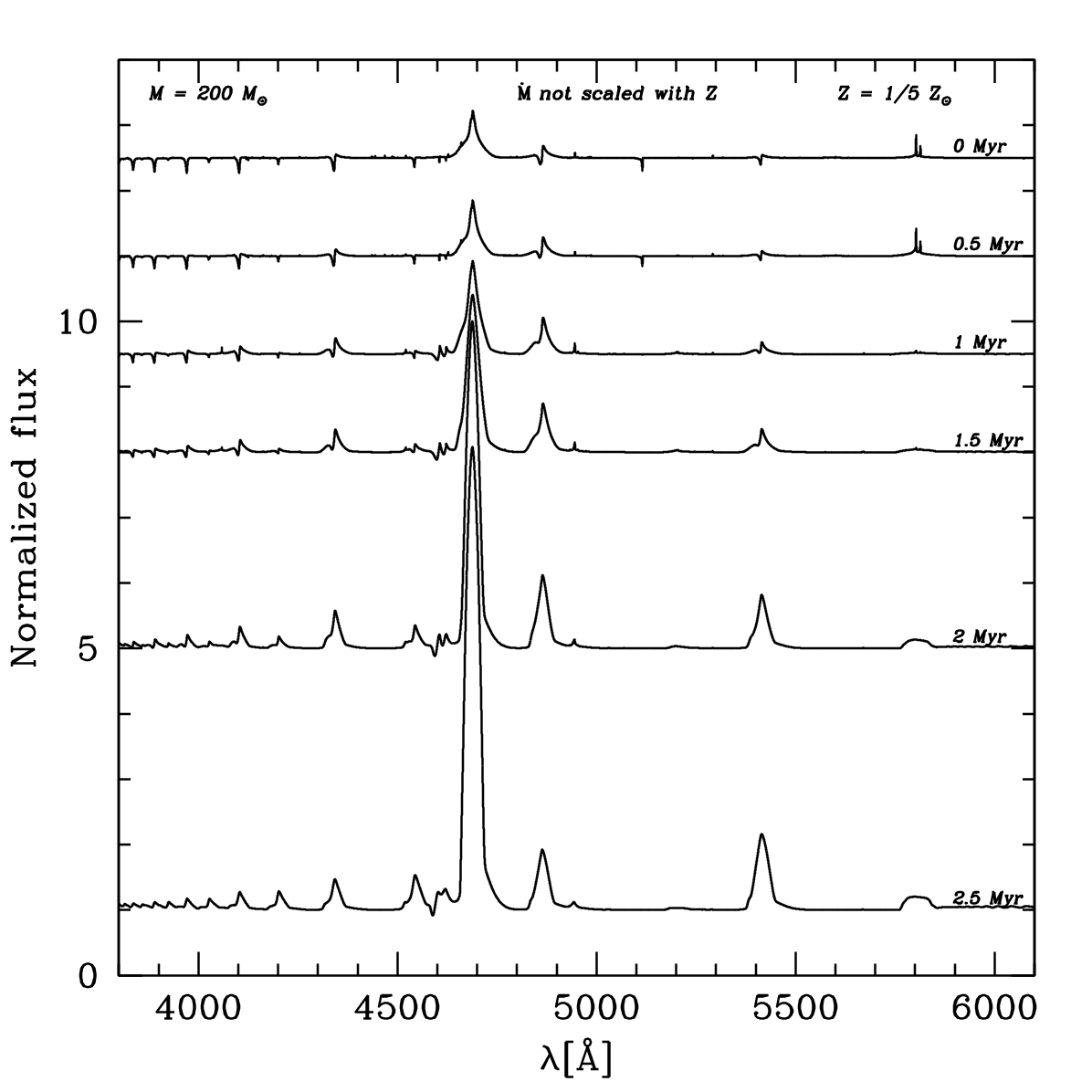}
\includegraphics[width=0.49\textwidth]{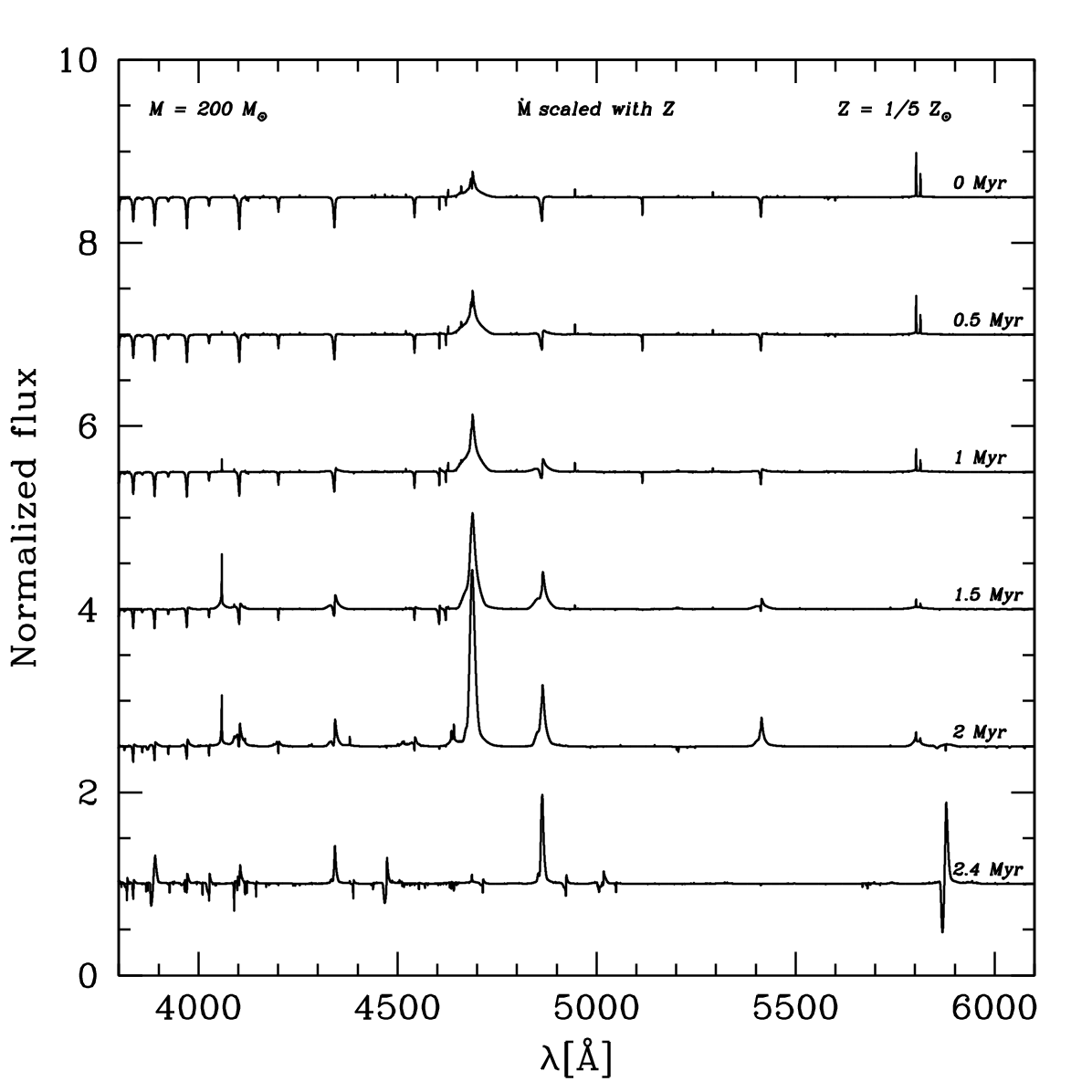}\\
\includegraphics[width=0.49\textwidth]{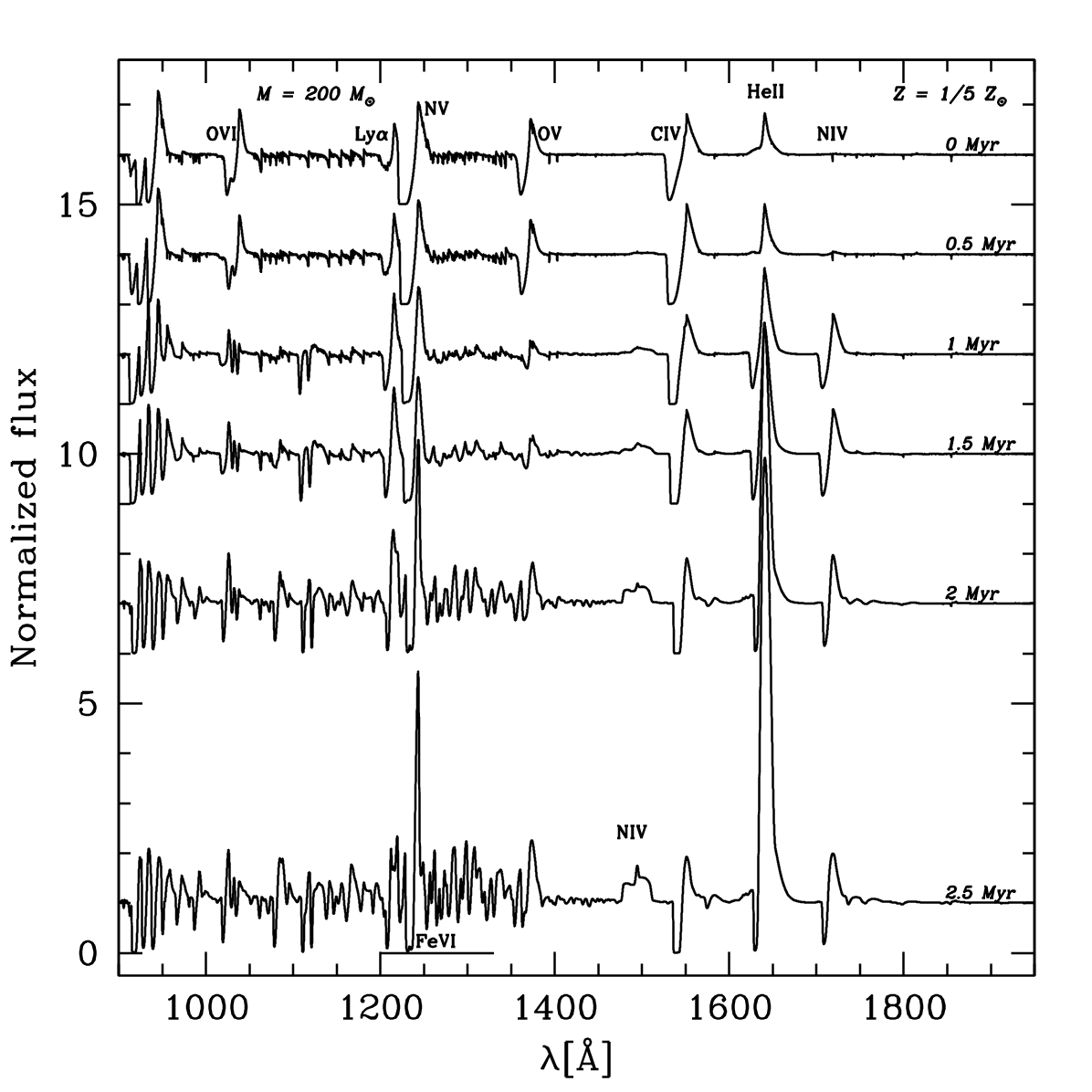}
\includegraphics[width=0.49\textwidth]{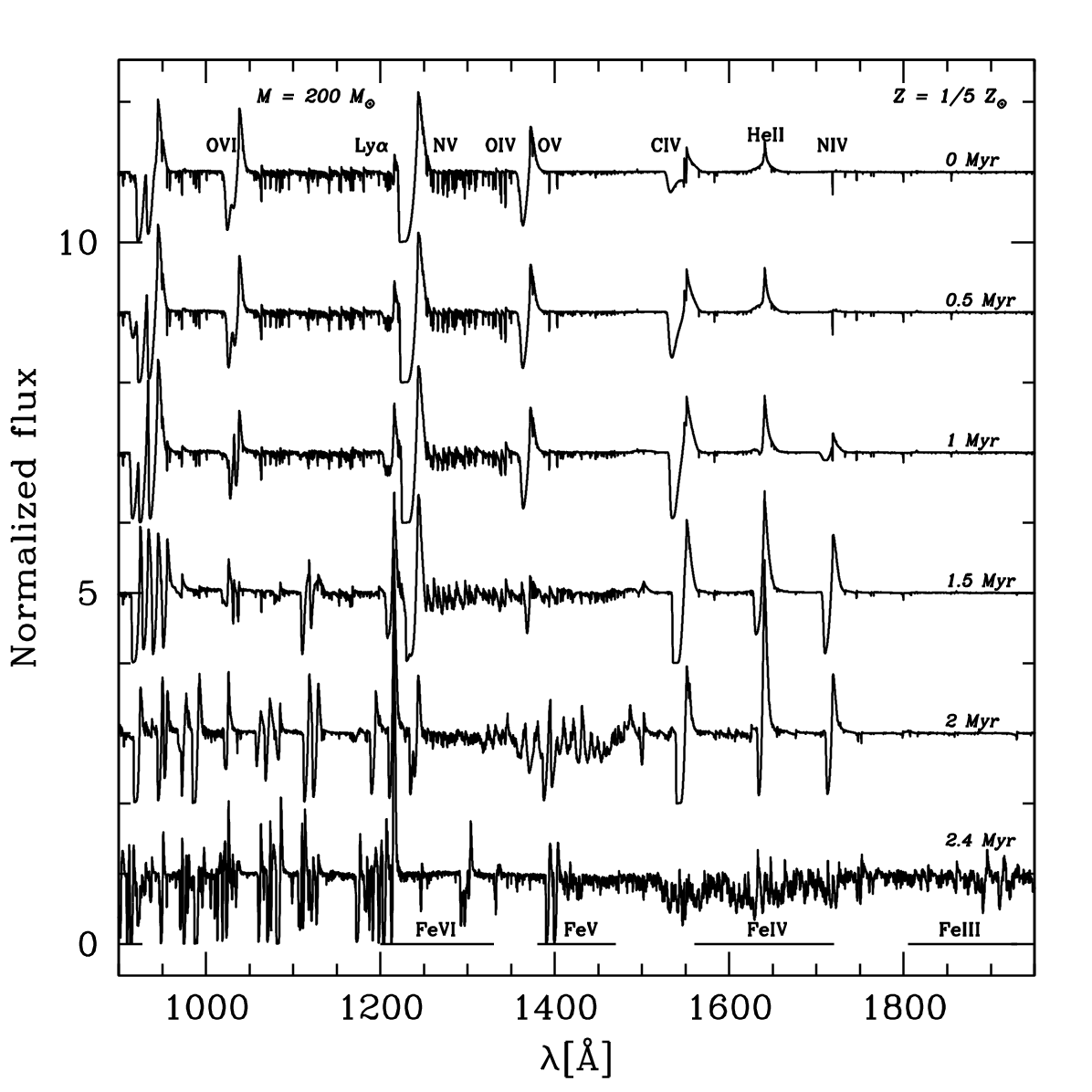}
\caption{Synthetic UV (bottom) and optical (top) spectra of the 200~\msun\ models at Z~=~0.2~\zsun. The left (right) panels correspond to models without (with) metallicity scaling of mass loss rates.}
\label{sv200_zsmc}
\end{figure*}

\begin{figure*}[h]
\centering
\includegraphics[width=0.49\textwidth]{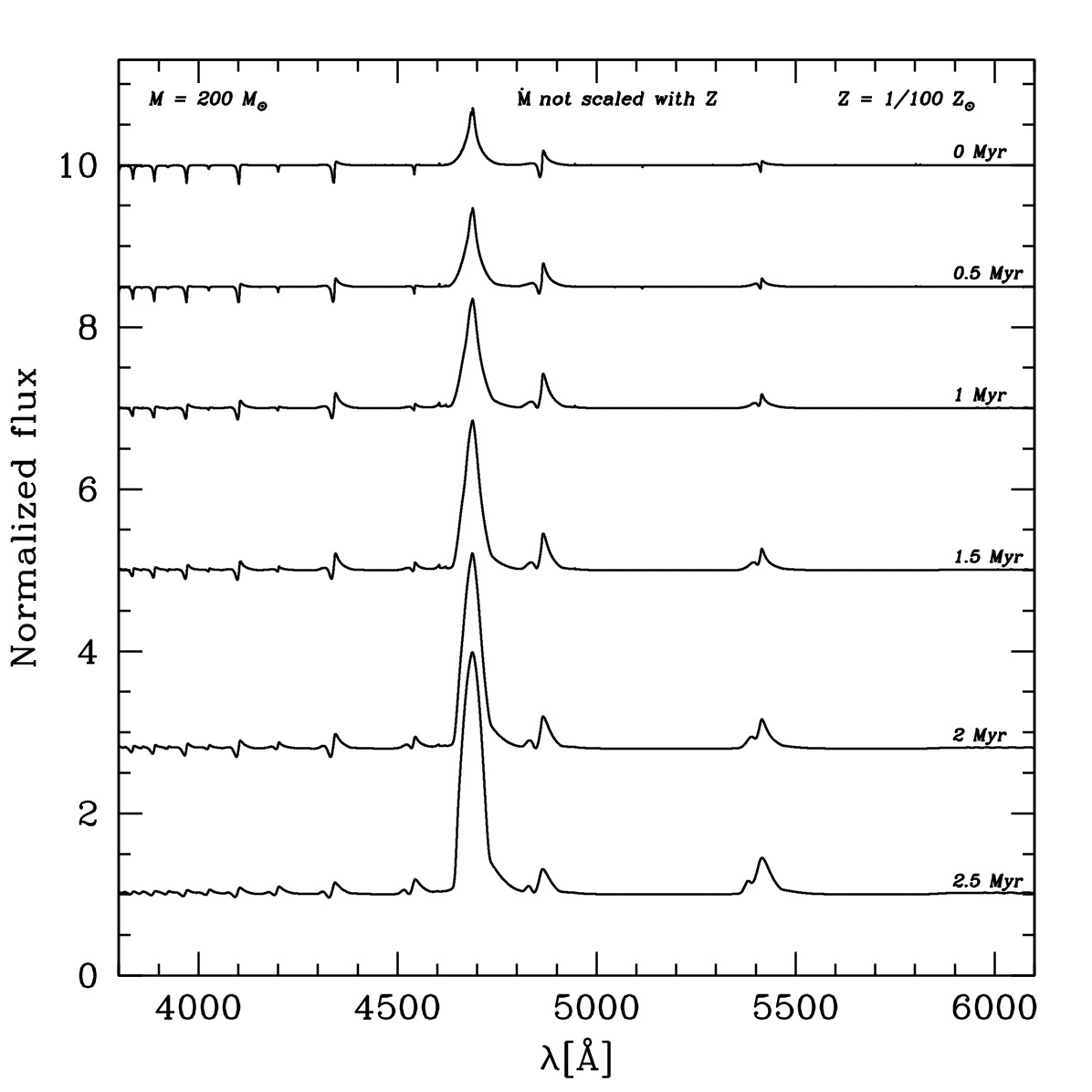}
\includegraphics[width=0.49\textwidth]{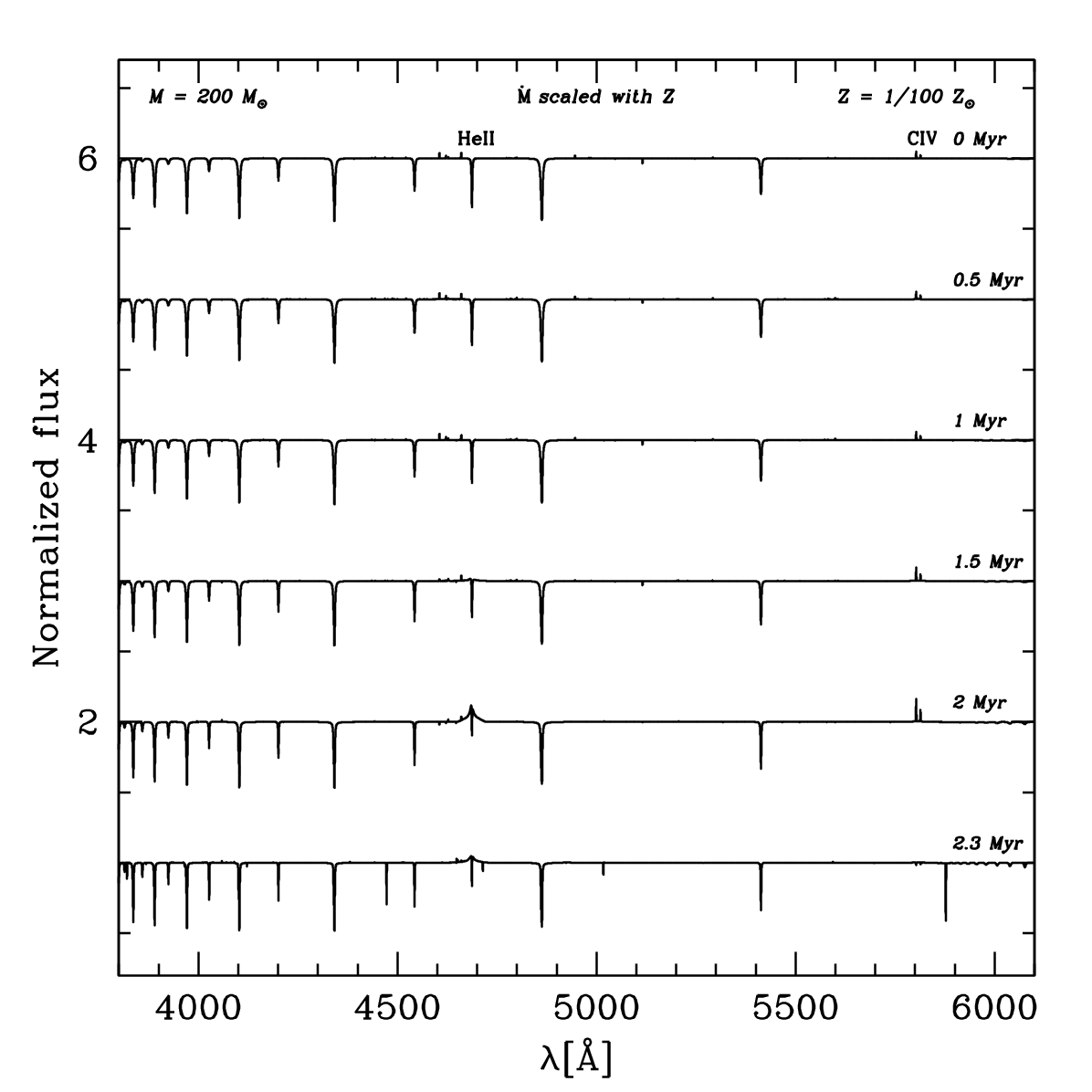}\\
\includegraphics[width=0.49\textwidth]{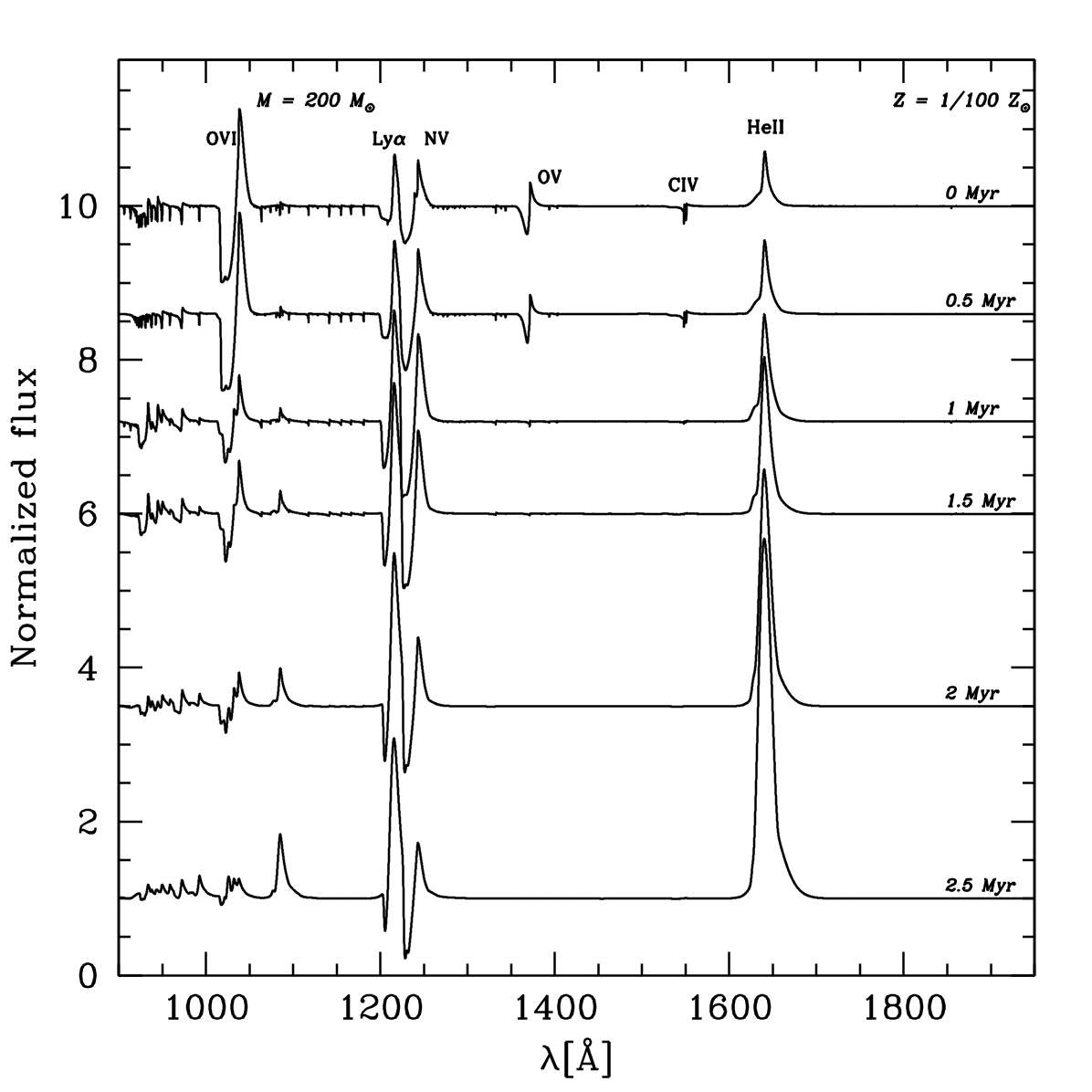}
\includegraphics[width=0.49\textwidth]{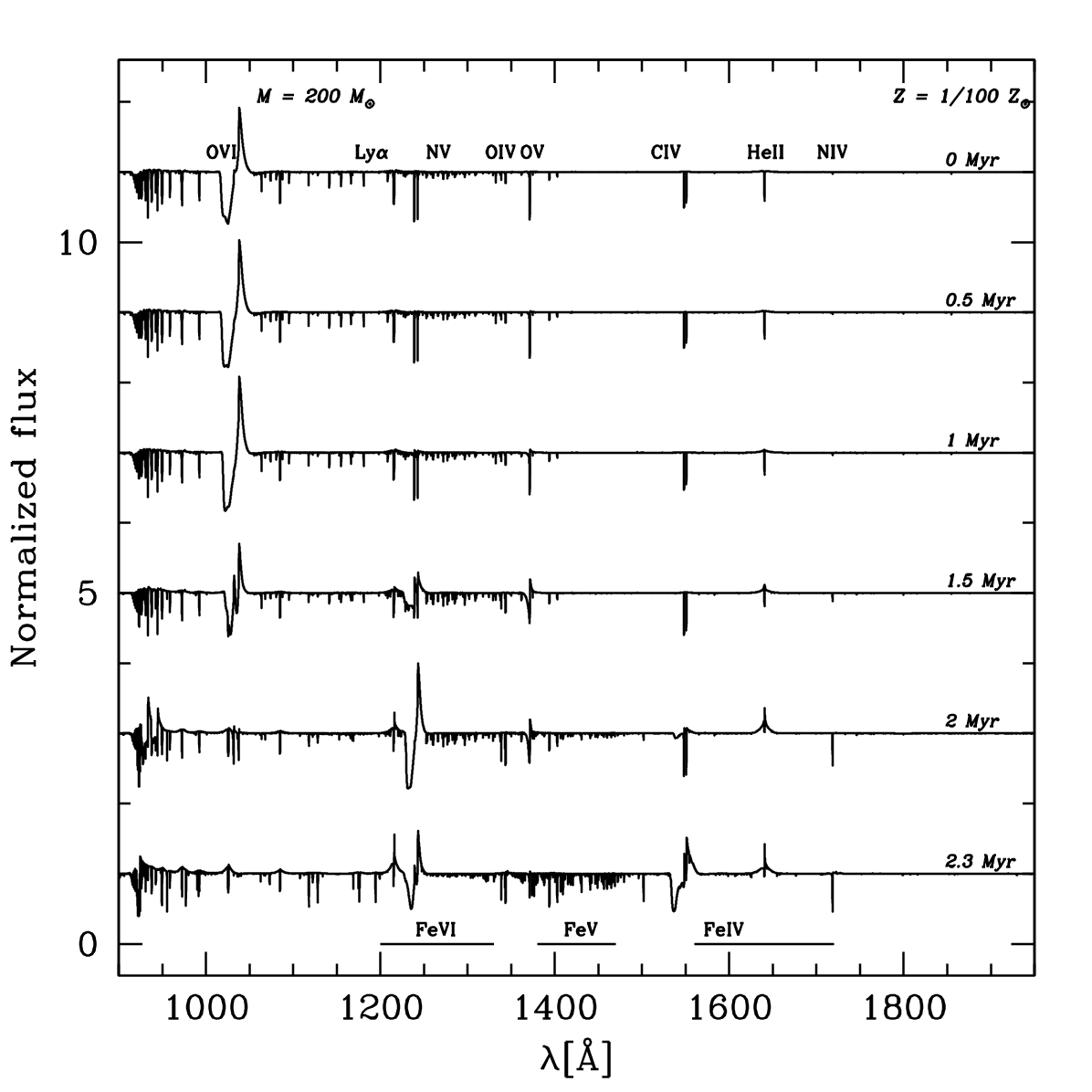}
\caption{Synthetic UV (bottom) and optical (top) spectra of the 200~\msun\ models at Z~=~0.01~\zsun. The left (right) panels correspond to models without (with) metallicity scaling of mass loss rates.}
\label{sv200_z0p01}
\end{figure*}

\FloatBarrier
\newpage

\end{appendix}

\end{document}